\documentclass[twocolumn,  % Format : preprint, twocolumn
 showpacs,  % Pacs : showpacs, noshowpacs
 preprintnumbers,  % Preprint: preprintnumbers,
 aps,  % Society: ...
 prd,  % Journal Style : pra, prb, prc, prd, pre,
 a4paper,  % Size : a4paper, ...
 superscriptaddress,  % Affiliation (Title) : groupedaddress,
 nofootinbib,  % Footnote: footinbib, nofootinbib
 tightenlines,  % Remove additional spaces in a line
 floats,floatfix  % Floating pictures and tables
 ]{revtex4}

\usepackage{graphicx}
\usepackage{subfigure}
\usepackage{latexsym}
\usepackage{amsmath,amssymb}        % amssymb includes amsfonts
\newcommand{\gsim}{\mbox{\raisebox{-1.ex}{$\stackrel
     {\textstyle>}{\textstyle\sim}$}}}
\newcommand{\lsim}{\mbox{\raisebox{-1.ex}{$\stackrel
     {\textstyle<}{\textstyle \sim}$}}}

\newcommand{\bk}{{\bf k}}

\newcommand{\T}{{\rm T}}
\newcommand{\beq}{\begin{equation}}
\newcommand{\eeq}{\end{equation}}
\newcommand{\beqa}{\begin{eqnarray}}
\newcommand{\eeqa}{\end{eqnarray}}
\newcommand{\bea}{\begin{array}}
\newcommand{\ena}{\end{array}}

\newcommand{\rd}{{\rm d}}

\input colordvi

\def\R{{\cal R}}
\def\S{{\cal S}}
\def\be{\begin{equation}}
\def\ee{\end{equation}}
\def\bea{\begin{eqnarray}}
\def\eea{\end{eqnarray}}

\def\4pig{\sfrac{4\pi G}{c^{4}}}

\def\hsp5{\hspace{5mm}}
\newcommand{\sfrac}[2]{{\textstyle{#1\over#2}}}
\def\case#1/#2{\textstyle\frac{#1}{#2}}

\begin{document}

\title{Inflation Dynamics and Reheating}

\author{Bruce A. Bassett}
\email{bruce.bassett@port.ac.uk}
\affiliation{Department of Physics, Kyoto University, Kyoto, Japan}
\affiliation{Institute of Cosmology and Gravitation, University of
Portsmouth,
Mercantile House, Portsmouth PO1 2EG, UK}

\author{Shinji Tsujikawa}
\email{shinji@nat.gunma-ct.ac.jp}
\affiliation{Department of Physics, Gunma National College of
Technology, Gunma 371-8530, Japan}
\email{shinji@nat.gunma-ct.ac.jp}

 \author{David Wands}
  \email{david.wands@port.ac.uk}
  \affiliation{Institute of Cosmology and Gravitation,
University of Portsmouth, Mercantile House, Portsmouth PO1 2EG, UK}

% v2*
%\date{\today}

\begin{abstract}
We review the theory of inflation with single and multiple fields paying
particular attention to the dynamics of adiabatic and entropy/isocurvature
perturbations which provide the primary means of testing inflationary
models.  We review the theory and phenomenology of reheating and
preheating after inflation providing a unified discussion of both the
gravitational and nongravitational features of multi-field inflation.
In addition we cover inflation in theories with extra dimensions and
models such as the curvaton scenario and modulated reheating which provide
alternative ways of generating large-scale density perturbations. Finally
we discuss the interesting observational implications that can result from
adiabatic-isocurvature correlations and non-Gaussianity.
\end{abstract}

\pacs{98.80.Cq
% v2*
 \hfill astro-ph/0507632}
\maketitle

%%%%%%%%%%%%%%%%%%%%%%%%%%
\section{Introduction}
%%%%%%%%%%%%%%%%%%%%%%%%%%

Inflation was
introduced \cite{Sta80,Guth81,Sato81p1,Sato81p2,Kazanas80}
as a way of addressing pressing problems that
were eating away at the foundations of the otherwise rather successful
big-bang model. It is a very flexible paradigm, based squarely in
semi-classical physics, and has provided a sturdy foundation linking
the classical cosmos and the quantum gravity
world \cite{Lindebook,LLbook,Kolbbook}.

Inflation can be viewed in many different ways. One approach is to
argue that
% v2 dw: COMMENTED OUT, like epicycles in the era before elliptical orbits,
inflationary models (of which there are hundreds) provide a convenient
method of parameterising the early universe but that, because they are
fundamentally semi-classical, are unlikely to be a true description of
the physics underlying the very early universe.
The other, probably more common, approach is to argue that an
inflationary phase did indeed occur at some stage in the early
universe and that the source of inflation is a scalar field whose
identity may be found by considering one of the extensions of the
standard model based on grand unified theories (GUT's), supergravity
or string theory. In the latter view, we can use inflation as a way
both to understand features of quantum gravity/string theory and of
particle physics beyond the standard model.

Taking this latter view, it is extremely natural to consider inflation
with many fields. As a simple example, consider a grand unified theory
(GUT) based on the group $SO(10)$. Such a GUT has
no pretensions to be a theory of everything
and yet it already implies the existence
of large numbers (of order $100$) scalar Higgs fields. Similarly
supersymmetry requires the existence of large numbers of
superpartners \cite{LRreview}
and string theory rather naturally has dynamical moduli fields
corresponding to the geometrical characteristics of compactified
dimensions \cite{LWCreview}.
If scalar fields are natural sources of inflation, then
modern particle physics is the perfect supplier.

The inflationary paradigm not only provides an excellent way to
solve flatness and horizon problems but also generates density
perturbations as seeds for large-scale structure in the universe
\cite{Mukhanov81,Guth82,Hawking82,Sta82}. Quantum fluctuations of
the field responsible for inflation -- called
% v2* typo
the {\it inflaton} -- are stretched on large scales by the
accelerated expansion. In the simplest version of the single-field
scenario the fluctuations are ``frozen'' after the scale of
perturbations leaves the Hubble radius during inflation. Long
after the inflation ends, the perturbations cross inside the
Hubble radius again. Thus inflation provides a causal mechanism
for the origin of large-scale structure in the universe. An
important prediction of inflation is that density perturbations
generally exhibit nearly scale-invariant spectra. This prediction
can be directly tested by the measurement of the temperature
anisotropies in Cosmic Microwave Background (CMB). Remarkably the
anisotropies observed by the Cosmic Background Explorer (COBE) in
1992 showed nearly scale-invariant spectra. Fortunately, all
existing and constantly accumulating data including WMAP
\cite{WMAP,Peiris03}, SDSS \cite{Tegmark03,Tegmarkdata} and 2dF
\cite{2dF} have continued to confirm the main predictions of the
inflationary paradigm within observational errors. We live in a
golden age for cosmology in which the physics of the early
universe can be probed from high-precision observations.

Recent progress in constructing particle-physics models of inflation
has shown us that a key question is ``how many light fields exist
during inflation?'' Here ``light'' is measured relative to the Hubble
constant (which has dimensions of mass). If there is only one light
field (typically required to get slow-roll inflation in the first
place), then inflation is effectively single-field dominated and the
cosmological consequences are rather well-understood. In the case of
multiple light fields, the situation is significantly more complicated
since the fields may interact and between each light field there will
typically be a dynamically important entropy/isocurvature perturbation
(we will use these two terms interchangeably).

Further, as the fields evolve, their effective mass can change,
renormalised by the expectation values of other fields. Since the
spectrum of fluctuations associated with any given perturbation mode
depend on its effective mass, there is a rich phenomenology of
possible effects associated with time-dependent effective masses.

In this review we will lay out the foundations of inflation and
cosmological perturbation theory appropriate for application to cases
involving many relevant fields. Our main aim is to provide the reader
with a unified framework and set of tools to begin practical
application in inflationary cosmology.
This review is complementary to the many excellent review on related
topics, given in
Refs.~\cite{Robert85,CKLLreview,Kolbbook,Liddle93,LLbook,LLKCreview,Lindebook,LRreview,Riotto02,NP91,Gio05}.

Our paper is organised as follows.
In Sec.~II we introduce the inflationary paradigm as a way to solve
several cosmological problems associated with standard big bang
cosmology. Inflationary models are classified into four different
types. In Sec.~III we review cosmological perturbation theory
using the gauge-invariant formalism.
Sec.~IV is devoted to the spectra of scalar and tensor
perturbations generated in single-field inflation.
In Sec.~V we present observational constraints on single-field
inflation from CMB and galaxy redshift surveys.
In Sec.~VI we review density perturbations generated in
higher-dimensional models including braneworld, pre-big-bang
and ekpyrotic/cyclic cosmologies.

In Sec.~VII the definition of adiabatic and entropy perturbations is
given with the field space rotation and we show how the correlation
between adiabatic and entropy perturbations emerges in the context of
two-field inflation. In Sec.~VIII we present general features in the
CMB from correlations.  In Sec.~IX we explain the elementary theory of
reheating after inflation.  Sec.~X is devoted to preheating in which
particles coupled to the inflaton are resonantly amplified by
parametric resonance.  In Sec.~XI we discuss the evolution of metric
perturbations during preheating and possible consequences from it.  We
review the curvaton scenario in Sec.~XII and the modulated reheating
scenario in Sec.~XIII to generate large-scale density perturbations as
alternative models of inflation. Summary and future outlook are given
in the final section.

%%%%%%%%%%%%%%%%%

%%%%%%%%%%%%%%%%%%%%%%%%%%%%%%%%%%%%%%%%
\section{Dynamics of inflation}
%%%%%%%%%%%%%%%%%%%%%%%%%%%%%%%%%%%%%%%%

\subsection{The standard big-bang cosmology}

The standard big-bang cosmology is based upon the cosmological
principle \cite{LLbook}, which requires that the universe is
homogeneous and isotropic on averaging over large volumes.  Then the metric takes
the Friedmann-Robertson-Walker (FRW) form:
\beqa
\hspace*{-1.5em} \rd s^2 &=& g_{\mu\nu}\rd x^{\mu}\rd x^{\nu} \nonumber \\
\hspace*{-1.5em} &=& -\rd t^2+a^2\left[\frac{\rd r^2}{1-Kr^2}+
r^2(\rd \theta^2+\sin^2\theta \rd \phi^2) \right].
\label{2_1_1}
\eeqa
Here $a(t)$ is the scale factor with $t$ being the cosmic time.  The
constant $K$ is the spatial curvature, where positive, zero, and
negative values correspond to closed, flat, and hyperbolic spatial sections
respectively.

The evolution of the universe is dependent on the material within it with a
key role played by the equation of state relating
the energy density $\rho(t)$ and the pressure $P(t)$.
For example we have
\beqa
P&=&\rho/3\,,~~~~~~{\rm radiation}\,, \\
P&=&0\,,~~~~~~~~~{\rm dust}\,.
\label{2_1_2}
\eeqa
The dynamical evolution of the universe is known once we
solve the Einstein equations of General Relativity:
\beqa
G_{\mu \nu} \equiv R_{\mu\nu}-\frac12 g_{\mu\nu} R=
8\pi G T_{\mu\nu}-\Lambda
g_{\mu\nu},
\label{2_1_3}
\eeqa
where $R_{\mu\nu}$, $R$, $T_{\mu\nu}$, and
$G$ are the Ricci tensor, Ricci scalar, energy-momentum tensor
and gravitational constant, respectively.  The Planck energy,
$m_{\rm pl}=1.2211 \times 10^{19}$\,GeV, is related to $G$ through
the relation $m_{\rm pl}=(\hbar c^5/G)^{1/2}$.  Here
$\hbar$ and $c$ are the Planck's constant and the speed of light,
respectively. Hereafter we use the units $\hbar=c=1$.
$\Lambda$ is the cosmological constant
originally introduced by Einstein to make the universe
static. In what follows we set the cosmological constant
to zero ($\Lambda=0$) unless otherwise stated,
preferring to include any non-zero vacuum energy density in the total
energy-momentum tensor.

{}From the Einstein equations (\ref{2_1_3})
for the background FRW metric (\ref{2_1_1}),
we obtain the field equations:
%They are related by  Bianchi identities and two of them
%are independent, which are given by \cite{Kolbbook}
%
\beqa
\label{2_1_4}
& &H^2=\frac{8\pi}{3m_{\rm pl}^2}\rho-\frac{K}{a^2} \,, \\
& &\frac{\ddot{a}}{a}=-\frac{4\pi}{3m_{\rm pl}^2} (\rho+3P) \,,
\label{2_1_6}
\eeqa
where a dot denotes the derivative with respect to $t$, and $H \equiv
\dot{a}/a$ is the Hubble expansion rate.  Eqs.~(\ref{2_1_4}) and
% v2 DW corrected eq.no.
(\ref{2_1_6})
are the Friedmann and Raychaudhuri equations,
respectively.  Combining these relations implies energy conservation,
\beqa
\dot{\rho}+3H(\rho+P)=0\,,
\label{2_1_5}
\eeqa
which is known as the continuity or fluid equation.

The Friedmann equation (\ref{2_1_4}) can be rewritten as
\beqa
\Omega-1=\frac{K}{a^2H^2}\,,
\label{2_1_7}
\eeqa
where
\beqa
\Omega \equiv \frac{\rho}{\rho_c},~~~{\rm with}
~~~\rho_c \equiv \frac{3H^2m_{\rm pl}^2}{8\pi}\,.
\label{2_1_8}
\eeqa
Here the density parameter $\Omega$ is the ratio of the energy
density to the critical density.
When the spatial geometry is flat ($K=0$; $\Omega=1$),
the solutions for Eqs.~(\ref{2_1_4}) and (\ref{2_1_5}) are
\beqa
{\rm Radiation}:~~a \propto t^{1/2}\,,~~~
\rho \propto a^{-4}\,, \\
{\rm Dust}:~~a \propto t^{2/3}\,,~~~\rho \propto a^{-3}\,.
\label{2_1_9}
\eeqa
In these simple cases, the universe exhibits a decelerated
expansion ($\ddot{a}<0$) as confirmed by Eq.~(\ref{2_1_6}).

%%%%%%%%%%%%%%%%%%%%%%%%%%%%%%%%%%%%%%%%%%%%%%%%%%%%%%%
\subsection{Problems of the standard big-bang cosmology}
%%%%%%%%%%%%%%%%%%%%%%%%%%%%%%%%%%%%%%%%%%%%%%%%%%%%%%%

\subsubsection{Flatness problem}

In the standard big-bang theory with $\ddot{a}<0$, the $a^2H^2
(=\dot{a}^2)$ term in Eq.~(\ref{2_1_7}) always decreases.  This means
that $\Omega$ tends to evolve away from unity with the expansion of
the universe.  However, since present observations suggest that
$\Omega$ is within a few percent of unity today
\cite{WMAP}, $\Omega$ is forced to be much closer to unity in the past.  For
example, we require $|\Omega-1| < {\cal O}(10^{-16})$ at the epoch of
nucleosynthesis and $|\Omega-1| < {\cal O}(10^{-64})$ at the Planck
epoch \cite{LLbook}.  This appears to be an extreme fine-tuning of
initial conditions.
Unless initial conditions are chosen very accurately, the
universe either collapses too soon, or expands too quickly before the
structure can be formed.  This is the so-called flatness problem.

\subsubsection{Horizon problem}

Consider a comoving wavelength, $\lambda$, and corresponding physical
wavelength, $a\lambda$, which at some time is inside the Hubble radius, $H^{-1}$
({\it i.e.,}~$a\lambda~\lsim~H^{-1}$).  The standard big-bang
decelerating cosmology is characterized by the cosmic evolution of $a \propto t^n$
with $0<n<1$.  In this case the physical wavelength
grows as $a\lambda
\propto t^n$, whereas the Hubble radius evolves as $H^{-1} \propto t$.
Therefore the physical wavelength becomes much smaller
than the Hubble radius at late times.
Conversely any finite comoving scale becomes
much larger than the Hubble scale at early times.  This means that a
causally connected region can only be a small fraction of the Hubble radius.

To be more precise, let us first define the particle horizon $D_H(t)$
which is the distance travelled by light since the beginning of the
universe, at time $t_*$,
\beqa
D_H(t)=a(t)d_H(t)\,,~~~{\rm with }
~~~d_H(t)=\int_{t_*}^t
\frac{\rd t'}{a(t')}\,.
\label{2_1_11}
\eeqa
Here $d_H(t)$ corresponds to the comoving particle horizon.  Setting
$t_*=0$, we find $D_H(t)=3t$ in the matter-dominant era and
$D_H(t)=2t$ in an early hot big bang. We observe photons in the cosmic
microwave background (CMB) which are last-scattered at the time of
decoupling.  The particle horizon at decoupling, $D_H(t_{\rm
  dec})=a(t_{\rm dec})d_H(t_{\rm dec})$, corresponds to the causally
connected region at that time.  The ratio of the comoving particle
horizon at decoupling, $d_H(t_{\rm dec})$, to the particle horizon
today, $d_H(t_0)$, can be estimated to be
\beqa
\frac{d_H(t_{\rm dec})}{d_H(t_0)} \approx \left(\frac{t_{\rm
dec}}{t_0}\right)^{1/3} \approx \left(\frac{10^5}{10^{10}}\right)^{1/3}
\approx
10^{-2}.
\label{2_1_12}
\eeqa
This implies that the causally connected regions at last scattering
are much smaller than the horizon size today. In fact causally
connected regions on the surface of last scattering corresponds to an
angle of order $1^{\circ}$.

% v2 DW: ADDED EXPLANATION RESPONDING TO REF1
This appears to be at odds with observations of
the cosmic microwave background (CMB) which
has the same temperature to high precision in all directions on the CMB sky.
% v2: DW and
Yet there is no way to establish thermal equilibrium if these points
were never been in causal contact before last-scattering.
This is the so-called horizon problem.

\subsubsection{The origin of large-scale structure in the universe}

Experiments which observe temperature anisotropies in the CMB
find that the amplitude of the anisotropies is
small and their power spectrum is close to scale-invariant on large
scales \cite{WMAP}.  These fluctuations are distributed on such a
large scale that it is impossible to generate them via causal processes
in a FRW metric in the time between the big
bang and the time of the last scattering. Hence, standard big-bang
models can neither explain the FRW metric nor explain
the deviations from FRW if a FRW background is assumed.

\subsubsection{Relic density problem}

The standard paradigm of modern particle physics is that physical laws were simpler in the early universe before gauge symmetries were broken.
The breaking of such symmetries leads to the production of many
unwanted relics such as monopoles, cosmic strings, and other topological
defects \cite{Lindebook}.
% v2 DW: sentence added
The existence of a finite horizon size leads to a maximum causal correlation length during any symmetry breaking transition and hence gives a lower bound on the density of defects.
In particular, any
% v2 DW: REFEREE1'S SUGGESTED CAVEAT
grand unified theory based on a simple Lie group
that includes the $U(1)$ of electromagnetism must produce monopoles.
String theories also predict supersymmetric
particles such as gravitinos, Kaluza-Klein particles, and weakly
coupled moduli fields.

If these massive particles exist in the early stages of the universe
then their energy densities decrease as a matter component ($\propto
a^{-3}$) once the temperature drops below their rest mass.  Since the
radiation energy density decreases $\propto a^{-4}$, these massive
relics
% v2 DW: REFEREE1'S RELATED COMMENT
if they are stable (or sufficiently long-lived)
could become the dominant matter in the early universe depending on their number density and therefore
contradict a variety of observations such as those of the light element
abundances.  This problem is known as the relic density
problem.

%%%%%%%%%%%%%%%%%%%%%%%%%%%%%%%%%%%%%%%%%%%
\subsection{Idea of inflationary cosmology}
%%%%%%%%%%%%%%%%%%%%%%%%%%%%%%%%%%%%%%%%%%%

The problems in the standard big bang cosmology lie in the fact that
the universe always exhibits decelerated expansion.  Let us assume
instead the existence of a stage in the early universe with an accelerated
expansion of the universe, {\it i.e.},
\beqa
\ddot{a}>0\,.
\label{2_2_1}
\eeqa
{}From the relation (\ref{2_1_6}) this gives the condition
\beqa
\rho+3P<0\,,
\label{2_2_2}
\eeqa
which corresponds to violating the strong energy condition.
The condition (\ref{2_2_1}) essentially means that $\dot{a}\,(=aH)$
increases during inflation and hence that the comoving Hubble radius,
$(aH)^{-1}$, decreases in the inflationary phase.  This property is
the key point to solve the cosmological puzzles in the standard
big-bang cosmology.

\subsubsection{Flatness problem}

Since the $a^2H^2$ term in Eq.~(\ref{2_1_7}) increases during
inflation, $\Omega$ is rapidly driven towards unity.  After the
inflationary period ends, the evolution of the universe is followed by
the conventional big-bang phase and $|\Omega-1|$ begins to increase
again.  But as long as the inflationary expansion lasts sufficiently
long and drives $\Omega$ very close to one, $\Omega$ will remain close
to unity even in the present epoch.

\subsubsection{Horizon problem}

Since the scale factor evolves approximately as $a \propto t^{n}$ with $n>1$ during
inflation, the physical wavelength, $a\lambda$, grows faster than the
Hubble radius, $H^{-1} (\propto t)$.  Therefore physical
wavelengths are pushed outside the Hubble radius during inflation which means that
causally connected regions can be much larger than the Hubble radius,
thus potentially solving the horizon problem.
% v2 DW: REFEREE1 REQUEST FOR CLARIFICATION
Formally the particle horizon, defined in Eq.~(\ref{2_1_11}) diverges as $a(t_*)\to0$
in an inflationary universe.

Of course the Hubble radius begins to grow faster than the physical
wavelength after inflation ends, during the subsequent radiation and
matter dominant eras.  In order to solve the horizon problem, it is
required that the following condition is satisfied for the comoving
particle horizon:
\beqa
\int_{t_*}^{t_{\rm dec}}\frac{\rd t}{a(t)} \gg \int_{t_{\rm
dec}}^{t_0}\frac{\rd t}{a(t)}\,.
\label{2_2_3}
\eeqa
This implies that the comoving distance that photons can travel before
decoupling needs to be much larger than that after the decoupling.  A
detailed calculation shows this is achieved when the universe expands
at least about $e^{70}$ times during inflation, or $70$ e-folds
of expansion \cite{Lindebook,LLbook,Riotto02}.

\subsubsection{The origin of the large-scale structure}

The fact that the Hubble rate, $H(t)$, is almost  constant
during inflation
means that it is possible to generate a nearly scale-invariant density
perturbation on large scales.  Since the scales of perturbations are
well within the Hubble radius in the early stage of inflation, causal
physics works to generate small quantum fluctuations.  On very small
scales we can neglect the cosmological expansion and perturbations can
be treated as fluctuations in flat spacetime. But after a scale is
pushed outside the Hubble radius ({\it i.e.}, the first Hubble radius
crossing) during inflation, we can no longer neglect the Hubble
expansion.

Fluctuations in a light field become over-damped on
long-wavelengths, leading to a squeezed state in phase-space, so that
the perturbations can effectively be described as classical on these
large scales. When the inflationary period ends, the evolution of the
universe follows the standard big-bang cosmology, and the
comoving Hubble radius begins to increase until the scales of
perturbations cross inside the Hubble radius again (the second Hubble
radius crossing).  The small perturbations imprinted during inflation
have amplitudes determined by the Hubble rate which is
approximately constant and hence leads to an almost scale-invariant
spectrum with constant amplitude on different scales.
In this way the inflationary paradigm naturally
provides a causal mechanism to generate the seeds of density
perturbations observed in the CMB anisotropies.

\subsubsection{Relic density problem}

During the inflationary phase ($\rho+3P<0$), the energy density of the
universe decreases very slowly.  For example, when the universe
evolves as $a \propto t^n$ with $n>1$, we have $H \propto t^{-1}
\propto a^{-1/n}$ and $\rho \propto a^{-2/n}$.  Meanwhile the energy
density of massive particles decreases much faster ($\propto a^{-3}$),
and these particles are red-shifted away during inflation, thereby solving
the monopole problem, as long as the symmetry breaking transition that produces the monopoles
occurs at least 20 or so e-foldings before the end of inflation.

We also have to worry about the possibility of producing
these unwanted particles {\it after} inflation.  In the process of
reheating followed by inflation, the energy of the universe can be
transferred to radiation or other light particles.  At this stage
unwanted particles must not be overproduced in order not to violate
the success of the standard cosmology such as nucleosynthesis.
Generally if the reheating temperature at the end of inflation is
sufficiently low, the thermal production of unwanted relics such as
gravitinos can be avoided \cite{KM95,moroi95}.

%%%%%%%%%%%%%%%%%%%%%%%%%%%%%%%
\subsection{Inflationary dynamics}
%%%%%%%%%%%%%%%%%%%%%%%%%%%%%%%

Scalar fields are fundamental ingredients in modern theories of particle physics.
We will consider a homogeneous single scalar field, $\phi$,
called the {\it inflaton}, whose potential energy can lead to the
accelerated expansion of the universe.
Neglecting spatial gradients, the energy density and the pressure
of the inflaton are given by
\beqa
\rho=\frac12\dot{\phi}^2+V(\phi)\,,
~~~P=\frac12\dot{\phi}^2-V(\phi)\,,
\label{rhoeq}
\eeqa
where $V(\phi)$ is the potential energy of the inflaton.  Substituting
Eq.~(\ref{rhoeq}) into Eqs.~(\ref{2_1_4}) and (\ref{2_1_5}),
we obtain
\beqa
\label{Heq}
& &H^2 = \frac{8\pi}{3m_{\rm pl}^2}
\left[\frac12 \dot{\phi}^2+V(\phi) \right]\,, \\
& &\ddot{\phi}+3H\dot{\phi}+V_\phi(\phi)=0\,,
\label{phieq}
\eeqa
where
$V_{\phi} \equiv {\rm d}V/{\rm d}\phi$.
The curvature term, $K^2/a^2$, is dropped in Eq.~(\ref{Heq})
since it adds nothing concrete to our discussion.

The condition for inflation (\ref{2_2_2}) requires
$\dot{\phi}^2<V(\phi)$ or classically that the potential energy of the
inflaton dominates over the kinetic energy.  Hence one requires a
sufficiently flat potential for the inflaton in order to lead to
sufficient inflation.  Imposing the slow-roll conditions:
$\dot{\phi}^2/2 \ll V(\phi)$ and $|\ddot{\phi}| \ll 3H|\dot{\phi}|$,
Eqs.~(\ref{Heq}) and (\ref{phieq}) are approximately given as
\beqa
\label{2_3_4}
& & H^2 \simeq \frac{8\pi V(\phi)}{3m_{\rm pl}^2}\,, \\
& & 3H\dot{\phi}\simeq -V_\phi(\phi)\,.
\label{2_3_5}
\eeqa
One can define the so-called slow-roll parameters
\beqa
\epsilon =
\frac{m_{\rm pl}^2}{16\pi}\left(\frac{V_\phi}{V}\right)^2\,,
~\eta = \frac{m_{\rm pl}^2V_{\phi \phi}}{8\pi V}\,,
~\xi^2 = \frac{m_{\rm pl}^4 V_{\phi} V_{\phi \phi \phi}}
{64\pi^2 V^2}\,.
\label{slowpara}
\eeqa
We can easily verify that the above slow-roll approximations
are valid when $\epsilon \ll 1$ and $|\eta| \ll 1$ for a prolonged
period of time.
% and $|\xi^2| \ll 1$.

The inflationary phase ends when $\epsilon$ and $|\eta|$ grow to of
order unity, though this does not, of itself, imply reheating
of the universe.
A useful quantity to describe the amount of inflation is the number
of e-foldings, defined by
\beqa
N \equiv \ln \frac{a_f}{a}=\int_{t}^{t_f} H {\rm d}t
\simeq \frac{8\pi}{m_{\rm pl}^2} \int_{\phi_f}^{\phi}
\frac{V}{V_\phi}
{\rm d}\phi\,,
\label{2_3_8}
\eeqa
where the subscript $f$ denotes evaluation of the quantity at the
end of inflation.

In order to solve the flatness problem, $\Omega$ is required to be
$|\Omega_f-1|~\lsim~10^{-60}$ right after the end of inflation.
Meanwhile the ratio $|\Omega-1|$ between the initial and final
phase of slow-roll inflation is given by
\beqa
\frac{|\Omega_f-1|}{|\Omega_i-1|} \simeq \left(\frac{a_i}{a_f}
\right)^2=e^{-2N_i}\,,
\label{2_3}
\eeqa
where we used the fact that $H$ is nearly constant during slow-roll
inflation.  Assuming that $|\Omega_i-1|$ is of order unity, the number
of e-foldings is required to be $N~\gsim~60$ to solve the flatness
problem.  This statement is a statement about the measure
on the space of initial conditions and is therefore properly in the domain of quantum gravity.
It is clear that for any fixed number of e-foldings
one can choose an infinite
number of $\Omega_i$ such that the flatness problem is not solved.
Nevertheless, inflation certainly mitigates the problem.
We require a similar number of e-foldings in order to
solve the horizon problem and hence $N > 60$ is taken
as a standard target minimum number of e-foldings
for any new model of inflation.

%%%%%%%%%%%%%%%%%
\subsection{Models of inflation}
%%%%%%%%%%%%%%%%%

So far we have not discussed the form of the inflaton potential,
$V(\phi)$.
The original ``old inflation'' scenario
\cite{Guth81,Sato81p1,Sato81p2} assumed the inflaton was trapped in a
metastable false vacuum and had to exit to the true vacuum via a
first-order transition.
As Guth pointed
out \cite{Guth81} this could occur neither gracefully
nor completely, problems avoided in the ``new inflation'' model where inflation ends via a second-order
phase transition after a phase of slow roll. We now have many varieties
of inflationary models : $R^2$, new, chaotic, extended, power-law,
hybrid, natural, supernatural, extra-natural, eternal, D-term, F-term,
brane, oscillating, trace-anomaly driven, k, ghost, tachyon,..., etc...

The different kinds of single-field inflationary models can be roughly classified in the
following
way \cite{Kolb99}. The first class (type I) consists of the ``large field"
models, in which the initial value of the inflaton is large and it slow rolls down
toward the potential minimum at smaller $\phi$. Chaotic inflation \cite{Linde83} is one of
the representative models of this class.
The second class (type II) consists of the ``small field" models,
in which the inflaton field is small initially and
slowly evolves toward the
potential minimum at larger $\phi$.
New inflation \cite{Linde82,Albre82} and natural
inflation \cite{Freese90} are the examples of this type.  In the
first class one usually has $V_{\phi \phi}> 0$,
whereas it can change the sign in the second
class.  The third class (type III) consists of the hybrid inflation
models \cite{Linde94}, in which
inflation typically ends by a phase transition
triggered by the presence of a second scalar field.
The fourth class (type IV) consists of the double inflation models
in which there exist two dynamical scalar fields leading to
the two stage of inflation.
A simple example is two light massive scalar fields
given in Ref.~\cite{Polarski92}.

We note that several models of inflation
can not be classified in the above four classes.
For example, some models do not have a potential minimum
such as quintessential inflation \cite{PV99} and
tachyon inflation \cite{FT02,Fe02,Paddy02,PST05,Sami02,SCQ02,TW05}.
Typically these scenarios suffer from a reheating
problem \cite{Kofman02},
since gravitational particle production is not efficient
compared to the standard non-gravitational particle production
by an oscillating inflaton field.
There exist other models of inflation in which an accelerated
expansion is realised without using the potential of the inflaton.
For example, k-inflation \cite{Picon99} and ghost
inflation \cite{Arkani03}
belong to this class. In this case inflation occurs
in the presence of higher-order kinematic terms of
a scalar field.
Inflation can also be realised when the higher-order curvature
terms are present \cite{Sta80,MO04,CTS05,NOZ00,NO00,NO03,HHR01,ellis99,BB89},
even without an inflaton
potential\footnote{We note that in the simple $R^2$ inflation
model \cite{Sta80} the system can be reduced to
a minimally coupled scalar
field with a large-field potential by making a conformal
transformation \cite{Maeda89}.
However this transformation is not generally easy in the presence of
more complicated higher-order curvature terms.}.
Apart from these models, let us
briefly review each class of inflationary models.

%%%%%%%%%%%%%%%%%%%%%%%%%%%%%%%%%%
\subsubsection{Large-field models}
%%%%%%%%%%%%%%%%%%%%%%%%%%%%%%%%%%

\begin{figure}
\begin{center}
\includegraphics[height=2.0in,width=3.0in]{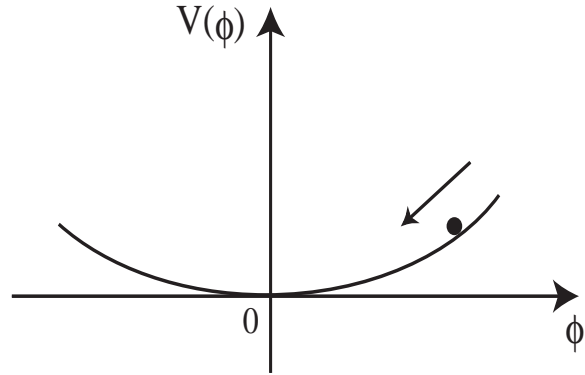}
\caption{\label{chaotic}
Schematic illustration of the potential of large-field models.}
\end{center}
\end{figure}

The large-field models are typically characterized by the monomial
potential
%%%%%%%%%%%%%%%%
\beqa
V(\phi)=V_0\phi^n\,.
\label{chaoticV}
\eeqa
%%%%%%%%%%%%%%%%
The quadratic and quartic potentials in chaotic inflation
correspond to $n=2$ and $n=4$, with inflation occurring for Planckian
scale values of $\phi$ (see Fig.~\ref{chaotic}).
Such models lend themselves naturally to randomly
distributed initial conditions with regions of
spacetime that initially have $\phi > m_{\rm pl}$ and are
homogeneous on the Hubble
scale undergoing inflation and therefore potentially giving rise to
our observable universe \cite{Lindebook}.

It is easy to get analytic forms of solutions under the slow-roll
approximation: $\epsilon \ll 1$ and $|\eta| \ll 1$.
For example in the case of the quadratic potential
($n=2$ and $V_0=m^2/2$)
we get the following relation by Eqs.~(\ref{2_3_4}) and (\ref{2_3_5}):
%%%%%%%%%%%%%%%%
\beqa
& &\phi \simeq \phi_i-\frac{m m_{\rm pl}}{2\sqrt{3\pi}}t\,, \\
& &a \simeq a_i \exp \left[2\sqrt{\frac{\pi}{3}}\frac{m}{m_{\rm pl}}
\left(\phi_i
t-\frac{mm_{\rm pl}}{4\sqrt{3\pi}}t^2\right) \right],
\label{phievo}
\eeqa
%%%%%%%%%%%%%%%%
where $\phi_i$ is an integration constant corresponding to the initial
value of the inflaton.
The relation (\ref{phievo}) implies that
the universe expands exponentially during the initial stage of
inflation.
The expansion rate slows down
with the increase of the second term in the square bracket of
Eq.~(\ref{phievo}).
We require the condition, $\phi_i~\gsim~3m_{\rm pl}$,
in order to have the number of e-foldings which is larger than
$N=60$.

%%%%%%%%%%%%%%%%
%%%%%%%%%%%%%%%%

\subsubsection{Small-field models}

The small-field models are characterized by the following
potential around $\phi=0$:
%%%%%%%%%%%%%%%%
\beqa
V(\phi)=V_0\left[1-\left(\frac{\phi}
{\mu}\right)^n\right]\,,
\label{smallfield}
\eeqa
%%%%%%%%%%%%%%%%
which may arise in spontaneous symmetry breaking.  The potential
(\ref{smallfield}) corresponds to a Taylor expansion about the origin,
but realistic small-field models also have a potential minimum at
some $\phi\neq0$ to connect to the reheating stage.

\begin{figure}
\begin{center}
\includegraphics[height=2.5in,width=3.5in]{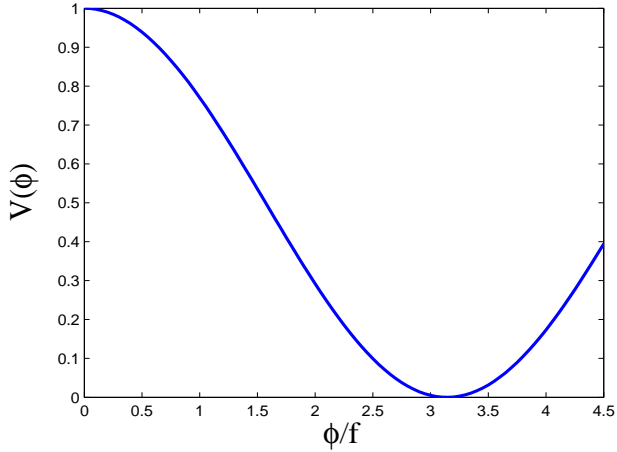}
\caption{\label{naturalfig} The schematic illustration of
the potential of small-field models.}
\end{center}
\end{figure}

For example let us consider natural inflation model
in which a Pseudo Nambu-Goldstone boson (PNGB)
plays the role of inflaton.
The PNGB potential is expressed as \cite{Freese90}
%%%%%%%%%%%%%%%%
\beqa
V(\phi)=m^4 \left[1+\cos \left(\frac{\phi}{f}\right) \right]\,,
\label{natural}
\eeqa
%%%%%%%%%%%%%%%%
where two mass scales $m$ and $f$ characterize
the height and width of the potential, respectively
(see Fig.~\ref{naturalfig}).
The typical mass scales for successful inflation are of order
$f \sim m_{\rm pl} \sim 10^{19}$ GeV and
$m \sim m_{\rm GUT}\sim 10^{16}$ GeV.
The potential (\ref{natural}) has a minimum
at $\phi=\pi f$.

One typical property in the type II model is that the second
derivative of the inflaton potential can change sign. In natural
inflation $V_{\phi \phi}$ is negative when inflaton evolves in the
region of $0 <\phi < \pi f/2$.  This leads to the enhancement of
inflaton fluctuations by spinodal (tachyonic) instability
\cite{Cormier99,Cormier00,Tsujispi,Felder01p1,Felder01p2}.
When the particle creation by spinodal instability is neglected,
the number of e-foldings is expressed by
%%%%%%%%%%%%%%%%
\beqa
N=\frac{16\pi f^2}{m_{\rm pl}^2}
{\rm ln} \left[ \frac{\sin (\phi_f/2f)} {\sin (\phi_i/2f)} \right]\,.
\eeqa
%%%%%%%%%%%%%%%%
In order to achieve a sufficient number of
e-foldings ($N~\gsim~60$), the initial value of the inflaton
is  required to be $\phi_{i}~\lsim~0.1 m_{\rm pl}$ for the mass scale
$f \sim m_{\rm pl}$.

%%%%%%%%%%%%%%%%
%%%%%%%%%%%%%%%%

\subsubsection{Hybrid inflation}

\begin{figure}
\begin{center}
\includegraphics[height=3.0in,width=3.5in]{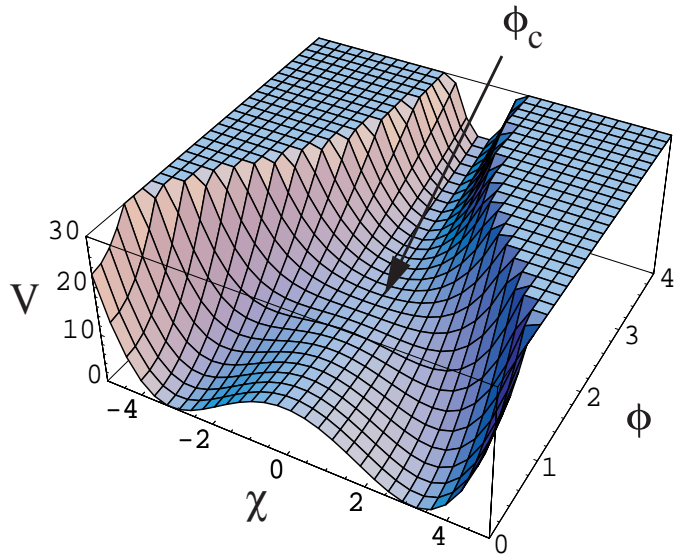}
\caption{\label{hybrid} Schematic illustration of
the potential of hybrid (or double) inflation models
given by Eq.~(\ref{hybridpo}). Here $\phi_c$ is the critical value of the
inflaton below which $\chi = 0$ becomes unstable due to
tachyonic instability ($m_{\chi}^2 < 0$).}
\end{center}
\end{figure}

Hybrid inflation models involve more than one scalar field.  This
scenario is particularly motivated from the viewpoint of particle
physics \cite{Cope94,LRreview,Linde97b}.
Inflation continues from an initial large
value of the inflaton which decreases until it reaches a bifurcation
point, after which the field becomes unstable and
undergoes a ``waterfall" transition towards a global minimum (see
Fig.~\ref{hybrid}).  During the initial inflationary phase the potential of
the hybrid inflation is effectively described by a single field:
%%%%%%%%%%%%%%%%
\beqa
V(\phi)=V_0\left[1+\left(\frac{\phi}
{\mu}\right)^n\right]\,.
\label{hyb}
\eeqa
%%%%%%%%%%%%%%%%

Consider the Linde hybrid inflation model
with potential \cite{Linde94}
%%%%%%%%%%%%%%%%
\beqa
V = \frac{\lambda}{4} \left(\chi^2-\frac{M^2}{\lambda}\right)^2
+\frac12 g^2 \phi^2 \chi^2 + \frac12 m^2\phi^2 \,.
\label{hybridpo}
\eeqa
%%%%%%%%%%%%%%%%
When $\phi^2$ is large the field rolls down toward the potential
minimum at $\chi=0$. Then we have
%%%%%%%%%%%%%%%%
\beqa
V \simeq \frac{M^4}{4\lambda} +\frac12 m^2\phi^2\,.
\label{hysingle}
\eeqa
%%%%%%%%%%%%%%%%
The mass-squared of $\chi$ becomes negative for $\phi<\phi_c \equiv M/g$ signifying
a tachyonic instability.  Then the field begins to roll down to one of the true minima at $\phi=0$
and $\chi=\pm M/\sqrt{\lambda}$ (and thereby creates domain walls). In this original version
of the hybrid inflation \cite{Linde94}
inflation soon comes to an end after the symmetry breaking
($\phi<\phi_c$) due to the rapid rolling of the field $\chi$.
In this case the number of e-foldings
can be approximately estimated by using the potential
(\ref{hysingle}):
\begin{eqnarray}
N \simeq \frac{2\pi M^4} {\lambda m^2m_{\rm pl}^2}\,{\rm ln}
\frac{\phi_i}{\phi_c} \,,
\label{Nhybrid}
\end{eqnarray}
where $\phi_i$ is the initial value of inflaton.

%%%%%%%%%%%%%%%%
%%%%%%%%%%%%%%%%

\subsubsection{Double inflation}

Double inflation can occur even for the potential (\ref{hybridpo})
depending on the model parameters.  When the condition, $M^2 \gg
\lambda m_{\rm pl}^2$, is satisfied, the mass of the field $\chi$ is
``light'' relative to the Hubble rate around $\phi=\phi_c$, thereby
leading to a second stage of inflation for $\phi<\phi_c$
\cite{Randall96,Garcia96hybrid,Tsuji03}.  This
corresponds to a genuine multi-field inflationary model, with more
than one light field, of the type that we will examine in later
sections.
More generally multi-field inflation may be naturally
realised near points of enhanced symmetry in moduli space
\cite{Kadota03,Kadota03v2}.
In any model where more than one scalar field is light during
inflation then there is no longer a unique attractor trajectory in
phase-space and such models can support isocurvature as well as
adiabatic perturbations about a particular background solution.

An alternative form of double inflation is also realised in the
following simple model:
%%%%%%%%%%%%%%%%
\beqa
V(\phi, \chi)=\frac12 m_\phi^2 \phi^2+
\frac12 m_\chi^2 \chi^2\,,
\eeqa
%%%%%%%%%%%%%%%%
as studied by Polarski and Starobinsky \cite{Polarski92}, and later by
Langlois \cite{Langlois99} who realised that the adiabatic and
isocurvature perturbations surviving at the end of inflation will in
general be correlated.

%%%%%%%%%%%%%%%%%%%%%%%%%%%%%%%%%%%%%%%%%%%%%%%%%%%
\section{Cosmological perturbations}
%%%%%%%%%%%%%%%%%%%%%%%%%%%%%%%%%%%%%%%%%%%%%%%%%%%
\label{sect:Pert}

Having undertaken a rapid tour of standard inflationary theory and models
we move to discussion of perturbations. The description of the Universe as a perfectly homogeneous and
isotropic FRW model is obviously an idealisation. In practice we are
interested in deviations from homogeneity and isotropy that enable
us to characterise different models. We will deal with small
perturbations, but we will assume that the distribution of
perturbations is statistically homogeneous and isotropic, which
is an alternative statement of the Copernican principle.

In particular we have so far considered only the dynamics of
homogeneous scalar fields driving inflation. But to investigate
inflation models in more detail, and to test theoretical predictions
against cosmological observations we need to consider inhomogeneous
perturbations. In this section we will define the variables and
notation used in subsequent sections to describe cosmological
perturbations generated by inflation.

We will consider only small perturbations about the
homogeneous fields
\begin{equation}
\phi = \phi_0(t) + \delta\phi(t,x) \,,
\end{equation}
and only keep terms to first-order in $\delta\phi$.

\subsection{Metric perturbations}

For an inhomogeneous matter distribution the Einstein equations
imply that we must also consider
inhomogeneous metric perturbations about the spatially flat FRW
metric. The perturbed FRW spacetime is described
by the line element
\begin{eqnarray}
\hspace*{-0.2em}\rd s^2 &=& - (1+2A)\rd t^2 +
2a(\partial_iB-S_i)\rd x^i\rd t
\nonumber\\
\hspace*{-0.2em}&& +a^2\left[ (1-2\psi)\delta_{ij} + 2\partial_{ij}E +
2\partial_{(i}F_{j)} + h_{ij} \right] \rd x^i \rd x^j\,, \nonumber\\
\end{eqnarray}
where $\partial_i$ denotes the spatial partial derivative
$\partial/\partial x^i$. We will use lower case latin indices to
run over the 3 spatial coordinates. Our metric perturbations
follow the notation of Ref.~\cite{Mukhanov90}, apart from our use
of $A$ rather than $\phi$ as the perturbation in the lapse
function.

The metric perturbations have been split into scalar, vector and
tensor parts according to their transformation properties on the
spatial hypersurfaces. The Einstein equations for the
scalar, vector and tensor parts then decouple to linear order.
We do not consider second-order cosmological
perturbations in this review \cite{Acqu03}.

\subsubsection{Scalar perturbations}

The four scalar metric perturbations $A$, $\partial_iB$,
$\psi\delta_{ij}$ and $\partial_{ij}E$ are constructed from 3-scalars,
their derivatives, and the background spatial metric.
The intrinsic Ricci scalar curvature of constant time hypersurfaces
is given by
\begin{equation}
^{(3)}R = \frac{4}{a^2} \nabla^2 \psi \,,
\end{equation}
where $\nabla^2 \equiv \delta^{ij}\partial_{ij}$ is the spatial
Laplacian, and hence we refer to $\psi$ as the curvature perturbation.
We can Fourier decompose an arbitrary scalar perturbation with respect to the complete set of
eigenvectors of the spatial Laplacian, $\nabla^2 \psi = -k^2\psi$, with
comoving wavenumber $k$ indexing the corresponding eigenvalues.

Under a scalar coordinate/gauge transformation
\begin{eqnarray}
t &\to& t+ \delta t\,, \\
x^i &\to& x^i + \delta^{ij} \partial_j \delta x\,,
\end{eqnarray}
$\delta t$ determines the time slicing and $\delta x$ the spatial
threading. The scalar metric perturbations then transform as
\begin{eqnarray}
A &\to& A -\dot{\delta t} \,, \\
B &\to& B + a^{-1} \delta t - a\dot{\delta x}  \,,\\
E &\to& E - \delta x \,,\\
\psi &\to& \psi + H\delta t \,.
\end{eqnarray}
Although $B$ and $E$ separately are spatially gauge-dependent, the
combination $\dot{E}-B/a$ is independent of spatial gauge and
describes the scalar potential for the anisotropic shear of
worldlines orthogonal to constant time hypersurfaces
\cite{Kodama84}.

We can construct a variety of gauge-invariant combinations of the
scalar metric perturbations. The longitudinal gauge corresponds to
a specific gauge-transformation to a (zero-shear) frame such that
$E=B=0$, leaving the gauge-invariant variables
\begin{eqnarray}
 \label{defPhi}
\Phi &\equiv& A - \frac{{\rm d}}
{{\rm d}t} \left[ a^2(\dot{E}-B/a)\right] \,,\\
 \label{defPsi}
\Psi &\equiv& \psi + a^2 H (\dot{E}-B/a) \,.
\end{eqnarray}

Matter perturbations are also gauge-dependent. Scalar field, density
and pressure perturbations all obey the simple transformation rule
\begin{equation}
\delta\rho \to \delta\rho - \dot\rho \delta t \,.
\end{equation}
The adiabatic pressure perturbation is defined to be
\begin{equation}
\delta P_{\rm ad} \equiv \frac{\dot P}{\dot\rho} \delta\rho \,,
\end{equation}
and hence the non-adiabatic part of the actual pressure
perturbation, or entropy perturbation, is a gauge-invariant
perturbation
\begin{equation}
 \label{defPnad}
\delta P_{\rm nad} = \delta P -\frac{\dot P}{\dot\rho}
\delta\rho\,.
\end{equation}
The scalar part of the 3-momentum is given by $\partial_i\delta q$ and
this momentum potential transforms as
\begin{equation}
\delta q \to \delta q + (\rho+P)\delta t \,.
\end{equation}
Thus we can obtain the gauge-invariant comoving density
perturbation~\cite{Bardeen80}
\begin{equation}
 \label{defrhom}
\delta\rho_m = \delta\rho - 3H \delta q \,.
\end{equation}

We can construct two further commonly used
gauge-invariant combinations in terms of matter and metric
perturbations.
The curvature perturbation on uniform-density
hypersurfaces is given by
\begin{equation}
 \label{defzeta}
- \zeta \equiv \psi + \frac{H}{\dot\rho}\delta\rho \,,
\end{equation}
first defined by Bardeen, Steinhardt and Turner~\cite{Bardeen83} (see
also Refs.~\cite{Bardeen88,Martin98,Wands00}).
The comoving curvature perturbation (strictly speaking the curvature
perturbation on hypersurfaces orthogonal to comoving worldlines)
\begin{equation}
 \label{eq:defR}
{\cal R} \equiv \psi - \frac{H}{\rho+P} \delta q \,.
\end{equation}
This has been used by Lukash \cite{Lukash80}, Lyth \cite{Lyth85}
and many others, including Mukhanov, Feldman and Brandenberger in
their review \cite{Mukhanov90}.
(Note that in their review the comoving curvature perturbation is
denoted by ``$\zeta$'' in Ref.~\cite{Mukhanov90} and defined
in terms of the metric perturbations in the longitudinal gauge,
but it is equivalent to our definition of ${\cal R}$
in a spatially flat background with vanishing anisotropic stress.)
The difference between the two curvature perturbations ${\cal R}$
and $-\zeta$ is proportional to the comoving density perturbation:
\begin{equation}
\label{eq:zetaR}
-\zeta = {\cal R} + \frac{H}{\dot\rho} \delta\rho_m \,.
\end{equation}

For single field inflation we have $\delta q=-\dot\phi\delta\phi$
and hence
\begin{equation}
{\cal R} \equiv \psi + \frac{H}{\dot\phi}\delta\phi \,.
\end{equation}
In slow-roll single-field inflation we have $\delta\rho/\dot\rho
\simeq \delta\phi/\dot\phi$ and hence $\delta\rho_m\simeq0$ and these
two commonly used curvature perturbations, ${\cal R}$ and $-\zeta$,
coincide.

Finally we note that another variable commonly used to describe scalar
perturbations during inflation is the field perturbation in the
spatially
flat gauge (where $\psi=0$).
This has the gauge-invariant definition \cite{Mukhanov85,Sasaki86}:
\begin{equation}
 \label{eq:defdphipsi}
 \delta\phi_\psi \equiv \delta\phi + \frac{\dot\phi}{H} \psi \,.
\end{equation}
In single field inflation this is simply a rescaling of the comoving
curvature perturbation $\R$ in (\ref{eq:defR}). We see that what
appears as a field perturbation in one gauge is a metric perturbation
in another gauge and vice versa.

\subsubsection{Vector perturbations}

The vector perturbations $S_i$ and $F_i$ can be distinguished
from scalar perturbations as they are solenoidal (divergence-free),
i.e., $\partial^iS_i=0$.

Under a vector coordinate/gauge transformation
\begin{eqnarray}
x^i &\to& x^i + \delta x^i\,,
\end{eqnarray}
the vector metric perturbations transform as
\begin{eqnarray}
S_i &\to& S_i+ a\dot{\delta x}_i \,,\\
F_i &\to& F_i - \delta x_i \,,
\end{eqnarray}
and hence $\dot{F}_i+S_i/a$ is the gauge-invariant vector shear
perturbation.

\subsubsection{Tensor modes}

The tensor perturbations $h_{ij}$ are transverse $\partial^i
h_{ij}=0$ and trace-free $\delta^{ij}h_{ij}=0$. They are
automatically independent of coordinate gauge transformations.

These are referred to as gravitational waves as they are the free
part of the gravitational field and evolve independently of
linear matter perturbations.

We will decompose arbitrary tensor perturbations into eigenmodes
of the spatial Laplacian, $\nabla^2e_{ij}=-k^2e_{ij}$, with
comoving wavenumber $k$, and scalar amplitude $h(t)$:
\begin{equation}
\label{eq:defh}
h_{ij} = h(t) e_{ij}^{(+,\times)}(x)\,,
\end{equation}
with two possible polarisation states, $+$ and $\times$.

\subsection{Field equations}

\subsubsection{Scalar perturbations}

By considering the perturbed Einstein equations
$\delta G_{\mu \nu}=8\pi G \delta T_{\mu \nu}$,
we find that the metric perturbations are related to matter
perturbations via the energy and momentum
constraints \cite{Mukhanov90}
\begin{eqnarray}
\label{eq:densitycon}
& &3H\left(\dot\psi+HA\right) +
\frac{k^2}{a^2}\left[\psi+H(a^2\dot{E}-aB)\right] \nonumber \\
& & =-4\pi G \delta\rho, \\
& & \dot\psi + HA = -4\pi G \delta q. \label{eq:mtmcon}
\end{eqnarray}
These can be combined to give the gauge-invariant generalisation of
the Poisson equation
\begin{equation}
 \label{eq:rhomcon}
\frac{k^2}{a^2} \Psi = - 4\pi G \delta\rho_m \,,
\end{equation}
relating the longitudinal gauge metric perturbation (\ref{defPsi}) to
the comoving density perturbation (\ref{defrhom}).

The Einstein equations also yield two evolution equations for the
scalar metric perturbations
\begin{eqnarray}
& &\ddot\psi + 3H\dot\psi +H\dot{A} + \left( 3H^2 + 2\dot{H}
\right)A \nonumber \\
& & = 4\pi G \left( \delta P - \frac{2}{3}k^2\delta\Pi \right) \,,\\
\label{eq:aniso}
& &\left( \dot{E}-B/a \right)^{\displaystyle{\cdot}}
+ 3H \left( \dot{E}-B/a \right) + \frac{\psi - A}{a^2} \nonumber \\
& &= 8\pi G\delta\Pi \,,
\end{eqnarray}
where the scalar part of the anisotropic stress is given by
$\delta\Pi_{ij}=[\partial_i\partial_j + (k^2/3)\delta_{ij}]\Pi$.
Equation~(\ref{eq:aniso}) can be written in terms of the longitudinal
gauge metric perturbations, $\Phi$ and $\Psi$ defined in
Eqs.~(\ref{defPhi}) and~(\ref{defPsi}), as the constraint
\begin{equation}
\Psi - \Phi = 8\pi G a^2 \delta\Pi \,,
\end{equation}
and hence we have $\Psi=\Phi$
in the absence of anisotropic stresses.

Energy-momentum conservation gives evolution equations
for the perturbed energy and momentum
\begin{eqnarray}
 \label{energycon}
& &\dot{\delta\rho} +3H \left( \delta\rho + \delta P \right)
\nonumber \\
& &=\frac{k^2}{a^2} \delta q + \left( \rho+P \right)
\left[ 3\dot\psi +
k^2 \left( \dot{E}+B/a \right) \right],\\
& &\dot{\delta q} +3H\delta q
= -\delta P + \frac{2}{3}k^2\delta\Pi -
(\rho+P)A.
\end{eqnarray}
Re-writing the energy conservation equation (\ref{energycon}) in terms
of the curvature perturbation on uniform-density hypersurfaces,
$\zeta$ in (\ref{defzeta}), we obtain the important result
\begin{equation}
\label{eq:dotzeta}
\dot\zeta = -H \frac{\delta P_{\rm nad}}{\rho+P} - {\Sigma} \,,
\end{equation}
where $\delta P_{\rm nad}$ is the
non-adiabatic pressure perturbation, defined in (\ref{defPnad}),
and $\Sigma$ is
the scalar shear along comoving worldlines~\cite{Lyth03p1},
which can be given relative to the Hubble rate as
\begin{eqnarray}
\hspace*{-0.5em} \frac{\Sigma}{H} &\equiv& - \frac{k^2}{3H} \left\{
\dot{E}-(B/a) + \frac{\delta
    q}{a^2(\rho+P)} \right\} \nonumber \\
\hspace*{-0.5em} &=& - \frac{k^2}{3a^2H^2} \zeta
 - \frac{k^2 \Psi}{3a^2H^2} \left[ 1 - \frac{2\rho}{9(\rho+P)}
     \frac{k^2}{a^2H^2} \right]. \nonumber \\
\end{eqnarray}
Thus $\zeta$ is constant for adiabatic perturbations on super-Hubble
scales ($k/aH\ll1$), so long as $\Psi$ remains finite, in which case
the shear of comoving worldlines can be neglected.

If we consider $N$ scalar fields  with  Lagrangian density
\begin{equation}
{\cal L} = -V(\varphi_1,\cdots,\varphi_N)
 -\frac{1}{2} \sum_{I=1}^{N} g^{\mu\nu}
 \varphi_{I,\mu}\varphi_{I,\nu}\,,
\label{lag}
\end{equation}
and minimal coupling to gravity, then the total energy, pressure and
momentum perturbations are given by
\begin{eqnarray}
\delta\rho &=& \sum_I\left[
 \dot\varphi_I \left( \dot{\delta\varphi}_I -\dot\varphi_I A \right)
 + V_{I}\delta\varphi_I \right] \,,
\label{eq:densityphi} \\
\delta P &=& \sum_I\left[
 \dot\varphi_I \left( \dot{\delta\varphi}_I -\dot\varphi_I A \right)
 - V_{I}\delta\varphi_I \right] \,,
\label{eq:pressurephi} \\
\delta q_{,i} &=& - \sum_I \dot{\varphi}_I
\delta\varphi_{I,i} \,,
\label{eq:mtmphi}
\end{eqnarray}
where $V_I \equiv \partial V/\partial \varphi_I$.
These then give the gauge-invariant comoving density perturbation
\begin{eqnarray}
\label{def:rhom}
 \delta\rho_m = \sum_I \left[ \dot\varphi_I \left(
\dot{\delta\varphi}_I -
\dot\varphi_I A \right) - \ddot\varphi_I \delta\varphi_I
\right]\,.
\end{eqnarray}
The comoving density is sometimes used to represent the total
matter perturbation but for a single scalar field it is
proportional to the non-adiabatic pressure (\ref{defPnad}):
\begin{equation}
\delta P_{\rm nad} = - \frac{2V_{,\varphi}}{3H\dot\varphi}
\delta\rho_m \,.
\end{equation}
{}From the Einstein constraint equation (\ref{eq:rhomcon}) this will
vanish on large scales ($k/aH\to0$) if $\Psi$ remains finite, and
hence single scalar field perturbations become adiabatic in this
large-scale limit.

The anisotropic stress, $\delta\Pi$, vanishes to linear order for
any number of scalar fields minimally coupled to gravity.

The first-order scalar field perturbations obey the wave equation
\begin{eqnarray}
\label{eq:pertKG}
\label{eq:scalareom}
\hspace*{-2.0em}&& \ddot{\delta\varphi}_I + 3H\dot{\delta\varphi}_I
 + \frac{k^2}{a^2} \delta\varphi_I + \sum_J V_{IJ}
\delta\varphi_J
  \nonumber\\
  \hspace*{-2.0em}&&~~{}=
-2V_{I}A + \dot\varphi_I \left[ \dot{A} + 3\dot{\psi} +
\frac{k^2}{a^2} (a^2\dot{E}-aB) \right]. \label{eq:perturbation}
\end{eqnarray}

\subsubsection{Vector perturbations}

The divergence-free part of the 3-momentum obeys the momentum
conservation equation
\begin{equation}
 \label{eq:evolvec}
\dot{\delta q}_i + 3H \delta q_i = k^2 \delta Pi_i \,,
\end{equation}
where the vector part of the anisotropic stress is given by
$\delta\Pi_{ij}=\partial_{(i}\Pi_{j)}$.
The gauge-invariant vector metric perturbation is then directly
related to the divergence-free part of the momentum via the constraint
equation
\begin{equation}
 \label{eq:convec}
k^2 \left( \dot{F}_i + S_i / a \right) = 16\pi G \delta q_i \,.
\end{equation}
Thus the Einstein equations constrain the gauge-invariant vector
metric perturbation to vanish in the presence of only scalar fields,
for which the divergence-free momentum necessarily vanishes.

Equation~(\ref{eq:convec}) shows that vector metric perturbations can
be supported only by divergence-free momenta, but even then equation
(\ref{eq:evolvec}) shows that the vector perturbations are redshifted
away by the Hubble expansion on large scales unless they are driven by
an anisotropic stress.

\subsubsection{Tensor perturbations}

There is no constraint equation for the tensor perturbations as these
are the free gravitational degrees of freedom (gravitational
waves). The spatial part of the Einstein equations yields a wave
equation for the amplitude, defined in Eq.~(\ref{eq:defh}),
of the tensor metric perturbations:
\begin{equation}
 \label{teneq}
\ddot{h} + 3H\dot{h} + \frac{k^2}{a^2} h = 0 \,.
\end{equation}
This is the same as the wave equation for a massless scalar field
(\ref{eq:pertKG}) in an unperturbed FRW metric.

\subsection{Primordial power spectra}

Around the epoch of primordial nucleosynthesis the universe
is constrained
to be dominated by radiation composed of photons and 3 species of
relativistic neutrinos. In addition there are non-relativistic
baryons, tightly coupled to the photons by Thomson scattering, and
cold dark matter which has decoupled.
There is probably also some form of
vacuum energy, or dark energy, which eventually comes to
dominate the
density of the universe at the present day. All of these different
components may have different density perturbations, $\delta\rho_i$.
These are usefully characterised by the gauge-invariant curvature
perturbations for each component:
\begin{equation}
\label{eq:defzetai}
\zeta_i \equiv -\psi - \frac{H}{\dot\rho_i} \delta\rho_i \,.
\end{equation}
These individual $\zeta_i$ remain constant on large scales
\cite{Wands00}
as a consequence of local
energy-conservation for photons, neutrinos,
baryons and cold dark matter, each of which has a well-defined
equation of state and hence $\delta P_{{\rm nad},i}=0$.
Even when energy is not separately conserved for each individual
component it may still be possible to define a conserved perturbation
on large scales with respect to some other locally conserved quantity,
such as the baryon number so long as the net baryon number is
conserved \cite{Lyth03p1}.
Perfect fluid models of non-interacting dark energy will also have
$\zeta_{\rm de}=$constant on large scales, but scalar field models of
dark energy do not in general have a well-defined equation of state
and hence $\zeta_{\rm de}$ is not necessarily constant on large scales
\cite{Doran,Malquarti,Malik04}.

The total curvature perturbation $\zeta$, defined in
Eq.~(\ref{defzeta}), is simply given by the weighted sum of the
individual curvature perturbations
\begin{equation}
\label{zetasum}
\zeta = \sum_i \frac{\dot\rho_i}{\dot\rho} \zeta_i \,.
\end{equation}
This is often referred to as the adiabatic density perturbation,
while the difference determines the isocurvature
density perturbations
\begin{equation}
\label{defSi}
\S_i \equiv 3 ( \zeta_i - \zeta_\gamma ) \,.
\end{equation}
By convention the isocurvature perturbations are defined with respect
to the photons, hence these are also referred to as entropy
perturbations. The factor of 3 arises so that $\S_B$ coincides with
the perturbation in the local baryon-photon ratio:
\begin{equation}
\S_B = 3 (\zeta_B - \zeta_\gamma) = \frac{\delta
  (n_B/n_\gamma)}{n_B/n_\gamma} \,.
\end{equation}

The relative isocurvature perturbation, $\S_i$, remains constant on
large scales as a consequence of the conservation of the individual
$\zeta_i$. The total curvature perturbation only remains constant on
large scales as the universe evolves from radiation to matter
domination for adiabatic perturbations with $\S_i=0$, in agreement
with Eq.~(\ref{eq:dotzeta}).

The primordial power spectrum of density perturbations in the
radiation-dominated era, after inflation but well before
matter-domination, is commonly given in terms of either $\zeta\simeq
\zeta_\gamma$, or the comoving curvature perturbation, $\R$ in
Eq.~(\ref{eq:defR}).
Combining Eqs.~(\ref{eq:rhomcon}) and (\ref{eq:zetaR}) we have
\begin{equation}
\R = -\zeta - \frac{2\rho}{9(\rho+P)} \left( \frac{k}{aH} \right)^2
\Psi \,,
 \end{equation}
and hence $\R$ and $-\zeta$ coincide on large scales.

The power on a given scale is given by the $k$-space
weighted contribution of modes with given wavenumber.
Thus the power
spectrum of scalar curvature perturbations, $\R$, is commonly given as
\begin{equation}
\label{eq:defPR}
{\cal P}_\R \equiv \frac{4\pi k^3}{(2\pi)^3} |\R^2| \,.
\end{equation}
This coincides with the definition of ${\cal P}_{\cal R}$ used in the
review by Lidsey et al \cite{LLKCreview}
and in the Liddle and Lyth book
\cite{LLbook}, and is denoted $\Delta_\R^2$
by the WMAP team \cite{Peiris03}.
An alternative notation widely used for the scalar power
spectrum is the
fractional density perturbation when adiabatic density perturbations
re-enter the Hubble scale during the matter dominated era
\cite{LLKCreview,LLbook}
\begin{equation}
\delta_H^2 \equiv A_{\rm S}^2
\equiv \frac{4}{25} {\cal P}_\R \,.
\end{equation}

An isocurvature power spectrum is naturally defined as
\begin{equation}
\label{eq:defPS}
{\cal P}_\S \equiv \frac{4\pi k^3}{(2\pi)^3} |\S^2| \,.
\end{equation}
The cross-correlation between adiabatic and isocurvature
perturbations can be given in terms of a correlation angle
$\Delta$:
\begin{equation}
{\cal C}_{\R\S} \equiv {\cal P}_\R^{1/2} {\cal P}_\S^{1/2}
\cos\Delta \,.
\end{equation}

The tensor power spectrum is denoted by
\begin{equation}
\label{eq:defPT}
{\cal P}_{\rm T} \equiv 2
\frac{4\pi k^3}{(2\pi)^3} |h^2| \,,
\end{equation}
where the additional factor of 2 comes from adding the 2 independent
polarisations of the graviton.
Again there is an alternative notation also
widely used \cite{LLKCreview,LLbook}
\begin{equation}
A_{{\rm GW}}^2 \equiv \frac{1}{100}
{\cal P}_{\rm T} \,.
\end{equation}

The scale dependence of the scalar power spectrum is given by the
logarithmic derivative of the power spectrum
\begin{equation}
\label{eq:defnR}
n_\R - 1 \equiv \frac{{\rm d}\ln {\cal P}_\R}
{{\rm d}\ln k} \biggr|_{k=aH} \,,
\end{equation}
which is evaluated at Hubble-radius crossing, $k=aH$.
We note that $n_\R=1$ for a scale-invariant spectrum
by convention.
Most authors refer to this as $n_s$ denoting the scalar spectrum.
We use $n_\R$ to distinguish this from the isocurvature spectrum:
\begin{equation}
\label{eq:defnS}
n_\S \equiv \frac{{\rm d}\ln {\cal P}_\S}
{{\rm d}\ln k} \biggr|_{k=aH}\,,
\end{equation}
where $n_\S=0$ for a scale-invariant spectrum. Similarly
$n_{\rm T}=0$ for a scale-invariant tensor spectrum.

The best way to distinguish multi-field models for the origin of
structure from other inflationary models be the statistical
properties of the primordial density perturbation.
Inflationary models start with small-scale vacuum fluctuations of an
effectively free scalar field, described by a Gaussian random field,
with vanishing three-point correlation function.
Simple deviations from Gaussianity in multi-field scenarios are
conventionally parameterised by a dimensionless parameter $f_{nl}$,
where \cite{Komatsu01,Komatsu03,BKMR04,Bartolo02,Bernar02}
\begin{equation}
 \label{deffnl1}
 \Phi = \Phi_{\rm Gauss}
 + f_{nl} \left( \Phi_{\rm Gauss}^2 - \langle \Phi_{\rm Gauss}^2 \rangle \right)
 \,,
\end{equation}
and $\Phi$ is the potential in the longitudinal gauge, defined in
Eq.~(\ref{defPhi}), on large scales in the matter-dominated era and
$\Phi_{\rm Gauss}$ is a strictly Gaussian distribution arising from
the first-order field perturbations.
For adiabatic perturbations on large scales in the matter dominated
era we have $\Phi=-3\zeta/5$ and hence this corresponds to
\begin{equation}
 \label{deffnl2}
 \zeta = \zeta_{\rm Gauss}
 - \frac35 f_{nl} \left( \zeta_{\rm Gauss}^2 -
\langle \zeta_{\rm Gauss}^2 \rangle \right)\,.
\end{equation}
This describes a ``local'' non-Gaussianity where the local curvature
perturbation, $\zeta$, is due to the local value of the first-order
field perturbation and the square of that perturbation. For example,
as we shall see, this naturally occurs in curvaton models and where
the local curvaton density is proportional to the square local value
of the curvaton field.

\subsection{$\delta N$ formalism}

A powerful technique to calculate the resulting curvature
perturbation in a variety of inflation models, including multi-field
models, is to note that the curvature perturbation $\zeta$ defined
in Eq.~(\ref{defzeta}) can be interpreted as a perturbation in the
local expansion \cite{Sasaki95}
\begin{equation}
\zeta = \delta N \,,
\end{equation}
where $\delta N$ is the perturbed expansion to uniform-density
hypersurfaces with respect to spatially flat hypersurfaces:
\begin{equation}
\delta N = - H \left. \frac{\delta\rho}{\dot\rho} \right|_\psi \,,
\end{equation}
and $\delta\rho$ must be evaluated on spatially flat ($\psi=0$)
hypersurfaces.

An important simplification arises on large scales where anisotropy
and spatial gradients can be neglected, and the local density,
expansion, etc, obeys the same evolution equations as the a
homogeneous FRW universe
\cite{Sasaki95,Sasaki98,Wands00,Lyth03p1,Rigopoulos03,Lyth05}.
Thus
we can use the homogeneous FRW solutions to describe the local
evolution, which has become known as the ``separate universe''
approach \cite{Sasaki95,Sasaki98,Wands00,Rigopoulos03}. In
particular we can evaluate the perturbed expansion in different
parts of the universe resulting from different initial values for
the fields during inflation using homogeneous background solutions
\cite{Sasaki95}. The integrated expansion from some initial field
values up to a late-time fixed density, say at the epoch of
primordial nucleosynthesis, is some function $N(\varphi_I)$. The
resulting primordial curvature perturbation on the uniform-density
hypersurface is then
\begin{equation}
\zeta = \sum_I \delta N_{I} \delta\varphi_I|_\psi \,,
\end{equation}
where $N_I\equiv \partial N/\partial \varphi_I$ and
$\delta\varphi_I|_\psi$ is the field perturbation on some initial
spatially-flat hypersurfaces during inflation.
In particular the power spectrum for the primordial density
perturbation in a multi-field inflation can be written in terms of
the field perturbations after Hubble-exit as
\begin{equation}
{\cal P}_\zeta = \sum_I (\delta N_I)^2 {\cal
P}_{\delta\varphi_I|_\psi} \,.
\end{equation}

This approach is readily extended to estimate the effect of
non-linear field perturbations on the metric perturbations
\cite{Sasaki98,Lyth03p1,Lyth05}. The curvature perturbation due to
field fluctuations up to second order is \cite{Rodriguez05,Seery05b}
\begin{equation}
\zeta \simeq \sum_I \delta N_{I} \delta\varphi_I|_\psi + \frac12
\sum_{I,J} \delta N_{IJ} \delta\varphi_I|_\psi \delta\varphi_J|_\psi
+ \ldots \,.
\end{equation}
We expect the field perturbations at Hubble-exit to be close to
Gaussian for weakly coupled scalar fields during inflation
\cite{Maldacena02,Seery05a,RS05,Seery05b}. Hence if
the contribution of only one field dominates the perturbed
expansion, this gives a non-Gaussian contribution to the curvature
perturbation of the ``local'' form (\ref{deffnl2}), where
\cite{Rodriguez05}
\begin{equation}
 f_{nl} = - \frac56 \frac{N_{II}}{N_I^2} \,.
\end{equation}

%%%%%%%%%%%%%%%%%%%%%%%%%%%%%%%%%%%%%%%%%%%%%%%%%%%%%
\section{The spectra of perturbations in single-field
inflation}
\label{sect:singlefieldps}
%%%%%%%%%%%%%%%%%%%%%%%%%%%%%%%%%%%%%%%%%%%%%%%%%%%%%

In this section we shall consider the spectra of scalar and tensor
perturbations generated in single-field inflation.

The perturbed scalar field equation of motion (\ref{eq:scalareom}) for
a single scalar field can be most simply written in the spatially flat
gauge (where $\psi=0$). Using the Einstein constraint equations to
eliminate the remaining metric perturbations one obtains the wave
equation
\begin{equation}
\label{eq:singlescalareom}
\ddot{\delta\phi}_\psi + 3H\dot{\delta\phi}_\psi
 + \left[ \frac{k^2}{a^2} + V_{\phi \phi}
 - \frac{8\pi G}{a^3} \frac{{\rm d}}{{\rm d}t}
 \left( \frac{a^3\dot\phi^2}{H} \right)
 \right] \delta\phi_\psi = 0,
\end{equation}
where a gauge-invariant definition of $\delta\phi_\psi$ is given in
(\ref{eq:defdphipsi}).

Introducing new variables, $v=a{\delta\phi}_\psi$ and $z=a\dot\phi/H$,
Eq.~(\ref{eq:singlescalareom}) reduces to \cite{Sasaki86,Mukhanov88}
%%%%%%%%%%%%%
\begin{eqnarray}
v''+\left(k^2-\frac{z''}{z}\right)v=0\,,
\label{veq}
\end{eqnarray}
%%%%%%%%%%%%%
where a prime denotes a derivative
with respect to conformal time $\tau \equiv \int a^{-1}dt$.
The effective mass term, $z''/z$, can be written as
\cite{Stewart93,LLKCreview,Hwang96}
%%%%%%%%%%%%%
\begin{eqnarray}
\frac{z''}{z}=
(aH)^2 \left[
% v2 DW z''/z completely rewritten
% using definitions of epsilon, eta and xi (see below)
% that coincide with our earlier definitions at leading order
% Would be worth checking!
% 2+2\epsilon-3\eta+2\epsilon^2-4\epsilon\eta+\eta^2+\xi^2
2+ 5\epsilon - 3\eta +9\epsilon^2-7\epsilon\eta +\eta^2 +\xi^2
 \right] \,,
\label{gra}
\end{eqnarray}
%%%%%%%%%%%%%
where
%%%%%%%%%%%%%
\begin{eqnarray}
 \epsilon \equiv -\frac{\dot{H}}{H^2}\,,~~
% v2 DW new definitions of eta and xi
% \eta \equiv \epsilon-\frac{\dot{\epsilon}}{2H\epsilon} \,,~~
\eta \equiv 2 \epsilon-\frac{\dot{\epsilon}}{2H\epsilon} \,,~~
% \xi^2 \equiv \left(\epsilon-\frac{\dot\eta}{2H\eta}\right)\eta
\xi^2 \equiv \left(2\epsilon-\frac{\dot\eta}{H\eta}\right)\eta \,.
 \label{ep}
\end{eqnarray}
%%%%%%%%%%%%%
These definitions of the slow-roll parameters coincide at leading
order in a slow-roll expansion \cite{Liddleetal94} with our
earlier definitions in Eq.~(\ref{slowpara}) in terms of the first,
second and third derivatives of the scalar field potential.

Neglecting the time-dependence of $\epsilon$ and $\eta$
during slow-roll inflation\footnote{Stewart \cite{Stewart01} has
  developed a generalised slow-roll approximation to calculate the
  spectrum of perturbations that drops the
  requirement that the slow-roll parameters are slowly varying.},
and other terms of second and higher order
in the slow-roll expansion, gives
\begin{equation}
\tau \simeq -\frac{1}{(1-\epsilon)aH}\,,
\end{equation}
and
%%%%%%%%%%%%%
\begin{eqnarray}
\frac{z''}{z} = \frac{\nu_\R^2-(1/4)}{\tau^2}\,,~~~
{\rm with}~~~
% v2 DW: following order of Bessel function now consistent both with
% preceding equations and usual slow-roll parameters
\nu_\R \simeq \frac32 + 3\epsilon - \eta
 \,.
\label{gra2}
\end{eqnarray}
%%%%%%%%%%%%%
The general solution to Eq.~(\ref{veq}) is then expressed
as a linear combination of Hankel functions
%%%%%%%%%%%%%
\begin{eqnarray}
v \simeq \frac{\sqrt{\pi|\tau|}}{2} e^{i(1+2\nu_\R)\pi/4}
\left[ c_1 H_{\nu_\R}^{(1)}(k|\tau|) +c_2 H_{\nu_\R}^{(2)}(k|\tau|)
\right]\,. \nonumber \\
\label{han}
\end{eqnarray}
%%%%%%%%%%%%%

The power spectrum for the scalar field perturbations is given by
\begin{equation}
{\cal P}_{\delta\phi} \equiv \frac{4\pi k^3}{(2\pi)^3} \left|
  \frac{v}{a} \right|^2\,.
\end{equation}
Imposing the usual Minkowski vacuum state,
%%%%%%%%%%%%%
\begin{eqnarray}
 \label{vacnorm}
v \to \frac{e^{-ik\tau}}{\sqrt{2k}}\,,
\end{eqnarray}
%%%%%%%%%%%%%
in the asymptotic past
% v2 DW eta corrected to tau
($k\tau\to-\infty$)
corresponds to the choice
$c_1=1$ and $c_2=0$ in Eq.~(\ref{han}).
The power spectrum on small scales ($k\gg aH$) is thus
\begin{equation}
{\cal P}_{\delta\phi}
 \simeq
\left( \frac{k}{2\pi a} \right)^2 \,,
\end{equation}
and on the large scales ($k\ll aH$) we have
\begin{equation}
{\cal P}_{\delta\phi} \simeq
\left( (1-\epsilon)
 \frac{\Gamma(\nu_\R)}{\Gamma(3/2)} \frac{H}{2\pi} \right)^2
 \left( \frac{|k\tau|}{2} \right)^{3-2\nu_\R}
\label{latetimes}
 \,,
\end{equation}
where we have made use of the relation
$H_{\nu}^{(1)} (k|\tau|) \to -(i/\pi)
\Gamma(\nu) (k|\tau|/2)^{-\nu}$ for $k\tau\to0$ and
$\Gamma(3/2)=\sqrt{\pi}/2$.
In particular for a massless field in de Sitter ($\epsilon=\eta=0$ and
hence $\nu_\R=3/2$) we recover the well-known result
\begin{equation}
{\cal P}_{\delta\phi} \to
\left( \frac{H}{2\pi} \right)^2
\quad {\rm for}~~\frac{k}{aH} \to 0 \,.
\end{equation}

One should be wary of using the exact solution (\ref{han}) at late
times as this is really only valid for the case of constant slow-roll
parameters. At early times (on sub-Hubble scales) this does not matter
as the precise form of $z''/z$ in Eq.~(\ref{veq}) is unimportant for
$k^2\gg z''/z$. Thus Eq.~(\ref{latetimes}) should be valid some time
after Hubble-exit, $k\simeq aH$, where $\nu_\R$ can be taken to be
evaluated in terms of the slow-roll parameters around Hubble-exit,
as these vary only slowly with respect to the Hubble time.
At later times we need to use
a large-scale limit which is most easily derived in terms of the
comoving curvature perturbation, $\R$.

{}From the definition of the comoving curvature perturbation
(\ref{eq:defR}) we see that $\R=(H/\dot\phi)\delta\phi_\psi$.  The
equation of motion (\ref{eq:singlescalareom}) in terms of the comoving
curvature perturbation ${\cal R}$ becomes
%%%%%%%%%%%%%
\begin{eqnarray}
\frac{1}{a^3\epsilon} \frac{{\rm d}}{{\rm d}t}
\left(a^3 \epsilon \dot{\cal R}\right)
+\frac{k^2}{a^2}{\cal R}=0\,.
\label{Req}
\end{eqnarray}
%%%%%%%%%%%%%
In the large-scale limit ($k \to 0$) we obtain the following solution
%%%%%%%%%%%%%
\begin{eqnarray}
{\cal R}=C_1+C_2 \int \frac{{\rm d}t}{a^3\epsilon}\,,
\label{Rsolu}
\end{eqnarray}
%%%%%%%%%%%%%
where $C_1$ and $C_2$ are integration constants.
In most single-field inflationary scenarios (and in all slow-roll
models), the second term can be identified as a decaying mode and
rapidly becomes negligible after the Hubble-exit.  In some
inflationary scenarios with abrupt features in the potential the
decaying mode can give a non-negligible contribution after Hubble-exit
(see Ref.~\cite{Sta92,Leach01,Leach02}), but in this report we will
not consider such cases.

Thus the curvature perturbation becomes constant on super-Hubble
scales and, using Eq.~(\ref{latetimes}) to set the initial amplitude
shortly after Hubble-exit we have
\begin{equation}
{\cal P}_\R = \left( \frac{H}{\dot\phi} \right)^2 {\cal
  P}_{\delta\phi}\simeq
 \left( \frac{H^2}{2\pi\dot\phi} \right)^2_{k=aH} \,,
\end{equation}
to leading order in slow-roll parameters. This can be written in terms
of the value of the potential energy and its first derivative at
Hubble-exit as
\begin{equation}
{\cal P}_\R
= \left( \frac{128\pi}{3m_{\rm pl}^6}\frac{V^3}{V_\phi^2}
\right)_{k=aH} \,.
\label{AS2}
\end{equation}

Since the curvature perturbation is conserved on large scales in
single-field inflation, one can equate the value (\ref{AS2}) at the
first Hubble radius crossing (Hubble exit during inflation) with the
one at the second Hubble radius crossing (Hubble entry during
subsequent radiation or matter-dominated eras).  The COBE
normalization \cite{Bunn96} corresponds to
${\cal P}_\R \simeq 2 \times 10^{-9}$ for the mode
which crossed the Hubble radius about 60
e-folds before the end of inflation.  One can determine the energy
scale of inflation by using the information of the COBE normalization.
For example let us consider the quadratic potential $V(\phi)=\frac12
m_\phi^2 \phi^2$.  Inflation ends at $\epsilon \simeq 1$, giving
$\phi_e\simeq m_{\rm pl}/\sqrt{4\pi}$.  The field value
60 e-folds before the end of inflation
is $\phi_{60} \simeq 3m_{\rm pl}$.
Substituting this value for Eq.~(\ref{AS2}) and using
${\cal P}_\R\simeq 2 \times 10^{-9}$,
the inflaton mass $m_\phi$ is found to be $m_{\phi} \simeq
10^{-6}m_{\rm pl}$.

The spectral index, $n_\R$, is given by
%%%%%%%%%%%%%
\begin{eqnarray}
n_\R-1 =3-2\nu_\R\,.
\label{index}
\end{eqnarray}
%%%%%%%%%%%%%
To leading order in the slow-roll parameters we therefore have
%%%%%%%%%%%%%
\begin{eqnarray}
n_\R
=1-6\epsilon+2\eta\,.
\label{slowns}
\end{eqnarray}
%%%%%%%%%%%%%

Since the parameters $\epsilon$ and $\eta$ are much smaller than unity
during slow-roll inflation, scalar perturbations generated in standard
inflation are close to scale-invariant ($n_{\cal R} \simeq 1$).
When $n_\R<1$ or $n_\R>1$, the power spectrum rises on long
or short
wavelengths we refer to the spectrum as being red or blue,
respectively.  For example in the case of the chaotic inflation with
potential given by Eq.~(\ref{chaoticV}), one has
%%%%%%%%%%%%%
\begin{eqnarray}
n_\R = 1-\frac{n(n+2)}{8\pi}\left(\frac{m_{\rm pl}}
{\phi}\right)^2\,,
\label{nchaotc}
\end{eqnarray}
%%%%%%%%%%%%%
which is a red spectrum.
The hybrid inflation model is able to give rise to a blue spectrum.
In fact evaluating the slow-roll parameters for the potential
(\ref{hysingle}) with the condition $V_0 \equiv M^4/4\lambda \gg
\frac12 m^2\phi^2$, we get the spectral index
%%%%%%%%%%%%%
\begin{eqnarray}
n_\R=1+ \frac{m^2m_{\rm pl}^2}{4\pi V_0}
\left(1-\frac32 \frac{m^2\phi^2}{V_0}\right)\,,
\label{nhybrid}
\end{eqnarray}
%%%%%%%%%%%%%
which gives $n_\R>1$.

We define the running of the spectral tilt as
%%%%%%%%%%%%%
\begin{eqnarray}
\alpha_{\R} \equiv \frac{{\rm d}n_\R}{{\rm d}\rm ln\,k}
\biggr|_{k=aH}\,.
\end{eqnarray}
%%%%%%%%%%%%%
Then $\alpha_\R$ can be written in terms of the slow-roll
parameters defined in (\ref{slowpara}):
%%%%%%%%%%%%%
\begin{eqnarray}
\alpha_\R=16\epsilon \eta- 24\epsilon^2-2\xi^2\,.
\label{slow}
\end{eqnarray}
%%%%%%%%%%%%%
In evaluating this it is useful to note that
the derivative of a quantity $x$
in terms of ${\rm ln}\,k$ can be re-written in terms of the
time-dependence of quantities at Hubble-exit:
%%%%%%%%%%%%%
\begin{equation}
\frac{{\rm d}\,x}{{\rm d}{\rm ln}\,k}\biggr|_{k=aH}
=\left( \frac{{\rm d}\,x}{{\rm d}\,t} \right)
\left(\frac{{\rm d}\,t}{{\rm d} {\rm ln}\,a}\right)
\left(\frac{{\rm d} {\rm ln}\,a}{{\rm d}{\rm
ln}\,k}\right)\biggr|_{k=aH}
= \frac{\dot{x}}{H} \biggr|_{k=aH} \,,
\label{xderi}
\end{equation}
%%%%%%%%%%%%%
where ${\rm d} {\rm ln}\,a/{\rm d}{\rm ln}\,k|_{k=aH} \simeq 1$,
since the variation of $H$ is small during inflation.  Since
Eq.~(\ref{slow}) is second-order in slow-roll parameters, the running
is expected to be small in slow-roll inflation.

As noted in Section \ref{sect:Pert} linear vector perturbations are
constrained to vanish in a scalar field universe. However tensor
perturbations can exist and describe the propagation of free
gravitational waves.
The wave equation for tensor perturbations (\ref{teneq})
can be written in terms of $u=ah/2\sqrt{8\pi G}$, where $h$ is
the amplitude
of the gravitational waves defined in Eq.~(\ref{eq:defh}), as
\begin{equation}
u'' + \left( k^2 - \frac{a''}{a} \right) u = 0 \,.
\end{equation}
This is exactly the same form as the scalar equation (\ref{veq}) where
instead of $z''/z$ given by Eq.~(\ref{gra}) we have
\begin{equation}
\frac{a''}{a} = (aH)^2 (2-\epsilon) \,.
\end{equation}
In the slow-roll approximation this corresponds to
\begin{eqnarray}
\frac{a''}{a} \simeq \frac{\nu_{\rm T}^2-(1/4)}
{\tau^2}\,,~~{\rm with}~~
\nu_\T \simeq \frac32 + \epsilon
 \,.
\end{eqnarray}
Hence neglecting the time dependence of $\epsilon$ and using the same
vacuum normalisation (\ref{vacnorm}) for small-scale modes in the
asymptotic past, we get the tensor power spectrum (\ref{eq:defPT}) on
large scales ($k\ll aH$) to be
\begin{equation}
{\cal P}_{\rm T}\simeq
\frac{64 \pi }{m_{\rm pl}^2} \left( (1-\epsilon)
 \frac{\Gamma(\nu_T)}{\Gamma(3/2)} \frac{H}{2\pi} \right)^2
 \left( \frac{|k\tau|}{2} \right)^{3-2\nu_T}\,.
\label{PT}
\end{equation}

As in the case of scalar perturbations, we can use the exact
solution to the wave equation (\ref{teneq}) in the
long-wavelength limit
\begin{eqnarray}
h = D_1 + D_2 \int \frac{{\rm d}t}{a^3} \,,
\label{hsolu}
\end{eqnarray}
where the constant amplitude, $D_1$, of gravitational waves on
super-Hubble scales is set by Eq.~(\ref{PT}) shortly after
Hubble-exit.
Thus to leading order in slow-roll we have
\begin{eqnarray}
\label{tenamplitude}
{\cal P}_{\rm T} \simeq
\frac{64\pi}{m_{\rm pl}^2}
\left(\frac{H}{2\pi}\right)^2_{k=aH} \simeq
\frac{128}{3} \left( \frac{V}{m_{\rm pl}^4}
\right)_{k=aH}\,.
\end{eqnarray}

The spectral index of tensor perturbations,
$n_{\rm T} \equiv {\rm d}{\rm ln}{\cal P}_{\rm T}
/{\rm d}{\rm ln}k$, is given by
\begin{eqnarray}
n_{\rm T}=-2\epsilon\,,
\label{nT}
\end{eqnarray}
which is a red spectrum.
The running of the tensor tilt,
$\alpha_{\rm T} \equiv
{\rm d}n_{\rm T}/{\rm d}{\rm ln} k$,
is given by
\begin{eqnarray}
\alpha_{\rm T}=-4\epsilon(2\epsilon-\eta)\,.
\label{alT}
\end{eqnarray}

An important observational quantity is the
tensor to scalar ratio which is defined as
%%%%%%%%%%%%%
\begin{eqnarray}
r \equiv \frac{{\cal P}_{\rm T}}
{{\cal P}_\R} \simeq 16\epsilon\,.
\label{ratio}
\end{eqnarray}
%%%%%%%%%%%%%
Note that the definition of $r$ is the same as in
Refs.~\cite{Peiris03,Barger03,Tegmark03}
but differs from the ones in
Refs.~\cite{Kinney04,Leach03}.
Since $\epsilon \ll 1$, the amplitude of tensor perturbations is
suppressed
relative to that of scalar perturbations.
{}From Eqs.~(\ref{nT}) and (\ref{ratio}) one gets the relation between
$r$ and $n_{\rm T}$, as
%%%%%%%%%%%%%
\begin{eqnarray}
r=-8n_{\rm T}\,.
\label{consistency}
\end{eqnarray}
%%%%%%%%%%%%%
This is the so-called consistency relation \cite{LLKCreview} for
single-field slow-roll inflation.
The same relation is known to hold in some braneworld models of
inflation \cite{Huey01} as well as the 4-dimensional dilaton gravity and
generalized Einstein theories \cite{TsujiBu}.
But this is also modified in the case of multifield inflation
\cite{Sasaki95,Garcia96,Bartolo01p2,Wands02,Tsuji03},
as we shall see later.

%%%%%%%%%%%%%%%%%%%%%%%%%%%%%%%%%%%%%%%%%%%%%%%%%%%%%
\section{Observational constraints on single-field
inflation from CMB}
%%%%%%%%%%%%%%%%%%%%%%%%%%%%%%%%%%%%%%%%%%%%%%%%%%%%%

\subsection{Likelihood analysis of inflationary model parameters}

In this section we place constraints on single-field slow-roll
inflation using a compilation of observational data.
As outlined in the previous subsection, we have 6 inflationary
parameters, i.e., $A_\R^2$, $r$, $n_\R$, $n_{\rm T}$,
$\alpha_\R$ and $\alpha_{\rm T}$.
Since the latter 5 quantities are written in terms of the slow-roll
parameters $\epsilon$, $\eta$ and $\xi$, we have 4 free parameters
($A_\R^2, \epsilon, \eta, \xi$).
Let us introduce horizon flow parameters defined by
\cite{Leach02}
\begin{equation}
\epsilon_0=\frac{H_{\rm inf}}{H}\,,~~~
\epsilon_{i+1}=\frac{{\rm d ln} |\epsilon_i|}{{\rm d}N }\,,~~~
(i \ge 0)\,,
\label{hflow}
\end{equation}
where $H_{\rm inf}$ is the Hubble rate at some chosen time
and in terms of the slow-roll parameters defined in Eq.~(\ref{ep}) we have
\begin{equation}
\epsilon_1 = \epsilon \,, \quad
\epsilon_2 = 4\epsilon - 2\eta
 \,.
\end{equation}
Then the above inflationary observables may be rewritten as
\begin{eqnarray}
& & n_\R = 1-2\epsilon_1-\epsilon_2,~~n_{{\rm T}} =
-2\epsilon_1,~~r= 16\epsilon_1,~~ \nonumber \\
& &\alpha_\R= -2\epsilon_1 \epsilon_2-
\epsilon_2 \epsilon_3\,,~~
\alpha_{\rm T}= -2\epsilon_1 \epsilon_2\,.
\label{lowslow}
\end{eqnarray}
These expressions are convenient when we compare them
with those in braneworld inflation.

Various analyses of the four parameters $A_\R^2$, $r$, $n_\R$ and
$\alpha_\R$ have been done using different sets of observational data.
The availability of the WMAP satellite CMB data revolutionised studies
of inflation \cite{Peiris03}.  Analysis is
typically carried out using using the Markov Chain Monte Carlo method
\cite{mcmc,mcmc2} which allows the likelihood distribution to be
probed efficiently even with large number of parameters where direct
computation of the posterior distribution is computationally
impossible. User-friendly codes such as CosmoMc (Cosmological Monte
Carlo)\footnote{http://cosmologist.info/cosmomc/}
\cite{antony00,antony02} and the C++ code
CMBEASY\footnote{http://www.cmbeasy.org} \cite{cmbeasy,cmbeasy2} has
made it easy to compare model predictions for the matter power
spectrum and the CMB temperature and polarisation spectra with the
latest data.

Examples of such analyses applied to inflation include study of first
year WMAP data only \cite{Peiris03,Barger03,Kinney04}, WMAP + 2df
\cite{Leach03}, 2df + WMAP + SDSS \cite{Tegmark03,Tsuji04,TsujiBu}.
Each new data set provides incremental improvements.  For example, the
$2\sigma$ upper limits of $\epsilon_1$ and $\epsilon_2$ are currently
$0<\epsilon_1 <0.032$ and $-0.15<\epsilon_2<0.08$ in
Ref.~\cite{Leach03}. As of late 2005, the parameter $\epsilon_3$ is
poorly constrained and is currently consistent with zero, which means
that the current observations have not reached the level at which the
consideration of higher-order slow-roll parameters is necessary.

%
% Here I removed the figure 1 and modified several sentences. by S.T.
%

\begin{figure}
\begin{center}
\includegraphics[height=3.5in,width=3.5in]{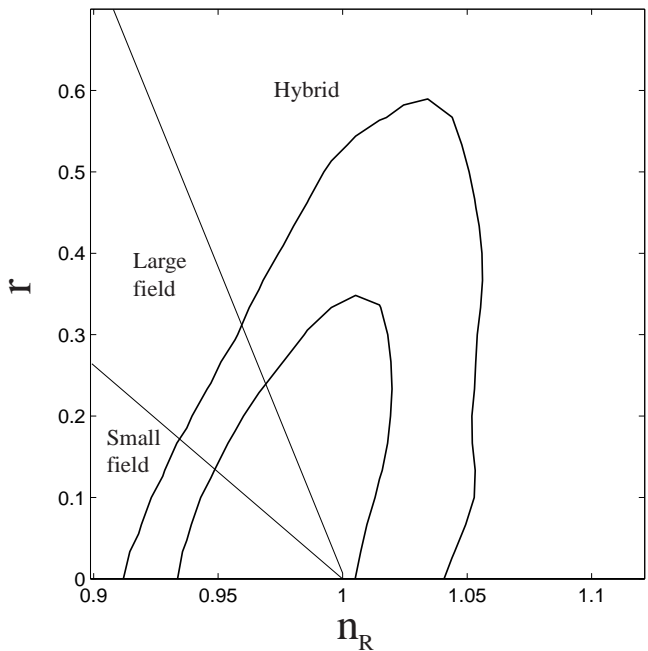}
\caption{\label{classification} Classification of inflationary models
in the $n_\R$-$r$ plane in the low-energy limit.
The line $r=(8/3)(1-n_\R)$
marks the border of large-field and small-field models,
whereas the border of large-field and hybrid models
corresponds to $r=8(1-n_\R)$. }
\end{center}
\end{figure}

In Fig.~\ref{classification} we show the $1\sigma$ and
$2\sigma$ observational contour bounds for $n_\R$ and $r$
found in an analysis which includes the  four
inflationary variables ($n_\R$, $r$, $\epsilon_3$,
$A_\R$) and four cosmological parameters
($\Omega_b h^2$, $\Omega_c h^2$, $Z=e^{-2\tau}$, $H_0$).
Here $\Omega_b h^2$ and $\Omega_c h^2$ are the baryon and
dark matter density, $\tau$ is the optical depth, and
$H_0$ is the Hubble constant.
We assume a flat, $\Lambda$CDM universe and use the SDSS + 2df + first year
WMAP data.

Note that we used the relation (\ref{lowslow}), which gives
the values of $n_{\rm T}$, $\alpha_\R$ and
$\alpha_{\rm T}$ in terms of $n_\R$, $r$ and $\epsilon_3$.
The amplitude of scalar perturbations is
distributed around $A_\R^2=24 \times 10^{-10}$,
which corresponds to the COBE normalization
mentioned in the previous section.
The spectral index $n_\R$ and the tensor to scalar ratio $r$
are consistent with the prediction of the slow-roll limit in
single-field inflation ($n_\R=1$ and $r=0$).

The amplitude of scalar perturbations can be written as
$A_\R^2=\frac{1}{\pi \epsilon_1}\left(\frac{H}
{m_{\rm pl}}\right)^2$.
We can use the constraint
$A_\R^2 \simeq 2.4 \times 10^{-9}$ and
$\epsilon_1<0.032$ to obtain an upper limit on the energy scale
of inflation:
\begin{equation}
\frac{H}{m_{\rm pl}}<1.55 \times 10^{-5}\,.
\label{Hlimit}
\end{equation}
Intriguingly,
the $n_{\cal R} = 1$, pure Harrison-Zel'dovich value (corresponding to $\epsilon_1=0=\epsilon_2$),
is still consistent with the data. A clear, unambiguous detection of non-zero $\epsilon_1$ will immediately set the scale for inflation and will be a crucial step forward in building realistic inflationary models.

%%%%%%%%%%%%%%%%
%%%%%%%%%%%%%%%%

While there is no signature in CMB data of statistically
significant deviations from the predictions of
the single-field inflationary paradigm, the suppressed
quadrupole \cite{WMAP} is
rather unexpected. Although the lack of power on the largest scales may be purely due to cosmic
variance and hence statistically insignificant \cite{Efs03}, theoretically motivated explanations are not ruled out; see, e.g.
Refs.~\cite{Yokoyama99,Tsujiloop,Tsuji03noncom,Piao03,Piao04,Contaldi03,Feng03,Kawasaki03,BFM03,LMMR04,AS03,SP05}
for a number of attempts to explain this loss of power on the largest scales.

%%%%%%%%%%%%%%%%%%%%%%%%%%%%
\subsection{Classification of inflation models
in the $n_\R$-$r$ plane}
%%%%%%%%%%%%%%%%%%%%%%%%%%%%

By Eqs.~(\ref{slowns}) and (\ref{ratio}) the general relation
between $n_\R$ and $r$ is
\begin{eqnarray}
r=\frac83(1-n_\R)+\frac{16}{3}\eta\,.
\label{Rnrelation}
\end{eqnarray}
The border of large-field and small-field models is given
by the linear potential
\begin{equation}
V=V_0\phi\,.
\label{linear}
\end{equation}
Since $V_{\phi \phi}$ vanishes in this case (i.e., $\eta=0$),
we have $n_\R-1=-6\epsilon$ and
\begin{eqnarray}
r= \frac83(1-n_\R)\,.
\label{nandR}
\end{eqnarray}

The exponential potential \cite{LM85,Yokoyama88}
\begin{equation}
V=V_{0}\exp \left(-\sqrt{\frac{16\pi}{\alpha}}
\frac{\phi}{m_{{\rm pl}}}\right)\,,
\label{exp}
\end{equation}
characterizes the $n\to\infty$ limit of the large field models in
Eq.(\ref{chaoticV}) and hence the border between large-field and hybrid models.
In this case we have $\eta=2\epsilon=2/\alpha$ and
\begin{eqnarray}
r=8(1-n_\R)\,.
\label{lowcon}
\end{eqnarray}
Then we can classify inflationary
models such as (A) large-field ($0<\eta \le 2\epsilon)$,
(B) small-field ($\eta<0$), and (C) hybrid models ($\eta>2\epsilon$).
This is illustrated in Fig.~\ref{classification}.
The allowed range of hybrid models is wide
relative to large-field and small-field models.
We note that double inflation models are not categorised
in above classes, since the discussion of density perturbations
in the single-field case is not valid.
We shall discuss this case separately in a later section.

The large-field potential (\ref{chaoticV}) involves only one
free parameter, $V_0$, for a given value of $n$.
The small-field potential (\ref{smallfield})
has two parameters $V_0$ and $\mu$.
The hybrid model involves more free parameters, e.g.,
$g, \lambda, M, m$ (4 parameters)
for the potential (\ref{hybridpo}).
This implies that the small-field and hybrid models are difficult
to be constrained relative to the large-field models,
since these have additional freedom to be compatible with
observational data.
In fact the large-field models are severely constrained from
current observations
\cite{Barger03,Kinney04,Leach03,WMAP},
while it is not so for small-field and hybrid models
due to additional model parameters.
In the next subsection we shall discuss the observational
constraint on large-field models.

%%%%%%%%%%%%%%%%%%%%%%%%%%%%%%
\subsection{Observational constraints on large-field inflation}
%%%%%%%%%%%%%%%%%%%%%%%%%%%%%%

\begin{figure}
\begin{center}
\includegraphics[height=3.5in,width=3.5in]{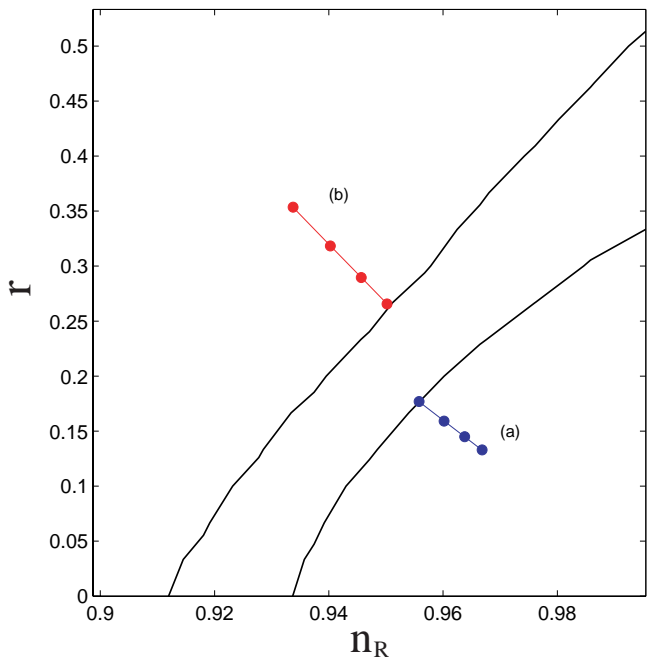}
\caption{\label{largef}
Theoretical prediction of large-field models
together with the $1\sigma$ and $2\sigma$
observational contour bounds from first year WMAP data.
Each case corresponds to (a) $n=2$ and (b) $n=4$
with e-foldings $N=45, 50, 55, 60$ (from top to bottom)
showing how models with few e-foldings
are under severe pressure from observations.}
\end{center}
\end{figure}

Let us consider the monomial potential (\ref{chaoticV}).
In this case the number of e-foldings is given as
$N=4\pi/(nm_{\rm pl}^2)(\phi^2-\phi_f^2)$
with $\phi_f=nm_{\rm pl}/4\sqrt{\pi}$ being the value of
inflaton at the end of inflation. Then the spectral
index $n_\R$ and the tensor to scalar ratio $r$
are written in terms of the function of $N$:
\begin{eqnarray}
n_\R=1-\frac{2(n+2)}{4N+1}\,,~~~
r=\frac{16n}{4N+1}\,.
\label{nsR}
\end{eqnarray}
Note that these are independent of the energy
scale $V_0$. In Fig.\,\ref{largef} we plot
the theoretical values (\ref{nsR})
for the quadratic ($n=2$) and the quartic ($n=4$)
potentials with several different values of $N$.
The predicted points for the quadratic potential are
within the $1\sigma$ observational contour bound
for the e-foldings greater than $N=45$,
thus preferable observationally.
The quartic potential is outside of the
$2\sigma$ contour bound for
the e-foldings less than $N=60$.
Therefore the $n=4$ case is under strong
observational pressure even with first year WMAP data
unless the number of
e-foldings is sufficiently large\footnote{
For quartic potential the number of e-foldings corresponding
to the scale at which observable perturbations are
generated is estimated to be $N \sim 64$
by assuming instant transitions between several
cosmological epochs \cite{Liddle03efold}.} ($N>60$).
This situation is improved if the inflaton is
coupled to gravity with a negative non-minimal
coupling \cite{KF99,TsujiBu}.

%%%%%%%%%%%%%%%%
%%%%%%%%%%%%%%%%

The small-field and hybrid models
involve more than two parameters, so we have a freedom
to fit the model parameters so that it matches with the
observational constraints. In this sense we can not currently rule out
these models, although some of the model parameters can
be constrained.

%%%%%%%%%%%%%%%%%%%%%%%%%%%%%%%%%%%%%%%%%%%
\section{Perturbations generated in higher-dimensional
models}
%%%%%%%%%%%%%%%%%%%%%%%%%%%%%%%%%%%%%%%%%%%

There has been a lot of interest in the construction of early
universe scenarios in higher-dimensional models motivated by
string/M-theory. A well-known example is the Randall-Sundrum (RS)
braneworld scenario \cite{RSI,RSII}, in which our four-dimensional
brane is embedded in a five-dimensional bulk spacetime (see
Ref.\,\cite{Maa03,BVD} for reviews). In this scenario standard
model particles are confined on the brane, while gravitons
propagate in the bulk spacetime. Since the effect of the extra
dimension induces additional terms at high energies, e.g., a
quadratic term of energy density \cite{BDL,SMS}, this can lead to
a larger amount of inflation relative to standard inflationary
scenarios \cite{Maa99,Cope01}.

In conventional Kaluza-Klein theories, extra dimensions are
compactified on some internal manifold in order to obtain a
four-dimensional effective gravity. A simple cosmological model
using toroidal compactifications is the Pre-Big-Bang (PBB)
scenario \cite{Vene91,Gas93},
which is based upon the
low-energy, tree-level string effective
action (see also Refs.~\cite{LWCreview,Gas02}). In this scenario
there exist two branches--one of which is a dilaton-driven super
inflationary stage and another is the Friedmann branch with a
decreasing curvature. It is possible to connect the two branches
by taking into account string loop and derivative corrections to
the tree-level action \cite{Bru97,Foffa99,Cartier99}. If we transform the
string-frame action to the four-dimensional Einstein frame, the
universe exhibits a contraction with $a \propto (-t)^{1/3}$ in the
PBB phase. Therefore  the Pre-Big-Bang (PBB) scenario can be
viewed as a bouncing cosmological model in the Einstein frame.

The ekpyrotic \cite{Khoury01} and cyclic \cite{Stein02} models
have a similarity to the PBB scenario in the sense that the
universe contracts before a bounce. In ekpyrotic/cyclic scenarios
the collision of two parallel branes embedded in an
extra-dimensional bulk signals the beginning of the hot,
expanding, big bang of standard cosmology.

Models with a cosmological bounce potentially provide an
alternative to inflation in addressing the homogeneity problem of
big-bang cosmology and in yielding a causal mechanism of structure
formation.  In this sense it is important to evaluate the spectra
of density perturbations in order to make contact with
observations and distinguish these models from the inflationary
scenario.

More recently there has been a lot of effort to construct more
conventional inflationary models in string theory using D-branes
(and anti D-branes) with a flux compactification in a warped
geometry to give rise to de Sitter solutions in four-dimensions.
We do not have enough space to review this emerging field, but
refer readers to the other papers
\cite{Dvali99,Quevedo02,Kachru03,Bur04,Blanco04,Garousi04,KSW05}.
In principle we can
evaluate the spectra of perturbations using the method in the
previous sections once the effective potential of the inflaton is
known in an effective 4-dimensional theory in four-dimensional
gravity.

In the rest of this section we shall review brane-world, PBB and
ekpyrotic/cyclic models in separate subsections.

%%%%%%%%%%%%%%
\subsection{Braneworld}
%%%%%%%%%%%%%%

In the RSII model \cite{RSII} the Einstein  equations on our
3-brane can written as \cite{SMS}:
\beq
^{(4)}G_{\mu\nu}=-\Lambda_4 g_{\mu\nu}+\frac{8\pi}{m_{{\rm pl}}^2}
T_{\mu\nu}+ \left(\frac{8\pi}{M_5^3}\right)^2
\pi_{\mu\nu}-E_{\mu\nu}\,,
\label{geq}
\eeq
where $T_{\mu\nu}$ and $\pi_{\mu\nu}$ represent the
energy--momentum tensor on the brane and a term quadratic in
$T_{\mu\nu}$, respectively. $E_{\mu\nu}$ is a projection of the
5-dimensional Weyl tensor, which carries information about the
bulk gravity.  The 4- and 5-dimensional Planck masses, $m_{{\rm
pl}}$ and $M_5$, are related via the 3-brane tension, $\lambda$,
as
\beq
\lambda=\frac{3}{4\pi}\frac{M_5^6}{m_{{\rm pl}}^2}\,.
\label{M5}
\eeq
In what follows the 4-dimensional cosmological constant
$\Lambda_4$ is assumed to be zero.

The Friedmann equation in the flat FRW background becomes
\beq
H^2 \equiv \left(\frac{\dot{a}}{a}\right)^2=
\frac{8\pi}{3m_{{\rm pl}}^2}
\rho \left(1+\frac{\rho}{2\lambda}\right)\,,
\label{hubblebra}
\eeq
where $\rho$ is the energy density of the matter on the brane.
At high energies the $\rho^2$ term can play an important role in
determining the evolution of the Universe.
We neglected the contribution of the so-called ``dark radiation'',
$E_{\mu\nu}$, which decreases as $\sim a^{-4}$ during
inflation. However we caution that this may be important in
considering perturbations at later stages of cosmological
evolution \cite{Koyama03,Rho03}.

The inflaton field $\phi$, confined to the brane, satisfies the 4D
Klein--Gordon equation given in Eq.~(\ref{phieq}). The quadratic
contribution in Eq.~(\ref{hubblebra}) increases the Hubble
expansion rate during inflation, which makes the evolution of the
inflaton slower than in the case of standard General Relativity.
Combining Eq.~(\ref{phieq}) with Eq.~(\ref{hubblebra}), we obtain
\cite{Maa99,TMM01}
\beqa
\frac{\ddot{a}}{a}=\frac{8\pi}{3m_{{\rm pl}}^2}
\left[(V-\dot{\phi}^2)+\frac{\dot{\phi}^2+2V}{8\lambda}
(2V-5\dot{\phi}^2)\right]\,.
\label{ddota}
\eeqa
The condition for inflation is $\ddot{a}>0$, which reduces to the
standard expression $V>\dot{\phi}^2$ for
$(\dot{\phi}^2+2V)/8\lambda \ll 1$. In the high-energy limit, this
condition corresponds to $V>(5/2)\dot{\phi}^2$.

It was shown in Ref.~\cite{Wands00} that the conservation of the
curvature perturbation, ${\cal R}$, holds for adiabatic
perturbations irrespective of the form of gravitational equations
by considering the local conservation of the energy-momentum
tensor. One has $|{\cal R}| \simeq (H/\dot{\phi})\delta \phi
\simeq (H/\dot{\phi})(H/2\pi)$ after the Hubble-radius crossing,
as in the case of standard General Relativity discussed in Sec.
IV. Then we get the amplitude of scalar perturbations, as
\cite{Maa99}
\begin{eqnarray}
\label{scala}
 {\cal P}_\R  \simeq \frac{H^2}{\dot{\phi}^2}
\left(\frac{H}{2\pi}\right)^2
 \simeq \frac{128 \pi}{3 m_{{\rm pl}}^6} \,
\frac{V^3}{V_\phi^2} \, \left( 1 + \frac{V}{2\lambda} \right)^3
\,,
\end{eqnarray}
which is evaluated at the Hubble radius crossing, $k=aH$. Note
that it is the modification of the Friedmann equation that changes
the form of ${\cal P}_\R$ when it is expressed in terms of the
potential.

Tensor perturbations in cosmology are more involved since
gravitons propagate in the bulk. The equation for gravitational
waves in the bulk corresponds to a partial differential equation
with a moving boundary, which is not generally separable. However
when the evolution on the brane is de Sitter, it is possible to
make quantitative predictions about the evolution of gravitational
waves in slow-roll inflation. The amplitude of tensor
perturbations was evaluated in Ref.~\cite{LMW00}, as
\begin{eqnarray}
\label{tensa} {\cal P}_{\rm T}
= \frac{64\pi}{m_{\rm Pl}^2}
\left(\frac{H}{2\pi}\right)^2 F^2(x)\,,
\end{eqnarray}
where $x=Hm_{{\rm pl}}\sqrt{3/(4\pi\lambda)}$ and
\begin{eqnarray}
F(x) =\left[\sqrt{1+x^2} - x^2 \sinh^{-1}(1/x)
\right]^{-1/2}\,.
\label{F}
\end{eqnarray}
Here the function $F(x)$ appeared from the normalization
of a zero-mode.

The spectral indices of scalar and tensor perturbations are
\begin{eqnarray}
\label{ntiltS}
n_{{\cal R}} - 1 = -6 \epsilon+ 2 \eta\,,~~~
n_{{\rm T}} =-\frac{2}{N_\phi}
\frac{x_\phi}{x}
\frac{F^2}{\sqrt{1+x^2}}\,,
\label{ntiltT}
\end{eqnarray}
where the modified slow-roll parameters are defined by
\begin{eqnarray}
\epsilon & \equiv & \frac{m_{{\rm pl}}^2}{16\pi} \, \left(
        \frac{V_\phi}{V} \right)^2 \; \frac{1 + V/\lambda}{\left(1 +
        V/2\lambda \right)^2} \,, \\
\eta & \equiv & \frac{m_{{\rm pl}}^2}{8\pi} \,
          \frac{V_{\phi \phi}}{V} \;
        \frac{1}{1+V/2\lambda}  \,,
\label{eta}
\end{eqnarray}
together with the number of e-foldings
\begin{equation}
\label{efolds}
N \simeq - \frac{8\pi}{m_{{\rm pl}}^2} \int^{\phi_f}_{\phi}
\frac{V}{V_\phi}
\left( 1+\frac{V}{2\lambda} \right) {\rm d} \phi \,.
\end{equation}

By Eqs.~(\ref{scala}), (\ref{tensa}) and (\ref{ntiltT}), one can
show that the same consistency relation Eq.~(\ref{consistency})
relates the tensor-scalar ratio to the tilt of the gravitational
wave spectrum, independently of the brane tension
$\lambda$ \cite{Huey01} (see also Refs.~\cite{Cal03,Cal03v2,Ramirez03}).
This degeneracy of the consistency relation means that to lowest order
in slow-roll parameters it is not possible to observationally
distinguish perturbations spectrum produced by braneworld
inflation models from those produced by 4D inflation with a
modified potential \cite{LiddleTaylor}. If one uses horizon-flow
parameters defined in Eq.~(\ref{hflow}), we obtain in the high-energy
($V\gg\lambda$) limit \cite{Tsuji04,Calshinji}
\begin{eqnarray}
\label{highslow}
n_{{\cal R}} &=& 1-3\epsilon_1-\epsilon_2,~n_{{\rm T}} =
-3\epsilon_1,~
r=24\epsilon_1\,, \nonumber \\
\alpha_{{\cal R}} &=& -3\epsilon_1 \epsilon_2-
\epsilon_2 \epsilon_3\,,~
\alpha_{\rm T}=-3\epsilon_1 \epsilon_2
\quad ({\rm for}~~V/\lambda \gg 1)\,.  \nonumber \\
\end{eqnarray}
We note that these results are identical to those given for 4D general
relativity in Eq.~(\ref{lowslow}) if one replaces $2\epsilon_1$ in
Eq.~(\ref{lowslow}) with $3\epsilon_1$ in Eq.~(\ref{highslow}).

This correspondence suggests that a separate likelihood analysis of
observational data is not needed for the braneworld scenario, as
observations can be used to constrain the same parameterisation
of the spectra produced. Therefore the observational contour
bounds in Fig.~\ref{classification}
can be used in braneworld as well.
However, when those constraints are then interpreted in terms of the
form of the inflationary potential,
differences can be seen depending on the regime we are in.
In what follows we will obtain observational constraints on
large-field potentials (\ref{chaoticV}) under the
assumption that we are in the high-energy regime ($\rho \gg \lambda$).

One can estimate the field value at the end of inflation by setting
$\epsilon (\phi_f)=1$. Then by Eqs.~(\ref{highslow}) and
(\ref{efolds}), we get
\begin{eqnarray}
\label{nbrane}
n_{{\cal R}}-1 &=& -\frac{2(2n+1)}{N(n+2)+n}\,, \\
r &=& \frac{24n}{N(n+2)+n}\,.
\label{Rbrane}
\end{eqnarray}
Since $N(n+2) \gg n$ for the e-folds $N>50$,
one can neglect the second term as in Ref.~\cite{Tsuji04}.
For a fixed value of $n$, $n_{\cal R}$ and $r$
are only dependent on $N$.

\begin{figure}
\includegraphics[scale=0.6]{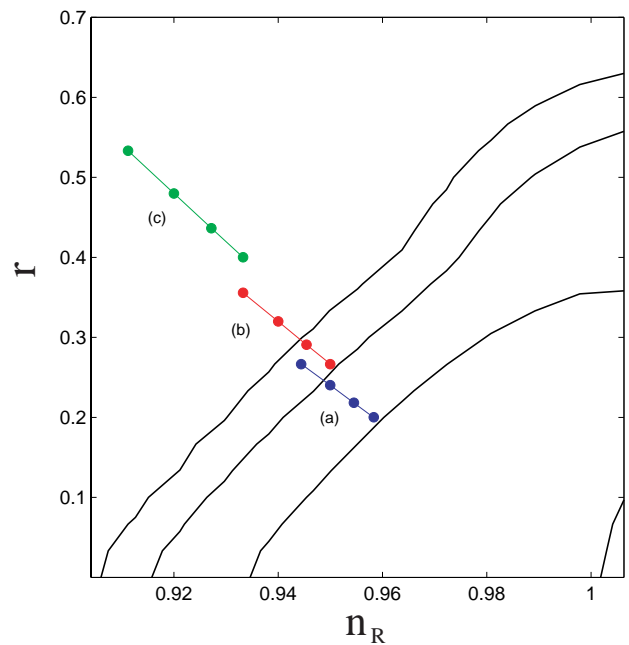}
\caption{
Constraints on large-field models in the context of
braneworld \cite{Tsuji04}.
We show theoretical predictions
together with the $1\sigma$, $2\sigma$
and $3\sigma$ observational contour bounds.
Each case corresponds to
(a) $n=2$, (b) $n=4$
and (c) $n \to \infty$ (exponential potentials),
respectively, with e-foldings
$N=45, 50, 55, 60$ (from top to bottom).
}
\label{braneconst}
\end{figure}

The quadratic potential ($n=2$) is within the $2\sigma$
observational contour bound for $N>50$ as found from
Fig.~\ref{braneconst}. The quartic potential is outside the
$2\sigma$ bound for $N<60$, which means that this model is under
strong observational pressure. Note that the theoretical points
tend to be away from the point $n_{\cal R}=1$ and $r=0$ compared
to the standard General Relativistic inflation. Exponential
potentials correspond to the limit $n \to \infty$, in which case
we have $n_{\cal R}-1=-4/N$ and $r=24/N$ from Eqs.~(\ref{nbrane})
and (\ref{Rbrane}). This case does not lie within the $2\sigma$
bound unless $N > 90$. Therefore steep inflation \cite{Cope01}
driven by an exponential potential is excluded
observationally \cite{Liddle03,Tsuji04},
unless other effects coming from a higher-dimensional bulk
modifies the spectra of perturbations.

%%%%%%%%%%%%%%%%%%%%
%%%%%%%%%%%%%%%%%%%%

This situation changes if we consider the Gauss-Bonnet (GB)
curvature invariant \cite{Lidsey03} in five dimensional gravity,
arising from leading-order quantum corrections of the low-energy
heterotic gravitational action \cite{TSM04}. One effect of the GB
term is to break the degeneracy of the standard consistency
relation \cite{DLMS04}. Although this does not lead to a
significant change for the likelihood results of inflationary
observables, the quartic potential is rescued from marginal
rejection for a wide range of energy scales \cite{TSM04}.

\begin{figure}
\includegraphics[scale=0.6]{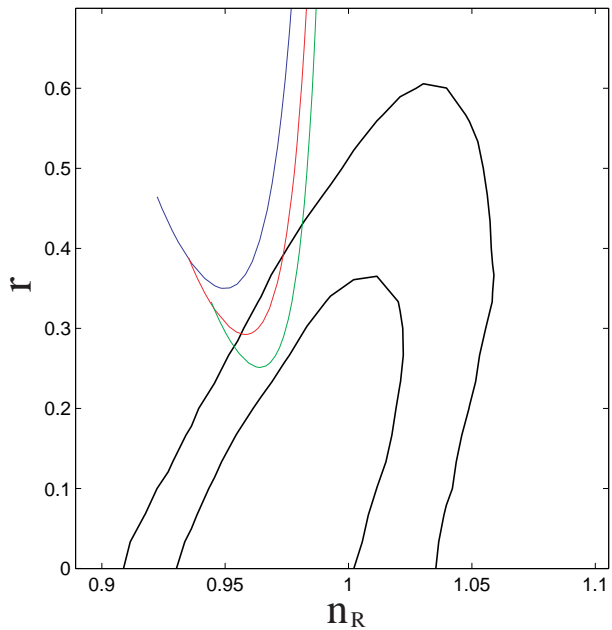}
\caption{
Observational constraints on
exponential potentials in a Gauss-Bonnet braneworld
for the e-foldings $N=50, 60, 70$ (from top to bottom).
The theoretical curves are inside the $2\sigma$
contour bound for the e-folds $N>55$.
}
\label{gaussbon}
\end{figure}

Even steep inflation exhibits marginal compatibility for a sufficient
number of e-foldings. This property is illustrated in
Fig.~\ref{gaussbon}. In Gauss-Bonnet (GB) braneworld the
background equation is given as $H^2 \propto \rho^{2/3}$ in a
high-energy regime, whereas the RS regime is characterised by $H^2
\propto \rho^2$. In both regions, the ratio $r$ is larger than in
the case of General Relativity ($H^2 \propto \rho$). The tensor to
scalar ratio $r$ has a minimum in the intermediate energy region
between the Gauss-Bonnet (extreme right) and Randall-Sundrum
(extreme left) regimes \cite{TSM04}. As seen in
Fig.~\ref{gaussbon} exponential potentials tend to enter the
$2\sigma$ contour bound for $N>55$, thus showing the observational
compatibility. (see Refs.~\cite{DLMS04,TSM04} for more details).

%%%%%%%%%%%%%%%%%%%%
%%%%%%%%%%%%%%%%%%%%

Finally we note that braneworld effects on the evolution of
perturbations after the second Hubble radius crossing can
potentially leave signatures on the temperature anisotropies in
CMB, but the techniques for calculating these signatures are still
under development \cite{Koyama03,Rho03}. While it is generally
complicated to fully solve the perturbation equations in a
higher-dimensional bulk coupled to matter perturbations on the
brane \cite{Kodama00,Koyama00,vande00,BMW,KMW05,Mukoh00},
it is of great interest to see how the effect of five-dimensional
gravity affects the CMB power spectra by solving
the bulk geometry consistently.

%%%%%%%%%%%%%%%%%%%%%%%%%%%%%%
\subsection{Pre-big-bang and ekpyrotic/cyclic cosmologies}
%%%%%%%%%%%%%%%%%%%%%%%%%%%%%%

The PBB scenario can be characterized in four-dimensions by an
effective action in the string
frame \cite{Vene91,Gas93}:
\be
 \label{effactions}
{\cal S}_S \, = \, \int \rd^4x \sqrt{-g} e^{-\phi}
\left[ \frac12 R+\frac12 (\nabla \phi)^2-V_S(\phi) \right]\,,
\ee
where $\phi$ is the dilaton field with potential $V_S(\phi)$.
Note that we neglect here additional modulus fields corresponding to
the size and shape of the internal space of extra dimensions.
The potential for the dilaton
vanishes in the perturbative string effective action.
The dilaton evolves from a weakly coupled regime ($e^{\phi} \ll 1$)
toward a strongly coupled region during which the Hubble
parameter grows (superinflation).
This PBB branch connects to a Friedmann one with a decreasing
Hubble rate if the singularity can be avoided
leading to a maximum value for the Hubble parameter.

If we make a conformal transformation
\be
 \label{conformalframe}
\hat{g}_{\mu\nu}=e^{-\phi}g_{\mu\nu}\,,
\ee
the action in the Einstein frame can be written as
\be \label{effactione}
{\cal S}_E \, = \, \int \rd^4x \sqrt{-\hat{g}}\left[ \frac12 \hat{R}
- \frac14 (\hat{\nabla} \phi)^2-V_E(\phi) \right]\,,
\ee
where $V_E(\phi) \equiv e^{\phi}V_S(\phi)$.
Introducing a rescaled field $\varphi=\pm \phi/\sqrt{2}$,
the action (\ref{effactione}) reads
\be \label{effactione2}
{\cal S}_E \, = \, \int \rd^4x \sqrt{-\hat{g}}\left[ \frac12 \hat{R}
-\frac12 (\hat{\nabla} \varphi)^2-V_E(\phi(\varphi)) \right]\,.
\ee
Then the action (\ref{effactions}) can be used to describe both the
PBB model in the Einstein frame, as well as the ekpyrotic scenario
\cite{Durrer02}.

In the original version of the ekpyrotic scenario \cite{Khoury01}, the
Einstein frame is used where the coupling to the Ricci curvature is
fixed, and the field $\phi$ describes the separation of a bulk brane
from our four-dimensional orbifold fixed plane.  In the case
of the second version of the ekpyrotic scenario \cite{Khoury02} and in
the cyclic scenario \cite{Stein02}, $\phi$ is the modulus field
denoting the size of the orbifold (the separation of the two orbifold
fixed planes).

The ekpyrotic scenario is characterized by a negative
exponential potential \cite{Khoury01}
\be \label{einpoten}
V_E = - V_{0} \exp
\left(-\sqrt{\frac{2}{p}}\,\varphi\right)\,,
\ee
with $0 < p \ll 1$.
The branes are initially widely separated but are approaching each
other, which means that $\varphi$ begins near $+\infty$ and is
decreasing toward $\varphi = 0$.
The universe exhibits a contraction in this phase
in the Einstein frame.
In the PBB scenario the dilaton starts out from a weakly coupled
regime with $\phi$ increasing from $-\infty$.
If we want the potential (\ref{einpoten}) to describe a
modified PBB scenario with a dilaton potential which is important
when $\phi \to 0$ but negligible for $\phi \to - \infty$, we
have to use the relation $\varphi=-\phi/\sqrt{2}$ between the field
$\varphi$ in the ekpyrotic case and the dilaton $\phi$
in the PBB case.

In the flat FRW background the system with the exponential
potential (\ref{einpoten}) has the following exact
solution \cite{LM85,Lyth02ekp1,Durrer02,TBF02,Hwang02eky,Tsujiekp,HW02}
\be \label{ekysolution}
a_E \propto |t_E|^p,~H_E=\frac{p}{t_E},~
V_E=-\frac{p(1-3p)}{t_E^2},~\dot{\varphi}=
\frac{\sqrt{2p}}{t_E} \,,
\ee
where $p>0$ and the subscript ``{\it E}\,'' denotes the quantities
in the Einstein frame.
The solution for $t_E<0$ describes the contracting universe
prior to the collision of branes.
Note that the PBB scenario corresponds to $p=1/3$, in which case
the potential of the dilaton is absent.
The ekpyrotic scenario corresponds to a slow contraction
with $0<p \ll 1$.
In the string frame we have \cite{Durrer02,TBF02}
\be
\label{ekysolution2}
 a_S \propto (-t_S)^{-\sqrt{p}},~~~
\phi=-\frac{2\sqrt{p}}{1-\sqrt{p}}
\ln \left[ -(1-\sqrt{p})t_S\right] \,.
\ee
This illustrates the super-inflationary solution with growing dilaton.

Let us evaluate the spectrum of scalar perturbations generated in
the contracting phase given by Eq.~(\ref{ekysolution}).
In this case we have $\epsilon=1/p$ and $\nu_\R^2=[(3p-1)/(1-p)]^2$
in Eq.~(\ref{gra2}). Then by using Eq.~(\ref{index})
we obtain the spectral index of curvature
perturbations \cite{Wands98,Lyth02ekp1}
(see also Refs.~\cite{Lyth02ekp2,Bran01,Hwang02eky,Tsujiekp,TBF02,Allen04,Finelli02a}):
\begin{eqnarray}
\label{speekp}
n_{\cal{R}}-1 &=& \frac{2}{1-p}~~~({\rm for}~~0<p \le 1/3)\,, \\
                      &=& \frac{4-6p}{1-p}~~~({\rm for}~~1/3 \le p
<1)\,.
\label{tilt_po}
\end{eqnarray}
We can obtain the above exact result of the perturbation spectra
for exponential potentials without using slow-roll approximations.
We see that a scale-invariant spectrum with $n_\R=1$ is obtained
either as $p\to\infty$ in an expanding universe, corresponding to
conventional slow-roll inflation, or for $p=2/3$ during collapse
\cite{Star79,Wands98}.
In the case of the PBB cosmology ($p=1/3$) one has
$n_{\cal{R}}=4$, which is a highly blue-tilted spectrum.
The ekpyrotic scenario corresponds to a slow contraction
($0<p \ll 1$), in which case we have $n_{\cal{R}} \simeq 3$.

The spectrum (\ref{tilt_po}) corresponds to the one
generated before the bounce.
In order to obtain the final power spectrum at sufficient late-times
in an expanding branch, we need to connect the contracting
branch with the Friedmann (expanding) one.
In the context of PBB cosmology, it was realized
in Ref.~\cite{Gas96,Bru97} (see
also \cite{Rey96,Foffa99,Cartier99}) that
loop and higher derivative corrections (defined in the string frame)
to the action induced by inverse
string tension and coupling constant corrections can yield a
nonsingular background cosmology.
This then allows the
study of the evolution of cosmological perturbations without having
to use matching prescriptions. The effects of the higher
derivative terms in the action on the evolution of fluctuations in
the PBB cosmology was investigated
numerically in \cite{Cartier01,TBF02}. It was found
that the final spectrum of fluctuations is highly blue-tilted
($n_{\cal R} \simeq 4$)  and the result
obtained is the same as what follows from the
analysis using matching conditions between two Einstein Universes
\cite{Bru94,Der95} joined along a constant scalar field
hypersurface.

In the context of ekpyrotic scenario nonsingular cosmological
solutions were constructed in Ref.~\cite{TBF02} by implementing
higher-order loop and derivative corrections analogous to
the PBB case. A possible set of corrections include terms of the
form \cite{Gas96,Bru97,TBF02}
\begin{eqnarray}
 {\cal L}_c = -\frac12 \alpha' \lambda
  \xi(\phi) \left[ c R_{\rm GB}^2+ d
 (\nabla \phi)^4 \right]\,,
\label{lagalpha}
\end{eqnarray}
where $\xi(\phi)$ is a general function of $\phi$ and $R_{\rm GB}^2
=R^2-4R^{\mu\nu}R_{\mu\nu}+
R^{\mu\nu\alpha\beta}R_{\mu\nu\alpha\beta}$ is
the Gauss-Bonnet term.
The corrections ${\cal L}_c$ are the sum of the tree-level
$\alpha'$ corrections and the quantum $n$-loop corrections
($n=1, 2, 3,\cdots$), with the function $\xi(\phi)$ given by
\beqa
\xi(\phi)=-\sum_{n=0} C_n e^{(n-1)\phi} \,,
\label{xifunction}
\eeqa
where $C_n$ ($n \ge 1$) are coefficients of
$n$-loop corrections, with $C_0=1$.
Nonsingular bouncing solutions that connect to a Friedmann
branch can be obtained by accounting for the corrections up
to two-loop with a negative coefficient ($C_2<0$).
See Ref.~\cite{TBF02} for a detailed analysis on
the background evolution.

It was shown in Ref.~\cite{TBF02} that the spectrum of
curvature perturbations long after the bounce is given as
$n_{\cal R} \simeq 3$ for $0<p \ll 1$ by numerically
solving perturbation equations in a nonsingular background
regularized by the correction term (\ref{lagalpha}).
In particular comoving curvature perturbations are conserved
on cosmologically relevant scales much larger than
the Hubble radius around the bounce, which means that
the spectrum (\ref{speekp}) can be used
in an expanding background long after the bounce.

The authors in \cite{KOST02} showed that the spectrum of the
gravitational potential $\Phi$, defined in Eq.~(\ref{defPhi}),
generated before the bounce is nearly scale-invariant for $0<p \ll 1$,
i.e., $n_\Phi -1=-2p/(1-p)$.  A number of authors argued
\cite{Lyth02ekp1,Lyth02ekp2,Bran01,Hwang02eky,Hwang02eky2}
that this corresponds
to the growing mode in the contracting phase but to the decaying mode
in the expanding phase.
The authors in Ref.~\cite{Cartier03} studied several toy bouncing
models assuming non-singular second-order evolution equations for the
perturbations across the bounce. They found that the final power
spectrum is dependent on the details of the bounce model.
However Cartier \cite{Cartier04} recently performed detailed numerical
analysis using nonsingular perturbation equations and found that in
the case of the $\alpha'$-regularised bounce both $\Phi$ and ${\cal
  R}$ exhibit the highly blue-tilted spectrum (\ref{speekp}) long
after the bounce.  It was numerically shown that the dominant
mode of the gravitational potential is fully converted into the
post-bounce decaying mode.
Similar conclusions have also been reached in investigations of
perturbations in other specific non-singular models
\cite{Gasperini03,Gasperini04,Allen04,BG05,Bozza05}.
Arguments can given that the comoving curvature perturbation is
conserved for adiabatic perturbations on large scales under very
general conditions \cite{Lyth03p1,CNZ05}.

Nevertheless we have to caution that these studies are based on
non-singular four-dimensional bounce models and in the
ekpyrotic/cyclic model the bounce is only non-singular in a
higher-dimensional completion of the model \cite{Tolley02}.  The
ability of the ekpyrotic/cyclic model to produce a scale-invariant
spectrum of curvature perturbations after the bounce relies on this
higher-dimensional physics being fundamentally different from
conventional four-dimensional physics, such that the growing mode of
$\Phi$ in the contracting phase does not decay after the bounce
\cite{Tolley03}.

The main reason we obtain the blue-tilted spectrum for curvature
perturbations in a contracting universe is that the
system is characterised by a kinematic driven phase not by a slow-roll
phase. In the case of modulus-driven superinflation with a
Gauss-Bonnet term, the spectrum of scalar perturbations is
$n_{\cal R}=10/3$ \cite{Kawai99,Tsujiekp}, which is again highly blue-tilted.
This is contrast with slow-roll inflation in which nearly
scale-invariant spectra are automatically obtained when the slow-roll
conditions ($\epsilon \ll 1, |\eta| \ll 1$) are satisfied.  Therefore
slow-roll inflation is typically more ``stable'' than kinematic-driven
scalar field models to generate scale-invariant spectra in a generic
way.

This perturbation spectra can be changed by taking into account
a second scalar field \cite{Bozza02,Notari02,DiMarco03}.
A system with multiple scalar fields generally induces
isocurvature perturbations, which can be the source of adiabatic
perturbations as we will discuss in Section~\ref{sect:multifield}.
For instance, axion fields can acquire scale-invariant
perturbation spectra due to their non-minimal coupling to the dilaton
field in the PBB \cite{Copeland77,Sta01,DiMarco03}.
Although these isocurvature perturbations are decoupled from curvature
perturbation during the PBB collapse phase they could in principle
provide a source for curvature perturbations at the bounce
\cite{LLKCreview}, or through the decay of massive axions during the
expanding phase \cite{Enqvist02,Lyth02}, in an application of the
curvaton idea which we will discuss in Section~\ref{curvaton}.
The authors of ref.~\cite{Notari02} considered a specific
two-field system with a brane-modulus $\varphi$ and a dilaton $\chi$.
When the dilaton has a negative exponential potential with a
suppressed ekpyrotic potential for $\varphi$, the entropy ``field''
perturbation can be scale-invariant if the model parameters are
fine-tuned \cite{Notari02}.
It is certainly possible to construct non-singular multi-field
PBB/ekpyrotic/cyclic scenarios that provide flat power spectra at late
times independent of arguments over gauge choices for metric
perturbations.

%%%%%%%%%%%%%%%%%%%%%%%%%%%%%%%%%%%%%%%%%%%%%%%%%%%%%
\section{Adiabatic and entropy perturbations from inflation}
\label{sect:multifield}
\label{corrspectra}
%%%%%%%%%%%%%%%%%%%%%%%%%%%%%%%%%%%%%%%%%%%%%%%%%%%%%

Realistic inflationary models, embedded in GUT or supersymmetric
theories, must necessarily be theories of
multiple fields and it is fairly natural to have multiple phase of
inflation (e.g. \cite{ARS97,julien,subir,chain,Kadota03}).
In such models with more than one
scalar field we have to consider the effect upon the evolution of non-adiabatic fluctuations
in any light fields whose effective mass is less than the Hubble
scale.
%Isocurvature modes during inflation may not survive after
%inflation, but they can nonetheless have an important effect on the
%evolution of the total curvature perturbation.

In the presence of more than one light field the vacuum fluctuations
stretched to super-Hubble scales will inevitably include isocurvature
modes during inflation. It is important to emphasise that this does
not mean that the ``primordial'' density perturbation (at the epoch of
primordial nucleosynthesis) will contain isocurvature modes. In
particular, if the universe undergoes conventional reheating phase at
the end of inflation and all particle species are in thermal
equilibrium with their abundances determined by a single temperature
(with no non-zero chemical potentials) then the primordial
perturbations must be adiabatic \cite{Weinberg04b}.  It is these
primordial perturbations that set the initial conditions for the
evolution of radiation-matter fluid that determines the anisotropies
in the cosmic microwave background and large-scale structure in our
Universe and thus are directly constrained by observations.
We will see that while the existence of non-adiabatic perturbations
after inflation requires the existence of non-adiabatic perturbations
during inflation \cite{Weinberg04a}, it is not true that isocurvature
modes during inflation necessarily give primordial isocurvature modes
\cite{Weinberg04b}.

\subsection{Inflaton and entropy perturbations during inflation}

Following \cite{Gordon00} (see also \cite{Groot01,Rigopoulos})
we will identify the inflaton as the direction in field space
corresponding to the evolution of the background (homogeneous)
field. Thus for $n$ scalar fields $\varphi_I$,
where $I$ runs from $1$ to $n$, we have
\begin{equation}
\label{defsigma}
\sigma = \int \sum_I \hat\sigma_I \dot\varphi_I
{\rm d}t \,,
\end{equation}
where the inflaton direction is defined by
\begin{equation}
\label{defhatsigma}
\hat\sigma_I \equiv \frac{\dot\varphi_I}{\sqrt{\sum_J
\dot\varphi_J^2}}\,.
\end{equation}
Arbitrary field perturbations can be decomposed into adiabatic
perturbations along the inflaton trajectory and
$n-1$ entropy perturbations orthogonal
to the inflaton in field space:
\begin{eqnarray}
\label{defdeltasigma}
\delta\sigma &=& \sum_I \hat\sigma_I \delta\varphi_I \,, \\
\label{defdeltas}
\delta s_{I} &=& \sum_J \hat{s}_{IJ} \delta\varphi_J \,,
\end{eqnarray}
where $\sum_I \hat{s}_{JI}\hat\sigma_I=0$.
Without loss of generality we assume that the entropy fields are also
mutually orthogonal in field space. Note that we have assumed that the
fields have canonical kinetic terms, that is,
the field space metric is flat.
See Refs.~\cite{Groot01,DiMarco03} for the generalisation to
non-canonical kinetic terms.

The $n$ evolution equations for the homogeneous scalar fields
(\ref{eq:scalareom}) can then be written as the evolution for a single
inflaton field (\ref{phieq})
\begin{equation}
\ddot\sigma + 3H\dot\sigma + V_\sigma = 0 \,,
\end{equation}
where the potential gradient in the direction
of the inflaton is
\begin{equation}
V_\sigma \equiv \frac{\rd V}{\rd \sigma}
= \sum_I \hat\sigma_I V_I \,.
\end{equation}
The total energy density and pressure are given by
the usual single field result (\ref{rhoeq}).

Similarly the total momentum and pressure perturbation
(\ref{eq:mtmphi}) and (\ref{eq:pressurephi}) for $n$ scalar field
perturbations can be written as for a single inflaton field
\begin{eqnarray}
\delta q &=& - \dot\sigma \delta\sigma \,, \\
\delta P &=& \dot\sigma(\dot{\delta\sigma}-\dot\sigma A) -
V_\sigma\delta\sigma \,.
\end{eqnarray}
However the density perturbation is given by
\begin{equation}
\delta\rho = \dot\sigma(\dot{\delta\sigma}-\dot\sigma A) +
V_\sigma\delta\sigma + 2 \delta_s V \,,
\end{equation}
where the deviation from the single field result (\ref{eq:densityphi})
arises due to the perturbation of the potential orthogonal to the
inflaton trajectory:
\begin{equation}
\delta_s V \equiv \sum_I V_I \delta\varphi_I - V_\sigma \delta\sigma
 \,.
\end{equation}
The non-adiabatic pressure perturbation (\ref{defPnad})
is written as
\begin{equation}
\delta P_{\rm nad} = - \frac{2V_\sigma}{3H\dot\sigma} \delta\rho_m -
2\delta_s V \,,
\end{equation}
where the comoving density perturbation, $\delta\rho_m$, is given by
Eq.~(\ref{defrhom}).
Although Eq.~(\ref{eq:rhomcon}) requires the comoving density
perturbation to become small on large scales, as in the single field
case, there is now an additional contribution due to non-adiabatic
perturbations of the potential which
need not be small on large scales
(This generalises the result of Ref.~\cite{Gordon00} for two scalar
fields to the case of $n$ canonical fields).
We note that the dynamics of cosmological perturbations in
multi-field system was
investigated by a host of authors, see
Refs.~\cite{Bartolo01p1,Chiba97,Linde85,Salopek89,Sta94,Sta01,FRV04,Garcia95,Garcia96,HY05,Hwang00,Kadota03,Kadota03v2,Kana00p1,Kawasaki01,Polarski94,Sasaki95,Sasaki98,Langlois99,Langlois00,Mukhanov97,Kofman87,Kofman88,Lahiri}.

%%%%%%%%%%%%%%%%%%%%%%%%%%%%%%%%%%%%%%%%
%\section{Correlated spectra of adiabatic and entropy perturbations}
\subsection{Evolution of non-adiabatic perturbations}
%%%%%%%%%%%%%%%%%%%%%%%%%%%%%%%%%%%%%%%%

We will now consider the coupled evolution of two
canonical scalar fields, $\phi$ and $\chi$, during inflation and how
this can give rise to correlated curvature and entropy perturbations.
We will use the local rotation in field space defined by
Eq.~(\ref{defdeltasigma}) and (\ref{defdeltas}) to describe the
instantaneous adiabatic and entropy field perturbations.

Note that the inflaton field perturbation~(\ref{defdeltasigma}) is
gauge-dependent and thus we have to fix the gauge in order to obtain a
gauge-invariant variable. We will choose to work with the inflaton
perturbation in the spatially flat ($\psi=0$) gauge:
\begin{equation}
\delta\sigma_\psi \equiv \delta\sigma + \frac{\dot\sigma}{H}\psi \,.
\end{equation}
By contrast the orthogonal entropy perturbation (\ref{defdeltas}) is
automatically gauge-invariant.

The generalisation to two fields of the evolution equation for the
inflaton field perturbations in the spatially flat gauge, given in
Eq.~(\ref{eq:singlescalareom})
for a single field, is \cite{Gordon00}
\begin{eqnarray}
\label{eq:twoscalareom}
\hspace*{-1.3em}
&& \ddot{\delta\sigma}_\psi + 3H\dot{\delta\sigma}_\psi
 + \left[ \frac{k^2}{a^2} + V_{\sigma\sigma}
  - \dot{\theta}^2
  - \frac{8\pi G}{a^3} \frac{{\rm d}}
  {{\rm d}t} \left( \frac{a^3\dot\sigma^2}{H} \right)
 \right] \delta\sigma_\psi \nonumber\\
 \hspace*{-1.3em}
 && =  ~~~ 2\frac{{\rm d}}{{\rm d}t}
(\dot\theta\delta s) - 2\left(
{V_\sigma \over \dot\sigma} + {\dot{H}\over H} \right)
\dot\theta \delta s\,,
\end{eqnarray}
and the entropy perturbation obeys
\begin{equation}
\label{eq:entropyeom}
\ddot{\delta s} + 3H\dot{\delta s} + \left(\frac{k^2}{a^2}
  + V_{ss} + 3\dot{\theta}^2 \right) \delta s =
{\dot\theta\over\dot\sigma} {k^2 \over 2\pi G a^2} \Psi\,,
\end{equation}
where $\tan \theta=\dot{\chi}/\dot{\phi}$ and
\beqa
\hspace*{-1.0em}
V_{\sigma \sigma} &\equiv& (\cos^2 \theta) V_{\phi \phi} +(\sin
2\theta)V_{\phi \chi}+(\sin^2 \theta) V_{\chi \chi},\\
\hspace*{-1.0em}
V_{ss} &\equiv& (\sin^2 \theta) V_{\phi \phi}
-(\sin 2\theta)V_{\phi \chi}+(\cos^2 \theta) V_{\chi \chi}.
\label{Vdd}
\eeqa

We can identify a purely adiabatic mode where $\delta s=0$ on large
scales. However a non-zero entropy perturbation does appear as a
source term in the perturbed inflaton equation whenever the inflaton
trajectory is curved in field space, i.e., $\dot{\theta} \ne 0$.
We note that $\dot{\theta}$ is given by \cite{Gordon00}
\begin{equation}
\dot\theta = - \frac{V_s}{\dot\sigma} \,,
\end{equation}
where $V_s$ is the potential gradient orthogonal to the inflaton
trajectory in field space.

The entropy perturbation evolves independently of the curvature
perturbation on large-scales. It couples to the curvature perturbation
only through the gradient of the longitudinal gauge metric potential,
$\Psi$. Thus entropy perturbations are also described as
``isocurvature'' perturbations on large scales.
Eq.~(\ref{eq:twoscalareom}) shows that the entropy perturbation
$\delta s$ works as a source term for the adiabatic perturbation.
This is in fact clearly seen if we take the time derivative of the
curvature perturbation \cite{Gordon00}:
\begin{equation}
\label{dotR}
\dot{\cal R}=\frac{H}{\dot{H}}\frac{k^2}{a^2}\Psi
+\frac{2H}{\dot{\sigma}}\dot{\theta}\delta s\,.
\end{equation}
Therefore ${\cal R}$ is not conserved even in the large-scale limit
in the presence of the entropy perturbation $\delta s$ with
a non-straight trajectory in field space ($\dot{\theta} \ne 0$).

Analogous to the single field case we can introduce slow-roll
parameters for light, weakly coupled fields \cite{Wands02}. At
first-order in a slow-roll expansion, the inflaton rolls directly down
the potential slope, that is $V_s\simeq0$. Thus we have only one slope
parameter
\begin{equation}
\epsilon \equiv -\frac{\dot{H}}{H^2} \simeq \frac{1}{16\pi G}
\left( \frac{V_\sigma}{V} \right)^2 \,,
\end{equation}
but three parameters, $\eta_{\sigma\sigma}$, $\eta_{\sigma s}$ and
$\eta_{ss}$,
describing the curvature of the potential, where
\begin{equation}
\eta_{IJ} \equiv \frac{1}{8\pi G} \frac{V_{IJ}}{V} \,.
\end{equation}

The background slow-roll solution is described by
\begin{equation}
 \label{twofieldbackground}
\dot\sigma^2 \simeq \frac23 \epsilon V \, , \quad
H^{-1} \dot\theta \simeq -\eta_{\sigma s} \,,
\end{equation}
while the perturbations obey
\begin{eqnarray}
\label{twofieldperturbations}
H^{-1} \dot{\delta\sigma} &\simeq&
 \left( 2\epsilon -\eta_{\sigma\sigma} \right) \delta\sigma
 - 2\eta_{\sigma s}\delta s \,,
  \nonumber\\
H^{-1} \dot{\delta s} &\simeq& -\eta_{ss}\delta s \,,
\end{eqnarray}
on large scales, where we neglect spatial gradients.
Although $V_s\simeq 0$ at lowest order in slow-roll, this does not
mean
that the inflaton and entropy perturbations decouple. $\dot\theta$
given by Eq.~(\ref{twofieldbackground}) is in general non-zero at
first-order and large-scale entropy perturbations do affect the
evolution of the adiabatic perturbations when $\eta_{\sigma s}\neq 0$.

While the general solution to the two second-order perturbation
equations (\ref{eq:twoscalareom}) and (\ref{eq:entropyeom}) has four
independent modes, the two first-order slow-roll equations
(\ref{twofieldperturbations}) give the approximate form of the
squeezed state on large scales. This has only two modes which we can
describe in terms of dimensionless curvature and isocurvature
perturbations:
\begin{equation}
\label{eq:defRS}
\R \equiv \frac{H}{\dot\sigma} \delta\sigma_\psi \,, \quad
\S \equiv \frac{H}{\dot\sigma} \delta s \,.
\end{equation}
The normalisation of $\R$ coincides with the standard definition of
the comoving curvature perturbation, Eq.~(\ref{eq:defR}). The
normalisation of the dimensionless entropy, $\S$, is chosen here
coincide with Ref.~\cite{Wands02}. It can be related to the
non-adiabatic
pressure perturbation (\ref{defPnad}) on large scales
\begin{equation}
\delta P_{\rm nad} \simeq - \epsilon
\eta_{\sigma s} \frac{H^2}{2\pi G} \S \,.
\end{equation}

The slow-roll approximation can provide a useful approximation to the
instantaneous evolution of the fields and their perturbations on large
scales during slow-roll inflation, but is not expected to remain
accurate when integrated over many Hubble times, where inaccuracies
can accumulate. In single-field inflation the constancy of the
comoving curvature perturbation after Hubble exit, which does not rely
on the slow-roll approximation, is crucial in order to make accurate
predictions of the primordial perturbations using the slow-roll
approximation only around Hubble crossing. In a two-field model we
must describe the evolution after Hubble exit in terms of
a general transfer matrix:
\begin{equation}
\label{defTransfer}
\left(
\begin{array}{c}
{\R} \\ {\S}
\end{array}
\right) = \left(
\begin{array}{cc}
1 & {T}_{\R\S} \\ 0 & {T}_{\S\S}
\end{array}
\right) \left(
\begin{array}{c}
\R \\\S
\end{array}
\right)_* \,.
\end{equation}
On large scales the comoving curvature perturbation still remains
constant for the purely adiabatic mode, corresponding to $\S=0$,
and adiabatic perturbations remain adiabatic. These general results
are enough to fix two of the coefficients in the transfer matrix, but
${T}_{\R\S}$ and ${T}_{\S\S}$ remain to be determined
either within a given theoretical model,
or from observations, or ideally by both.
The scale-dependence of the transfer functions depends upon
the inflaton-entropy coupling at Hubble exit during inflation and can
be given in terms of the slow-roll parameters as \cite{Wands02}
\begin{eqnarray}
\label{dTdlnk}
\frac{\partial}{\partial\ln k} {T}_{\R\S} &=& 2\eta_{\sigma s} +
 (2\epsilon - \eta_{\sigma\sigma} + \eta_{ss}) {T}_{\R\S}
\,,\nonumber\\
\frac{\partial}{\partial\ln k} {T}_{\S\S} &=&
 (2\epsilon - \eta_{\sigma\sigma} + \eta_{ss}) {T}_{\S\S} \,.
\end{eqnarray}

\subsection{Initial power spectra}

For weakly-coupled, light fields (with effective mass less than
the Hubble scale) we can neglect interactions on wavelengths below the
Hubble scale, so that vacuum fluctuations give rise to a spectrum of
uncorrelated field fluctuations on the Hubble scale ($k=aH$) during
inflation:
\begin{equation}
 \label{H2pi}
{\cal P}_{\delta\varphi_I} \simeq \left( \frac{H}{2\pi} \right)_*^2 \,,
\end{equation}
where we use a $*$ to denote quantities evaluated at Hubble-exit.
If a field has a mass comparable to the Hubble scale or larger then
the vacuum fluctuations on wavelengths greater than the effective
Compton wavelength are suppressed. In addition fluctuations in
strongly interacting fields may develop correlations before Hubble
exit. But in the slow-roll limit of weakly coupled, light fields the
vacuum fluctuations in orthogonal fields are independent at
Hubble-exit. This remains true under a local rotation in fields space
to another orthogonal basis such as the instantaneous inflaton and
entropy directions (\ref{defdeltasigma}) and (\ref{defdeltas}) in
field space.

The curvature and isocurvature power spectra at
Hubble-exit are given by
\begin{equation}
\left. {\cal P}_{\R} \right|_* \simeq
\left. {\cal P}_{\S} \right|_* \simeq
 \left( \frac{H^2}{2\pi\dot\sigma} \right)_*^2
 \simeq \frac{8}{3} \left( \frac{V}{\epsilon M_{\rm Pl}^4} \right)_*
 \,,
\end{equation}
while the cross-correlation is zero, at lowest order in slow-roll:
\begin{equation}
\left. {\cal C}_{\R\S} \right|_* \simeq 0 \,.
\end{equation}
The normalisation chosen for the dimensionless entropy perturbation in
Eq.~(\ref{eq:defRS}) ensures that the curvature and isocurvature
fluctuations have the same power at horizon exit \cite{Wands02}.
The spectral tilts at horizon-exit are also the same and are given by
\begin{equation}
\label{nR*}
n_\R|_* -1 \simeq n_\S|_* \simeq -6\epsilon + 2\eta_{\sigma\sigma} \,.
\end{equation}
where $n_\S\equiv \rd \ln{\cal P}_\S/\rd \ln k$.

The tensor spectrum is decoupled from scalar metric perturbations at
first-order and hence has the same form as in single field inflation,
described in section \ref{sect:singlefieldps}. Thus the power spectrum
of gravitational waves on super-Hubble scales during inflation is
given by
\begin{equation}
{\cal P}_{\rm T}
 \simeq \frac{16H^2}{\pi M_{\rm Pl}^2}
 \simeq \frac{128}{3} \frac{V_*}{M_{\rm Pl}^4} \,,
\end{equation}
and the spectral tilt is
\begin{equation}
n_{\rm T} \simeq -2 \epsilon \,.
\end{equation}

\subsection{Primordial power spectra}

The resulting primordial power spectra on large scales can be obtained
simply by applying the general transfer matrix (\ref{defTransfer}) to
the initial scalar perturbations. There scalar power spectra probed by
astronomical observations are thus given by \cite{Wands02}
\begin{eqnarray}
\label{PR}
{\cal P}_\R &=& (1+T_{\R\S}^2) {\cal P}_\R|_* \\
{\cal P}_\S &=& T_{\S\S}^2 {\cal P}_\R|_* \\
\label{CRS}
{\cal C}_{\R\S} &=& T_{\R\S} T_{\S\S} {\cal P}_\R|_* \,.
\end{eqnarray}
The cross-correlation can be given in terms of a dimensionless
correlation angle:
\begin{equation}
\label{defDelta}
\cos\Delta \equiv \frac{{\cal C}_{\R\S}}{\sqrt{{\cal P}_\R{\cal
      P}_\S}} = \frac{T_{\R\S}}{\sqrt{1+T_{\R\S}^2}} \,.
\end{equation}

We see that if we can determine the dimensionless correlation angle,
$\Delta$, from observations, then this determines the off-diagonal
term in the transfer matrix
\begin{equation}
\label{defTRS}
T_{\R\S} = \cot\Delta \,,
\end{equation}
and we can in effect measure the contribution of the entropy
perturbation during inflation to the resultant curvature perturbation.
In particular this allows us in principle to deduce from observations
the power
spectrum of the curvature perturbation at Hubble-exit during inflation
\cite{Wands02}:
\begin{equation}
{\cal P}_\R|_* = {\cal P}_\R  \sin^2\Delta \,.
\end{equation}

The scale-dependence of the resulting scalar power spectra
depends both
upon the scale-dependence of the initial power spectra and of the
transfer coefficients.
The spectral tilts are given from Eqs.~(\ref{PR}--\ref{CRS})
by
\begin{eqnarray}
 \label{gentilt}
n_\R &=& n_\R|_* + H_*^{-1} (\partial{T}_{\R\S}/\partial t_*) \sin
2\Delta \,,
 \nonumber\\
n_\S &=& n_\R|_* + 2 H_*^{-1} (\partial\ln{T}_{\S\S}/\partial t_*)
\,,\\
n_{\cal C} &=& n_\R|_* + H_*^{-1} \left[ (\partial{T}_{\R\S}/\partial
t_*) \tan\Delta + (\partial\ln{T}_{\S\S}/\partial t_*) \right] \,,
 \nonumber
\end{eqnarray}
where we have used Eq.~(\ref{defTRS}) to eliminate $T_{\R\S}$ in
favour of the observable correlation angle $\Delta$.
Substituting Eq.~(\ref{nR*}) for the tilt at Hubble-exit, and
Eqs.~(\ref{dTdlnk}) for the scale-dependence of
the transfer functions, we obtain \cite{Wands02}
\begin{eqnarray}
\label{srtilts}
n_{\R} &\simeq& 1 -(6-4\cos^2\Delta) \epsilon \nonumber\\
&&  + 2\left( \eta_{\sigma\sigma}\sin^2\Delta + 2\eta_{\sigma
 s}\sin\Delta\cos\Delta + \eta_{ss}\cos^2\Delta \right)
\,,\nonumber\\
n_\S &\simeq& -2\epsilon + 2\eta_{ss} \,,\\
n_{\cal C} &\simeq& -2\epsilon + 2\eta_{ss} + 2\eta_{\sigma
  s}\tan\Delta
 \nonumber\,.
\end{eqnarray}

Although the overall amplitude of the transfer functions are dependent
upon the evolution after Hubble-exit and through reheating into the
radiation era, the spectral tilts can be expressed solely in terms of
the slow-roll parameters at Hubble-exit during inflation and the
correlation angle, $\Delta$, which can in principle be
observed.

The gravitational wave power spectrum is frozen-in
on large scales, independent of the scalar perturbations,
and hence
\begin{equation}
{\cal P}_{\rm T}|_* = {\cal P}_{\rm T} \,.
\end{equation}
Thus we can derive a modified consistency relation
(\ref{consistency}) between observables applicable in the case of
two-field slow-roll inflation:
\begin{equation}
r=\frac{{\cal P}_{\rm T}}
{{\cal P}_\R} \simeq -8 n_{\rm T} \sin^2\Delta  \,.
\end{equation}
This relation was first obtained in Ref.~\cite{Bartolo01p2} at the
end of
two-field inflation, and verified in Ref.~\cite{Tsuji03} for
slow-roll models. But it was realised in Ref.~\cite{Wands02} that this
relation also applies to the observable perturbation spectra
some time after two-field slow-roll inflation has ended.

If there is another source of the scalar curvature perturbation, such
as from a third scalar field during inflation, then this could give an
additional contribution to the scalar curvature spectrum without
affecting the gravitational waves, and hence the more general result
becomes an inequality:
\begin{equation}
r \lesssim -8 n_{\rm T} \sin^2\Delta \,.
\end{equation}
%

%%%%%%%%%%%%%%%%%%%%%%%%%%%%%%%%%%%%%%%%%%%%%%
\section{Correlations and the CMB}
\label{correlation}
%%%%%%%%%%%%%%%%%%%%%%%%%%%%%%%%%%%%%%%%%%%%%%

The physical processes that drive inflation in the early universe
leave their mark in the perturbation spectra that are generated from
vacuum fluctuations. Single field models yield only adiabatic
perturbations on large scales during inflation, and adiabatic
perturbations stay adiabatic on large scales. In multi-field models we
have seen that perturbations orthogonal to the inflaton trajectory
describe non-adiabatic perturbations, $\S_*$ in
Eq.~(\ref{defTransfer}). These have two principal observational
effects as shown in Eq.~(\ref{defTransfer}). Firstly they can alter the large
scale curvature perturbation in the radiation era, through the
off-diagonal term $T_{\R\S}$ in the transfer matrix.  And secondly
they can yield a primordial isocurvature perturbation, through
$T_{\S\S}$. These relative perturbations between different components
of the cosmic energy-momentum tensor yield distinctive observational
features which enable the amplitude of such perturbations to be
tightly constrained.

Up until 1999 all studies of the effect of isocurvature modes only
considered isocurvature perturbations statistically independent of the
primordial curvature perturbation, i.e., uncorrelated. But the
off-diagonal term in the transfer matrix would give rise to
correlations between primordial curvature and isocurvature modes,
parameterised through the correlation angle, $\Delta$ in
Eq.~(\ref{defTRS}). Langlois \cite{Langlois99} pointed
out that isocurvature perturbations produced during inflation might
naturally be correlated with the adiabatic mode. Bucher et al
\cite{Bucher99} pointed out that the most general primordial
perturbations spectra could include several isocurvature modes
(including neutrino density and velocity perturbations) and these
would in general be correlated with the curvature perturbation and
with one another \cite{Trotta01}.

The contribution of isocurvature perturbations to the overall CMB
angular power spectrum is now tightly constrained due to the
distinctive peak structure of adiabatic versus isocurvature modes.
However in seeking observational signatures of isocurvature modes one
must include the effect of correlations which introduces a different
angular power spectrum. In effect one must include an additional term
in the CMB angular power spectrum which, in contrast to the
uncorrelated spectra, can be negative as well as positive (though the
resulting angular power spectrum must remain non-negative). For
instance this could actually decrease the angular power spectrum on
large angular scales due to correlated isocurvature perturbations in
some models of dark energy, as recently noted in
Refs.~\cite{MTde,GordonHu}.

\subsection{Matter isocurvature modes}

The most commonly considered isocurvature modes are perturbations
in the density of non-relativistic matter (either baryons or cold
dark matter) relative to the radiation energy density. This is
given from Eq.~(\ref{defSi}) as
\begin{equation}
\S_m = \frac{\delta\rho_m}{\rho_m} - \frac34
\frac{\delta\rho_\gamma}{\rho_\gamma} \,,
 \end{equation}
 and hence in effect reduces to the fractional matter density
perturbation deep in the primordial radiation dominated era when
$\rho_\gamma\gg\rho_m$. On large scales both the primordial
curvature perturbation, $\zeta_\gamma$, and the matter
isocurvature perturbation, $\S_m$, are conserved on large scales
\cite{Wands00}. In the rest of this section we shall adopt the
notation of section \ref{corrspectra} and use $\R=-\zeta_\gamma$ on
large scales to denote the primordial curvature perturbation.

After matter domination the matter perturbations come to dominate
the total curvature perturbation, $\zeta$ given in
Eq.~(\ref{zetasum}), which we can write in terms of the primordial
curvature and isocurvature perturbations as
\begin{equation}
\label{zetam}
\zeta_m = -\R + \frac13 \S_m \,,
\end{equation}
where for simplicity we neglect the neutrino density.
This is in turn related to the longitudinal gauge metric
perturbation, Eq.~(\ref{defPhi}), on large scales during the matter
dominated era \cite{Mukhanov90}
\begin{equation}
\Phi = \frac35 \R \,.
 \end{equation}
Temperature anisotropies in the CMB on large angular scales due to
the intrinsic temperature perturbation plus the Sachs-Wolfe effect are
given by \cite{Gordon02}
\begin{equation}
\frac{\delta T}{T} \simeq \zeta_\gamma + 2 \Phi \,,
\end{equation}
which can thus be written in terms of the primordial
curvature and isocurvature perturbations using Eq.~(\ref{zetam}) as
\begin{equation}
\frac{\delta T}{T} \simeq
\frac15 \left( \R - 2\S_m \right) \,.
\end{equation}

Isocurvature matter perturbations also produce acoustic peaks but
these are out of phase with those from adiabatic perturbations
\cite{Bucher99}. The success of the minimal model based on
scale-invariant, Gaussian and adiabatic perturbations in reproducing
the detailed structure of acoustic peaks in the angular power
spectrum means that models of structure formation based on
isocurvature primordial perturbations are now convincingly ruled out.
These models in any case required a steep blue
spectrum of isocurvature perturbations in order to overcome the
suppression of the contribution of isocurvature perturbations to the
matter power spectrum on smaller scales \cite{Peacock}. An almost
scale-invariant spectrum of isocurvature perturbations, e.g., from
fields obeying slow-roll conditions during inflation, gives a
relatively small effect on the CMB on sub-Hubble scales at the time
of last scattering \cite{Langlois00,Bucher99,Amendola02}.

There have been several different analyses of the observational
constraints on isocurvature matter perturbations incorporating the
first-year WMAP data and additional astronomical data on smaller
scales
\cite{Peiris03,Gordon02,Valiviita03,Crotty03,Parkinson04,Beltran04,Beltran05,Kurki-Suonio04}.

In a Bayesian analysis the posterior likelihood of quantities such as
the amplitude of isocurvature modes relative to curvature
perturbations inevitably depends both on the parameterisation
chosen for the isocurvature modes and prior distribution chosen
for those parameters.
For the slow-roll two-field inflation described in section
\ref{corrspectra} it is natural to adopt a power-law parameterisation
for the perturbations at Hubble-exit:
\begin{equation}
{\cal P}_\R|_* = {\cal P}_\S|_* = A_r^2
\left(\frac{k}{k_0}\right)^{n_1}\,,
\end{equation}
and the transfer functions
\begin{eqnarray}
T_{\R\S} &=&  T_r \left(\frac{k}{k_0}\right)^{\Delta n_r / 2} \,,\\
T_{\S\S} &=& T_s \left(\frac{k}{k_0}\right)^{\Delta n_s / 2} \,.
\end{eqnarray}
This gives the primordial power spectra (\ref{PR}--\ref{CRS})
\begin{eqnarray}
 \label{defArAs}
{\cal P}_\R &=& A_r^2 \left(\frac{k}{k_0}\right)^{n_1} + A_s^2
 \left(\frac{k}{k_0}\right)^{n_3} \,,\\
{\cal P}_\S &=& B^2 \left(\frac{k}{k_0}\right)^{n_2} \,,\\
{\cal C}_{\R\S} &=&
A_s B  \left(\frac{k}{k_0}\right)^{n_c}\,,
\end{eqnarray}
where
\begin{eqnarray}
A_s^2 = T_r^2 A_r^2 \,, \quad n_3 = n_1 + \Delta n_r \,,\\
B^2 = T_s^2 A_r^2 \,, \quad n_2 = n_1 + \Delta n_s \,,
\end{eqnarray}
and
\begin{equation}
 n_c = \frac{n_2+n_3}{2} \,.
\end{equation}

This coincides with the parameterisation used by Kurki-Suonio et
al~\cite{Kurki-Suonio04}, although they choose the opposite sign
convention for the primordial curvature perturbation and hence the
correlation angle. They find an upper limit (95\% c.l.) on the
allowed isocurvature fraction
(marginalising over other parameters)
\begin{equation}
f_{\rm iso} \equiv \sqrt{\frac{B^2}{A_r^2+A_s^2}} < 0.47 \,,
\end{equation}
where they use a pivot scale $k_0=0.01{\rm Mpc}^{-1}$.
(Note that results in \cite{Kurki-Suonio04} are given in terms of
$\alpha=f_{\rm iso}^2/(1+f_{\rm iso}^2)$ for which they choose a flat
prior.)
By contrast an uncorrelated subset with $\cos\Delta\simeq0$ yields a
weaker limit of only $f_{\rm iso}<0.53$ as uncorrelated models
have less effect on the CMB.
The best-fit model of Ref.~\cite{Kurki-Suonio04} has primordial power
spectra with $n_1=-0.012$, $n_2=-0.074$, $n_3=-0.612$ and
isocurvature fraction $f_{\rm iso}=0.044$
and correlation $\cos\Delta=0.82$. The principal effect of the
correlated isocurvature perturbations (small at the pivot scale of
$10$\,Mpc) is to reduce the power in the lowest multipoles for these
red primordial power spectra with $n_\R-1<0$.
Use of a larger pivot scale $k_0$ in the analysis tends to
favour these models \cite{Beltran04}.
But in general isocurvature models are not favoured.
Isocurvature modes do not produce the peak structure seen in the
current data.

Note that the WMAP team \cite{Peiris03} restricted their analysis to a
scale-invariant correlation which corresponds to $\Delta n_r=0$ above.
They found $f_{\rm iso}<0.33$ at 95\% c.l.~using a pivot scale of
$k_0=0.05{\rm Mpc}^{-1}$.

\begin{figure}
\includegraphics[scale=0.48]{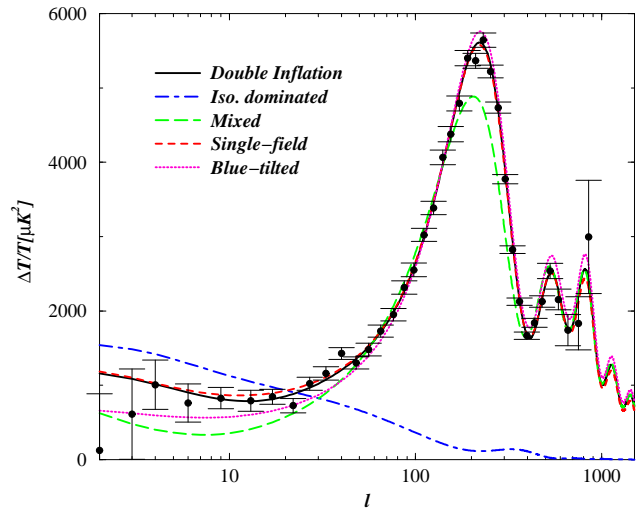}
\caption{The CMB angular power spectra for
the double inflation model (\ref{hybridpo})
with $g^2/\lambda=2$ \cite{Parkinson04}
for five different cases:\\
(i) the best-fit double inflation model,\\
(ii) isocurvature dominating over the adiabatic,\\\
(iii) the isocurvature is comparable to the adiabatic (mixed),\\
(iv) the best fit single-field model with potential
$V=(\lambda/4)(\chi^2-M^2/\lambda)^2$
and \\
(v) a model with blue-tilted spectrum ($n_{\cal R}>1$)
on large scales.
}
\label{cospectra}
\end{figure}

In Ref.~\cite{Parkinson04} the double inflation model (\ref{hybridpo})
was constrained by using the 1-st year WMAP data for the
supersymmetric case ($g^2/\lambda=2$).
It was found that the correlated isocurvature component can be at
most $7\%$ of the total contribution which is dominated by the
adiabatic spectrum.
In Fig.~\ref{cospectra} we plot the CMB power spectra for this model
for several different cases.
Clearly the spectra are significantly different
from the standard one when the isocurvature mode is dominant.
The best-fit power spectrum is not too much different from the one
for the single-field inflation with potential
$V=(\lambda/4)(\chi^2-M^2/\lambda)^2$, but
Akaike \& Bayesian model selection
criteria \cite{Akaike74,Schwarz78,Liddle04}
prefer single-field inflation over the double inflation model
(\ref{hybridpo})
as a result of 9 parameter likelihood analysis.

%%%%%%%%%%
%%%%%%%%%%

Physical models for the origin of the primordial perturbations may
give distinctive predictions for the amplitude and correlation of
isocurvature matter perturbations, such as the curvaton and modulated
reheating models that we will discuss later.

\subsection{Neutrino isocurvature modes}

The primordial cosmic fluid includes photons, baryons, cold dark
matter and neutrinos. The neutrinos may also have a density
perturbation relative to the photons, given from Eq.~(\ref{defSi}) as
 \begin{equation}
 \S_\nu = \frac34 \left( \frac{\delta\rho_\nu}{\rho_\nu} -
  \frac{\delta\rho_\gamma}{\rho_\gamma} \right) \,,
 \end{equation}
for relativistic neutrinos (which we shall assume hereafter). Like the
matter isocurvature perturbations this produces a series of acoustic
peaks out of phase with the adiabatic case.

Because neutrinos only decouple from the photons shortly before
primordial nucleosynthesis it is not easy to introduce an isocurvature
perturbation relative to the photons. It may only be possible in
models with a significant lepton asymmetry, $\xi_\nu$. In this case the neutrino
density before decoupling is determined not just by the photon
temperature but also a non-zero chemical potential. This is possible
in some curvaton models \cite{Lyth03p2}, but
the net lepton asymmetry is now tightly constrained by observations
% v2 DW: REFEREE1 REQUESTED VALUE
which require the neutrino degeneracy parameter $|\xi_\nu<0.07|$
for all flavours \cite{Dolgov02}.

For relativistic particles such as neutrinos it is also possible to
consider a relative velocity perturbation which is non-singular in the
early time limit in the synchronous gauge \cite{Bucher99} (though the
longitudinal gauge potential $\Phi$, for instance, is singular).
Because this is a relative velocity perturbation, not a relative
density perturbation, it produces acoustic oscillations out of phase
with the isocurvature density perturbations, but approximately in
phase with the adiabatic mode.

When one considers the most general primordial perturbation, including
all isocurvature modes and their correlations it becomes much harder
to exclude the possibility of a significant isocurvature contribution
to the CMB angular power spectrum \cite{Moodley04}. This is mainly
due to the neutrino isocurvature velocity mode and there is no theoretical
model of how a scale invariant spectrum for such a mode could be generated.

%%%%%%%%%%%%%%%%%%%%%%%%%%%%%%%%%%%
\section{Reheating the universe after inflation}
\label{oldreheat}
%%%%%%%%%%%%%%%%%%%%%%%%%%%%%%%%%%%

At the end of inflation the universe is typically, but not always
\footnote{An exception is the warm inflation scenario where there is
  particle production during inflation \cite{warm,warm2}.}, in a
highly non-thermal state - and a very cold one at that.  The key
ability of inflation to homogenise the universe also means that it
leaves the cosmos at effectively zero temperature and hence any
successful theory of inflation must also explain how the cosmos
was reheated - or perhaps defrosted - to the high temperatures we
require for the standard hot Big Bang picture.  At the very least
this must include baryogenesis and nucleosynthesis.
% v2*
Baryogenesis requires energies greater than the electroweak scale
but is very model-dependent and is known to require
out-of-equilibrium processes. But primordial nucleosynthesis
requires that the universe is close to thermal equilibrium at a
temperature around 1 MeV.

One of the key realisation of the past few years has been that the process
of reheating can have a profound impact on the
cosmological predictions of the preceding inflationary phases,
as we will discuss in sections \ref{metricpre} and \ref{curvaton}.
In addition our understanding of the process
by which inflation ends and reheating takes place has undergone
significant advances recently, which we now review.

The so-called ``old'' theory of reheating, developed in the immediate
wake of the first inflationary
theories \cite{abbott82,DL82},
was based on the concept of single-body decays.
In this picture the inflaton field is a collection of scalar
particles
each with a finite probability of decaying, just as a free neutron
decay into a proton, electron and anti-neutrino.
Such decays can be treated by coupling the inflaton, $\phi$,
to other scalar ($\chi$) or fermion ($\psi$) fields through terms
in the Lagrangian such as $\nu \sigma \phi \chi^2$ and $h \phi
\overline{\psi} \psi$.
Here $\sigma$ has the dimensions of mass and $\nu$ and $h$
are dimensionless couplings.

Dimensional analysis allows one to estimate the tree-level decay
rates ($\Gamma$) for these two interactions
since $[\Gamma] =t^{-1} =m$.
When the mass of the inflaton is much larger than those of
$\chi$ and $\psi$ ($m_{\phi} \gg m_{\chi},m_{\psi}$),
the decay rate is \cite{DL82,Lindebook}
\bea
\Gamma_{\phi \rightarrow \chi\chi} &=& \frac{\nu^2 \sigma^2}
{8\pi m_{\phi}}\,, \\
\Gamma_{\phi \rightarrow \psi\overline{\psi}}
&=& \frac{h^2 m_{\phi}}{8\pi}\,.
\label{singlebody}
\eea
{}From this we can estimate the temperature at which the universe
will reach thermal equilibrium since until $\Gamma > H$
the expansion will not allow a thermal distribution to be reached.
This implies that an upper limit on the temperature
after inflation is given by solving
$\Gamma_{{\rm tot}} \equiv \Gamma_{\phi \rightarrow \chi\chi} +
\Gamma_{\phi\rightarrow \psi\overline{\psi}} = H =
(8\pi \rho/3m_{\rm pl}^2)^{1/2}$
for the temperature.
Assuming all the energy density $\rho$ of the universe
is in the form of relativistic matter with
$\rho \simeq g_* \pi^2 T^4/30$
where $g_*$ is the effective number of massless degrees of
freedom ($g_*=10^2$-$10^3$),
we obtain the reheat temperature, $T_{{\rm rh}}$:
\beq
T_{{\rm rh}} \simeq  0.2 \left(\frac{100}{g_*}\right)^{1/4}
\sqrt{\Gamma_{{\rm tot}} m_{{\rm pl}}}\,.
\label{trh}
\eeq
Note that if there is a significant amount of massive particles (with number density
evolving as $a^{-3}$), not in the form of radiation, then the reheat temperature
is modified since the dependence of $H$ on $T$ is altered.

Neglecting this case, let us impose the constraint that comes from
the normalisation  of the CMB on large scales; namely that $m_{\phi} \sim 10^{-6}
m_{{\rm pl}}$.  This ensures that the models do not over-produce anisotropies in the
CMB. Requiring that radiative corrections mediated by the couplings do not spoil the
flatness of the potential  limits the reheat temperature to be
below the GUT scale\footnote{The same result comes from requiring
$\Gamma < m_{\phi}$.}, $T_{{\rm rh}} < 10^{16} {\rm GeV}$,
which means  that the GUT symmetries are not restored and hence there is
not a second phase of production of monopoles that
inflation was introduced to solve in the first place!
However, this does not mean there are no problems. If one is building models of inflation in a
supergravity context then one must worry about the over-production of gravitinos, the
supersymmetric  partner of the graviton. For a wide range of gravitino masses the
reheat temperature  must be below $10^9\,{\rm GeV}$ in order not to ruin the standard
successes  with nucleosynthesis (e.g.,~\cite{ellis84,KM95,moroi95}). Similar
constraints come from other dangerous relics
which can overclose the universe or release unwanted
entropy by decaying around nucleosynthesis.

It is important to note that $T_{\rm rh}$ is not necessarily the largest
temperature  reached in the history of the universe and in some cases the
temperature can be much higher \cite{Giudice00} or plasma effects may renormalise
the masses of the decay products $\chi$ and $\psi$,
meaning that the inflaton may be kinematically forbidden
to decays \cite{Kolb03}.
In this case the reheat temperature is independent of the couplings
$h, \nu$ and depends only on $m_{\phi}$.
The lesson we learn from this is that
effective masses,
which differ from the bare particle mass either through classical
couplings or quantum
corrections, can have a powerful effect on the dynamics of the system.
This insight, together with the insight that effective masses can be
time and space dependent, is one of the main insights of the 1990's in
inflationary cosmology
and is at the heart of preheating.

%%%%%%%%%%%%%%%%%%%%%%%%%%%%%%%%%%%%%%%%%%%%%%%%%%%%%%%%%%%%%%%%%%%%%%%
\section{Preheating}\label{preheat}
%%%%%%%%%%%%%%%%%%%%%%%%%%%%%%%%%%%%%%%%%%%%%%%%%%%%%%%%%%%%%%%%%%%%%%%

The majority of the inflaton energy at the end of inflation is
homogeneous,  stored in the $k = 0$ mode of the inflaton.
If the inflaton potential has a minimum, such as in simple chaotic
inflation models
given by Eq.~(\ref{chaoticV}),
this energy oscillates perfectly coherently (at least at zero order)
in space. It is this coherence which is key to preheating.
Consider the archetypal massive, chaotic inflaton
potential:
\begin{equation}
V(\phi) = \frac12 m_{\phi}^2 \phi^2\,.
\label{massive}
\end{equation}
Under the influence of this potential, the homogeneous part of the
inflaton simply executes oscillations around $\phi=0$
which gradually decay due to the expansion of
the universe:
\begin{equation}
\phi(t) =\bar{\phi}(t) \sin(m_{\phi}t)\,,~~~
\bar{\phi}(t)=\frac{m_{\rm pl}}{\sqrt{3\pi}m_{\phi}t}\,.
\end{equation}
Here $\bar{\phi}(t)$ is the amplitude of inflaton
oscillations which decreases in time.
The end of inflation is estimated as
$\phi_f=m_{{\rm pl}}/2\sqrt{\pi}$
when the slow-roll parameter $\epsilon$ becomes unity.
The initial amplitude for the oscillation of the field $\phi$ is
slightly smaller than $\phi_{f}$, i.e.,
$|\phi_{0}| \sim 0.2m_{{\rm pl}}$ \cite{KLS97}.

Since the occupation number of the inflaton $k=0$ mode
(the homogeneous part of the inflaton) is very large at the end
of inflation it behaves essentially as a classical field.
One can therefore, to first approximation, treat the inflaton
as a {\em classical external force} acting on the {\em quantum}
 fields $\chi$ and $\psi$. Because the inflaton is time-dependent,
the effective masses of $\chi$ and $\psi$ change very rapidly.
Then this leads to the non-adiabatic excitation of the field
fluctuations by parametric resonance. As a result, the picture we
had before of the inflaton as a large collection of statistically
independent particles breaks down and the spatial and temporal
coherence of the inflaton can cause radical departures from the
``old'' theory of reheating described in the previous section.
This is the essence of preheating \cite{TB90,Dolgov82},
\cite{STB94,KLS94} (see also
Refs.~\cite{KT1,KT2,KT3,Yoshi,Boy96,Boy97,PR97,Son,Baacke97,KLS97}).

For simplicity, let us consider the coupling of the inflaton
to the scalar $\chi$ only, through an interaction term in the
Lagrangian of the form $(1/2)g^2 \phi^2 \chi^2$ where $g$ is
a dimensionless coupling that will play a key role in our analysis.
Classically this does not describe the single-body decay of the
inflaton, but rather the process in which two $\phi$ bosons interact and
decay into two $\chi$ particles.
The total effective potential for this system will be the sum of
the potential driving inflation which for simplicity we assume
is independent of $\chi$, $V(\phi)$, and the above interaction term:
\begin{equation}
V_{{\rm eff}}(\phi,\chi) = V(\phi) +
\frac12 g^2 \phi^2 \chi^2\,,
\label{pot2pc}
\end{equation}
which corresponds to a $\chi$ field with zero bare mass but with an
{\em effective mass} given by
\begin{equation}
m_{\chi, {\rm eff}}^2 \equiv
\frac{\partial^2 V_{{\rm eff}}(\phi,\chi)}{\partial
\chi^2} = g^2 \phi^2(t)\,.
\label{effmass}
\end{equation}
A base mass for the $\chi$ field, $m_{\chi}$ can be accommodated
simply into the above expression by adding
$m_{\chi}^2$. Equation (\ref{effmass}) is the appropriate
notion of effective mass for the $\chi$ field because, neglecting
metric perturbations for the moment, the Fourier modes of the $\chi$
field obey a modified Klein-Gordon equation
\bea
\ddot{\chi}_k +3 H \dot{\chi}_k +
\left[ \frac{k^2}{a^2} + g^2 \phi^2(t)\right]\chi_k = 0\,,
\label{chipreheat}
\eea
with $m_{\chi, {\rm eff}}^2$ playing the crucial role
of mass in the equation.

This is a radical point of view since we now are asking for the
quantum dynamics of the field with a time-dependent mass. From the
point of view of solving ordinary differential equations, this
equation resembles that of a damped ($H \neq 0$) harmonic oscillator
with a time-dependent mass.

{}From WKB theory we know that if the
frequency, $\omega_{k} \equiv [k^2/a^2 + g^2\phi^2(t)]^{1/2}$, is
varying slowly with time, then the solution to this equation is close
to those of the equation in which $\omega_{k}^2$ is constant.
In this case it is well known that the solutions $\chi_k(t)$ do not
grow, which corresponds physically to saying that there is no
production of $\chi$ particles. If, on the other hand, the effective
mass is changing rapidly, then WKB analysis breaks down.
This is quantified by the dimensionless ratio
\beq
R_{a} \equiv \frac{\dot{\omega}_{k}}
{\omega_{k}^2}\,.
\eeq
The regime $|R_{a}| \ll 1$ is often known
as the {\em adiabatic} region since in this case the particle number,
$n_k$, is an adiabatic invariant which does not change in time and
encapsulates the idea that there is no particle production.
In the region $|R_{a}| \gg 1$
the particle number is no longer an adiabatic
invariant and we can expect significant particle production. The
standard estimate of the {\em comoving} occupation number of bosons
in mode $k$ is \cite{KLS97}:
\beq
n_k = \frac{\omega_k}{2}\left(\frac{|\dot{X}_k|^2}{\omega_k^2} +
X_k^2\right) - \frac{1}{2}\,,
\label{numberden}
\eeq
where $X_k \equiv a^{3/2} \chi_k$.
Eq.~(\ref{numberden}) can be
justified, at least qualitatively, as the ratio of the energy in mode
$k$, (the sum of the ``kinetic energy'' $\dot{X}_k^2$ and the
``potential energy'' $\omega_k^2 X_k^2$), divided by the energy per
particle, $\omega_k$.  It clearly shows how the number of particles
is clearly linked to the amplitude of the mode, $X_k$.

For the interaction in (\ref{effmass}), and for long wavelengths
$k/aH \ll 1$, the dimensionless ratio $R_a$ is given approximately by
\beq
R_{a} \simeq \frac{\dot{\phi}}{g \phi^2}
\sim \frac{m_{\phi}}{g \phi} \,,
\label{R}
\eeq
where in the second equality we have assumed $\dot{\phi} \sim
m_{\phi} \phi$ which is suitable for most periodic oscillations of
the inflaton after inflation. The key point about this relation is
that $R_{a}$ diverges whenever $\phi \rightarrow 0$, i.e. at every
oscillation! Hence we can expect rampant particle production around
every oscillation of the inflaton.

{}From Eq.~(\ref{chipreheat}) the equation for $X_{k}$ can be formally
put in the form of the so-called Mathieu equation \cite{Mathieu}
\begin{eqnarray}
\frac{\rd^2 X_k}{\rd z^2} +
\left(A_k -2q \cos 2z \right) X_k=0\,,
\label{Mathieu}
\end{eqnarray}
where $z=m_{\phi}t$ is the natural dimensionless time and
\begin{eqnarray}
\label{Ak}
A_k= 2q + \frac{k^2}{m_{\phi}^2a^2}\,,~~~
q=\frac{g^2\bar{\phi}^2(t)}{4m_{\phi}^2}\,.
\end{eqnarray}

In deriving this we neglected the term
$-(3/4)(2\ddot{a}/a+\dot{a}^2/a^2)$, which is not important
relative to the $g^2\phi^2$ term during preheating.
{}From Eq.~(\ref{Ak}) the allowed range of $A_{k}$
and $q$ corresponds to $A_{k} \ge 2q$.

The strength of resonance depends upon the
variables $A_k$ and $q$, which is described by a
stability-instability chart of the Mathieu equation
\cite{Mathieu,KLS94}. Formally $A_k$ and $q$ should be
constant to use the Mathieu equation but as long as they
are not varying too rapidly the analogy is reasonable.

According to Floquet theory, when $A_{k}, q$ fall in
an instability band, the perturbation $X_k$ grows exponentially with a Floquet
index $\mu_k > 0$, i.e., $X_k \propto {\rm exp} (\mu_k z)$.
For small $q~(\lsim~1)$ the width of the instability band is small and
the expansion of the universe washes out the resonance.  On the other
hand, for the large $q~(\gg 1)$, broad resonance can occur for a wide range
of the parameter space and momentum modes.

Note that the initial amplitude $\phi_{0}$ of the inflaton and
the coupling $g$ play important roles in determining
whether resonance is efficient or not.  Since the inflaton mass
is constrained to be $m_\phi \sim 10^{-6}m_{\rm pl}$ by the COBE
normalization, large resonance parameters, $q \gg 1$, can be easily
achieved for the coupling $g~\gsim~10^{-4}$ with an initial
amplitude, $\phi_{0} \sim 0.2m_{\rm pl}$.

For $q \gg 1$, particle production only occurs near $\phi = 0$. Hence
we may Taylor expand the $\chi$ effective mass around this point and keep only the
quadratic \cite{F96,KLS97}, $\phi = \alpha (t-t_j)$ where
$\phi(t_j) = 0$ for $j=1,2,3...$ and $\alpha$ is a coefficient that
depends on the specific potential one is studying. For the quadratic
potential, Eq.~(\ref{chipreheat}) becomes the
equation of a particle scattering in a parabolic potential:
\beq
\frac{\rd^2 X_k}{\rd t^2} +
\left[\frac{k^2}{a^2} + g^2 \alpha^2
(t-t_j)^2\right]X_k = 0\,,
\label{parabolic}
\eeq
where $\alpha=m\bar{\phi}$.
The general solution to Eq.~(\ref{parabolic}) \cite{F96,KLS97} can be
written as a linear
combination of the parabolic cylinder functions
$W(-\kappa^2/2;\pm \sqrt{2} k_*(t-t_j))$ where
\beq
\kappa^2 \equiv \frac{k^2}{(ak_*)^2}\,,
\label{kappa}
\eeq
and $k_*^2 \equiv gm_{\phi} \bar{\phi}$.

Since the evolution of the inflaton is periodic, the problem is that
of repeated barrier penetration and we can use the exact solution to
estimate the Floquet index, $\mu^j_k$ (the exponent by which the
modes, $X_k$ grow, i.e. $\mu_k^j = \ln (\Delta X_k/\Delta t)$ at each
scattering, $j$)\cite{KLS97}:
\bea
\mu_k^j &=& \frac{1}{2\pi}\ln\left[1 + 2E - 2 \sin \theta^j_{tot}
\sqrt{E(1 + E)} \right]\,, \nonumber \\
E &\equiv& e^{-\pi \kappa^2}\,,
\label{stochres}
\eea
where $\kappa^2$ is the dimensionless wavenumber
defined by Eq.~(\ref{kappa}) and $\theta_{tot}^j$ is the phase of the
wavefunction which changes quantum mechanically at each
 scattering in
the parabolic potential. This expression shows how particle
production decreases exponentially with increasing frequency, $k$ and
how the phase can significantly alter the Floquet index
$\sin \theta_{tot}$.

In Ref.~\cite{KLS97} it was further noticed that while in
Minkowski spacetime the phase is independent of time, this is not
true in an expanding background\footnote{This is an interesting
example where, despite the particle production occurring on a very
short timescale, neglecting the expansion of the universe does not
provide a good approximation to the full result.} and in fact the
change in phase between successive scatterings, $\delta \theta_k
\simeq \sqrt{q}/\tilde{N}^2$ where $\tilde{N}$ is the number of
inflaton oscillations.
%
% v2* Charters reference added
(For more discussion of the phase dynamics see Ref.~\cite{CNM}.)
The requirement $\delta \theta_k > 2\pi$ defines a region of time
and parameter space during which the phase behaves as a
quasi-random number generator \footnote{Simple random number
generators are often of the form $A$ mod $B$ where $A$ is a large
number.}. Hence the value of the Floquet index changes effectively
stochastically from one oscillation of the inflaton to another.
This has become known as {\em stochastic resonance} \cite{KLS97}.
We caution that this is different from the term {\em stochastic
resonance} as used in control theory and condensed matter physics
which is now a well-established experimental field, e.g.,
\cite{stocres}.

{}From this formalism we can also illustrate a crucial point about the
non-perturbative nature of the particle production in preheating. If
decays of the inflaton are perturbative, it is obvious that an inflaton
boson at rest cannot decay to a particle which has more mass than the
inflaton, this is kinematically forbidden. In preheating this is not
true. Particles with masses larger than
the inflaton mass can be produced.

\begin{figure}
\includegraphics[height=2.7in,width=3.2in]{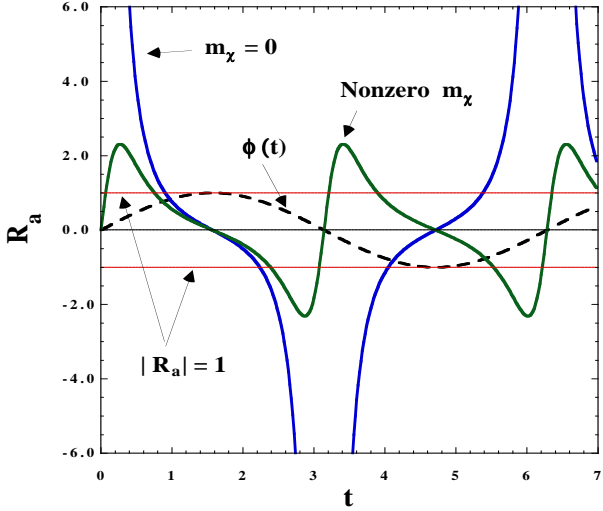}
\caption{$\phi(t)$ and the dimensionless ratio
$R_a \equiv \dot{\omega}_k/\omega^2_k$ for the massive model
for $m_{\chi} = 0$ and an arbitrary value of $m_{\chi} > 0$.
While $R_a$ diverges at $\phi=0$ for the massless case,
it vanishes at $\phi=0$ for the massive case.
Particle production takes place in the
non-adiabatic region characterised by $|R_a| > 1$.
Massive particles can be created provided that the maximum value
of $|R_a|$ is larger than 1.
}
\label{figR}
\end{figure}

To see this, consider the expression for $R_{a}$, in Eq. (\ref{R}), but
this time include a bare mass for the $\chi$ field, $m_{\chi}$.
Proceeding as before gives:
\beq
R_{a} \simeq \frac{m_{\phi} g^2 \phi^2}{(m^2_{\chi} + g^2
\phi^2)^{3/2}}\,.
\label{massiver}
\eeq
In this case (at least for $g^2 > 0$) $R_{a}$ can no longer
diverge, even in the $k\rightarrow 0$ limit and indeed now
vanishes at $\phi = 0$ (see Fig.~\ref{figR}).
However non-adiabatic particle creation occurs provided
$m_\chi<|g\phi|$ (similar properties hold in
producing particles at large momentum, $k$).
When $g^2 < 0$, $R_{a}$ can still diverge formally and
in this case production of
extremely massive particles is possible, although care must be taken
so that the total potential is bounded from below.

\subsection{The conformally invariant case}

There is an interesting exactly solvable special model worth
mentioning in which the expansion of the universe can be transformed away and exact
Floquet theory is applicable. This is the conformally invariant potential
\beq
V_{\rm eff}(\phi,\chi) = \frac{1}{4}\lambda\phi^4 +
\frac12 g^2\phi^2\chi^2\,.
\label{conformal}
\eeq
In this model the universe rapidly becomes radiation dominated
($a \propto t^{1/2} \propto \tau$, where $\tau$ is a conformal time).
The homogeneous part of the inflaton obeys the
equation of motion
\beq
\frac{{\rm d}^2 \varphi}{{\rm d}x^2}+\lambda \varphi^3= 0\,,
\label{confback}
\eeq
where $\varphi \equiv a\phi$ and $x \equiv \sqrt{\lambda}
\varphi_0 \tau$ with $\varphi_0$ being
the initial amplitude of the oscillations
of the conformal field $\varphi$.
 Note that this equation is that of an harmonic
oscillator in flat space - the expansion of the universe has been
absorbed into the field and time redefinitions. The
solution to this is the Jacobi cosine function
\beq
\varphi = \varphi_0 \mbox{cn}
\left(x; \frac{1}{\sqrt{2}}\right)\,,
\label{cn}
\eeq
which is closely approximated by
$\varphi_0 \cos(0.8472x)$ \cite{GKLS97}.

One can show that the equation of motion for
$X_k \equiv a\chi_k$ in conformal time is also
just that of a field in Minkowski
spacetime with all the effects of the expansion absorbed into the
field and time redefinitions:
\beq
\label{Lame}
\frac{{\rm d}^2}{{\rm d}x^2} X_k+\left[\kappa^2
+\frac{g^2}{\lambda}{\rm cn}^2
\left( x; \frac{1}{\sqrt{2}}\right) \right]X_k=0\,,
\eeq
where $\kappa^2 \equiv k^2/(\lambda \varphi_0^2)$.
This equation is the so-called Lam\'e equation.
The advantage of converting the equations into Minkowski form is that
the coefficients appearing in Eq.~(\ref{Lame}) are now {\em
exactly} periodic in time and hence one can use the theorems of
Floquet theory to show that there must be exponentially growing
solutions, $X_{k} \propto e^{\mu x}$ where
the Floquet index varies between zero and a maximum value
$\mu_{max}
\simeq 0.238$ as a function of ${\kappa}$ and $g^2/\lambda$
\cite{Kaiser97,Kaiser98,GKLS97}.

\begin{figure}
\includegraphics[scale=0.75]{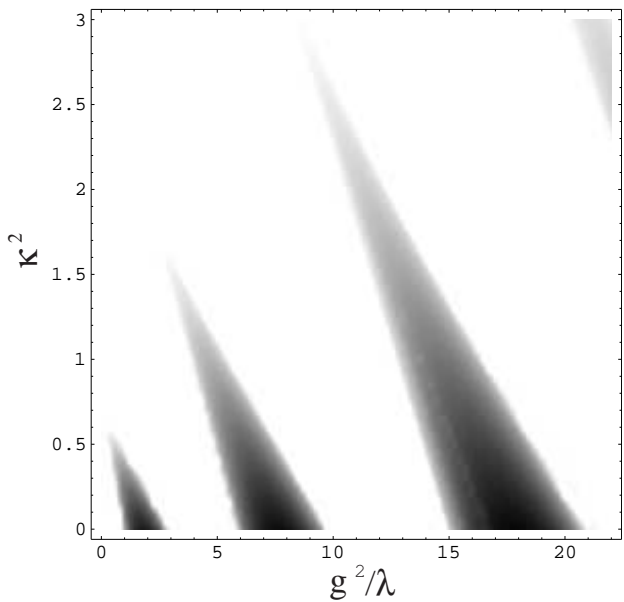}
\caption{
Density plot of the Floquet chart for the
Lam\'e equation (\ref{Lame})
for $0 \leq g^2/\lambda \leq 22$ and
$\kappa^2 \leq 3$.
The shaded regions correspond to parameter ranges in which
parametric resonance occurs, $\mu_k > 0$.
The Floquet index, $\mu_k$, takes larger values in the
darker shaded regions, and reaches its maxima  for
$g^2/\lambda=2n^2$ at $\kappa^2=0$.
}
\label{floquet}
\end{figure}

The structure of resonance is completely determined by the
value of the parameter $g^2/\lambda$.
One can expect an efficient particle production
even for small couplings, $g^2/\lambda \sim {\cal O}(1)$.
The long-wave modes ($\kappa \to 0$) are enhanced
in the intervals $n(2n-1)<g^2/\lambda<n(2n+1)$ with $n$
being a positive integer.
The centre of the resonance bands corresponds to
$g^2/\lambda=2n^2=2, 8, 18, \cdots$,
around which parametric resonance is efficient.
Figure \ref{floquet} shows that there are upper limits of the momenta
which are amplified by parametric resonance, depending on
the values of $g^2/\lambda$.

%%%%%%%%%%%%%%%%%%%%
%%%%%%%%%%%%%%%%%%%%

The perturbation equation for inflaton can be written as
\beq
\frac{{\rm d}^2}{{\rm d}x^2} \delta \varphi_{k}
+\left[\kappa^2+3{\rm cn}^2
\left( x; \frac{1}{\sqrt{2}}\right) \right]
\delta \varphi_{k}=0\,,
\eeq
where $\delta \varphi_{k} \equiv a\delta \phi_{k}$.
This corresponds to $g^2/\lambda=3$ in Eq.~(\ref{Lame}).
The perturbations $\delta \varphi_{k}$ grow provided that
the momenta exist in the range \cite{Kaiser97,Kaiser98,GKLS97}:
\beq
\label{subHubble}
\frac32<\kappa^2<\sqrt{3}\,.
\eeq
The maximum growth rate for $\delta \varphi_k$ is
found to be $\mu_{{\rm max}} \approx 0.03598$
at $\kappa^2 \approx 1.615$.
It is interesting that preheating occurs for the quartic model
even in the absence of the $\chi$ field.
However the modes which are amplified are at sub-Hubble scales
(\ref{subHubble}) for the inflaton fluctuations. We will understand this later as
due to the fact that the entropy perturbation in single-field models
decays as $k^2$ on large scales.

\subsection{Geometric preheating}

In previous subsections, we considered the standard scenario of preheating
where inflaton $\phi$ is coupled to $\chi$ through an interaction,
$(1/2) g^2\phi^2\chi^2$.
{}From the viewpoint of quantum field theories
in curved spacetime, non-minimal couplings naturally arise,
with their own nontrivial renormalization group flows.
The ultra-violet fixed point of these flows are often divergent,
implying that non-minimal couplings may be important
in the early universe. This provides an alternative - geometric - channel for resonance
\cite{BL97,TMT99} in which scalar fields
coupled to the scalar curvature $R$ which oscillates during reheating.

Let us consider an inflaton field $\phi$ interacting with
a scalar field $\chi$, which is non-minimally coupled
to gravity:
\beqa
{\cal L}=\sqrt{-g} \biggl[ \frac{m_{\rm pl}^2}{16\pi}R
&-& \frac12 (\nabla \phi)^2-\frac12 m_{\phi}^2 \phi^2
-\frac12 (\nabla \chi)^2 \nonumber \\
&-&\frac12 g^2\phi^2\chi^2
-\frac12 \xi R \chi^2 \biggr]\,.
\eeqa
Then the equation for the perturbation $\delta \chi_{k}$ reads
\beq
\ddot{\delta \chi}_{k}+3H\dot{\delta \chi}_{k}
+\left(\frac{k^2}{a^2}+g^2\phi^2+\xi R \right)
\delta \chi_k=0\,.
\eeq
The scalar curvature,
$R=6(2H^2+\dot{H})$, oscillates during reheating.
Making use of the time-averaged relation
$\langle m_{\phi}^2\phi^2 \rangle=
\langle \dot{\phi}^2 \rangle$, we find that
$R \sim 8\pi m_{\phi}^2\phi^2/m_{\rm pl}^2$.
Then the contribution of the $\xi R$ term becomes
more important than that of the $g^2\phi^2$ term
when the coupling satisfies the condition
\beq
\label{xig}
\frac{|\xi|}{g^2} \gtrsim
\frac{1}{8\pi} \left(\frac{m_{\rm pl}}{m_{\phi}}
\right)^2\,.
\eeq
When $g \gtrsim 10^{-4}$,
the condition (\ref{xig}) corresponds to $|\xi| \gtrsim 10^2$.
It was found that
parametric resonance occurs for negative non-minimal couplings
of order $|\xi| \sim 1$ provided that non-gravitational coupling
is smaller than of order $g=10^{-5}$.
Thus the geometric particle production can provide an alternative
scenario of preheating even when the non-gravitational interaction is negligible.

\subsection{Almost-periodic and random parametric resonance}

So far we have considered simple potentials with periodic evolution
of the $\chi$ effective mass modulated by powers of $a$.
What happens if we consider more general evolution
of the effective mass?
For example, what happens if the effective mass$^2$ behaves as:
\beq
m_{\chi, {\rm eff}}^2 \propto g_1^2 \cos(\omega_1 t)
+ g_2^2 \cos(\omega_2 t)\,,
\label{quasi}
\eeq
with $\omega_1/\omega_2$ irrationally related?
In this case the function is not exactly
periodic. Or what happens if the effective mass evolves with a random
component or stochastically in time?

Given our general criterion $R_{a}> 1$ in Eq.~(\ref{R}) for resonance we should suspect that exponential growth of $\chi_k$
modes should be
possible in certain cases, and this is indeed the case. One way to
show this is to note that there is a duality between the temporal
evolution of $\chi_k$ modes and the 1-dimensional time-independent
Schr\"odinger equation which relates the wavenumber $k$ to the
eigenvalues $\lambda$ and interchanges space and time. Then, the
modes $k$ which grow exponentially correspond to the complement of
the spectrum of allowed eigenvalues.

This immediately explains why the exponentially growing modes in the
periodic case belong to bands. These are, in the dual picture, just
the complement of the usual Bloch conduction bands that characterise
the allowed energy levels of periodic lattice structures in metals.

Using this insight, and results from the spectral theory of the
Schr\"odinger operator, we can show that in the case of
almost-periodic and similar evolution, the $\chi_k$ modes do not
experience resonance (have vanishing Floquet index $\mu_k = 0$) only
on a nowhere dense (Cantor) set \cite{BB98,Zanchin98,Zanchin99}.
Hence, generically resonance occurs at all wavelengths,
not just in certain resonance bands.

Similar results hold for random evolution of the $\chi$ effective
mass. In this case one can rigorously
show that for sufficiently random evolution\footnote{By sufficiently
random we mean that the temporal correlations $\langle \xi(t)\xi(t -
t')\rangle$ decay sufficiently rapidly with $|t-t'|$.} the Floquet
index $\mu_k$ is strictly positive for all $k$ except on a set of
measure zero \cite{BB98,Zanchin98,Zanchin99}.
In the case $m_{\chi, {\rm eff}}^2 = \kappa^2
+ q\xi(t)$ where $\kappa^2$ and $q$ are dimensionless constants and
$\xi(t)$ is a mean-zero, ergodic Markov process, perturbative
expansion gives \cite{BB98,Zanchin98,Zanchin99}
\beq
\mu_k = \frac{\pi}{4}\frac{q^2}{\kappa^2}
\hat{f}(2\kappa) + {\cal O}(q^3)\,,
\eeq
where $\hat{f}(\kappa)$ is the Fourier transform of the expectation
value of the two-time correlation function $\langle \xi(t)\xi(t -
t')\rangle$. Hence, a mode $k$ will grow exponentially if the Fourier
transform is non-zero at twice the frequency. In the case where the
noise is delta-correlated, then $\langle \xi(t)\xi(t - t')\rangle
\propto \delta(t-t')$ and $\hat{f}(\kappa) = {\rm const}.$ and all modes
grow exponentially. Temporal correlations (coloured noise) mean that
the Fourier transform has compact support and hence removes the
resonance for sufficiently large frequencies.

This result has a celebrate condensed matter analogue - Anderson
localisation, which states that with small random impurities,
eigenfunctions become exponentially localised.  In the case of
reheating, these results mean that periodic evolution is actually the
most modest. By studying the conformal case we can extend these
results to an expanding FRW background. Since white noise is a well-studied limit
of chaotic motion these results also provide insight into preheating
in cases where the inflaton evolves chaotically \cite{BT98}.

\subsection{Tachyonic preheating and the negative coupling instability}

% v2*

So far we have always considered the case $m_{\chi,{\rm eff}}^2 >
0$. However it was realised in Ref.~\cite{negci} that the coupling
$g^2 \phi^2 \chi^2/2$ in Eq.~(\ref{pot2pc}) could just as well be
replaced by $g \phi^2 \chi^2/2$ with $g<0$ if the potential was
supplemented by additional terms ($\propto \chi^4, \phi^4$) that
ensured that the full potential was bounded from below. In this
case, the $\chi$ effective mass squared is $m_{\chi,{\rm eff}}^2
\sim g \phi^2$ which can be negative, implying that there can be
tachyonic phases.
% v2* removed following sentence because I'm not sure what is meant.
% I think it can be left out, but put it back if you think it's needed:
%  This possibility has a profound impact on the resonance since
%  $\langle \chi \rangle > 0$ during inflation.
Crucially this negative coupling instability  implies that many
more modes are resonantly amplified compared with standard
preheating (with $g^2 > 0$) and the corresponding Floquet indices,
$\mu_k$ can be much larger, allowing for the production of very
massive $\chi$ particles that may be relevant for baryogenesis
\cite{negci}.

% v2* added 3 sentences
Tachyonic instabilities inevitably occur in models of spontaneous
symmetry breaking. In particular $m^2_{\chi,{\rm eff}}$ becomes
negative in the hybrid inflation model given by
Eq.~(\ref{hybridpo}) when the $\phi$ field drops below the
critical value $\phi_c=M/g$ at the end of inflation.
Long-wavelength modes initially grow due to the tachyonic
(spinodal) instability, but non-linear effects rapidly become
important \cite{Felder01p1,Felder01p2}.
The existence of non-perturbative features such as topological
defects has a profound impact on the time taken for backreaction
to end preheating. Instead of taking multiple oscillations it was
found that resonance ends very rapidly, after only ${\cal O}(1)$
oscillations; a process dubbed ``tachyonic preheating"
\cite{Felder01p1}. In the case of a $Z_2$ symmetry with two vacua
at $\phi=\pm v$, the universe is divided into regions of $\phi =
\pm v$, separated by domain walls. The gradient energy associated
with this non-perturbative field configuration is comparable to
the initial potential energy of the field before symmetry
breaking. The gradients contribute very strongly to the variance
$\langle \delta \phi^2 \rangle$ which quells the resonance very
rapidly. The reader is referred to \cite{Felder01p1,Felder01p2}
for more details.

\subsection{Fermionic preheating}

Quite soon after the initial studies of preheating of scalar fields
attention turned to the possibility of resonant production of
fermions. This is an important issue since many problematic
particles such as gravitinos, are fermions and resonant
production of them could have a profound impact
on dangerous relic abundances
\cite{Baacke98,GK98}.

Nevertheless fermions obey the exclusion principle which implies that
$n_k \leq 1$ so the system is severely constrained.
Consider the conformally coupled inflaton interacting
with a massless fermion
field $\psi$ through the interaction term
$h\phi(t) \overline{\psi}\psi$ where $h$ is the dimensionless Yukawa
coupling. The resulting Dirac equation to first order in a flat
FRW background is \cite{Baacke98,GK98} \footnote{Fermions have no homogeneous, classical component.}
\beq
[i\gamma_{\mu} \nabla^{\mu} - m_{\rm eff}]\psi = 0\,,
\eeq
where $\gamma_{\mu}$ are the Dirac matrices and the effective mass
is given by
\beq
m_{{\rm eff}}=m_{\psi}+h\phi(t)\,.
\eeq
Here $m_{\psi}$ is the bare mass of fermions.
As in the scalar case, the coupling to the homogeneous part of the inflaton acts as a
time-dependent effective mass $m_{{\rm eff}}$.

Let us consider the self-coupling inflaton potential
$V(\phi)=(1/4)\lambda \phi^4$.
We introduce a conformally rescaled field $\tilde{\psi}=a\psi$
and decompose the field $\tilde{\psi}$ into Fourier
components as
\begin{eqnarray}
\tilde{\psi} &=& \frac{1}{(2\pi)^{3/2}} \int d^3 k
\sum_{s} [a_s(k) \tilde{{\bf u}}_s(k,\eta)
e^{+i {\bf k} \cdot {\bf x}}  \\ \nonumber
& & +b_s^{\dagger}(k)
\tilde{{\bf v}}_s (k,\eta)e^{-i {\bf k} \cdot {\bf x}}]\,.
\label{8_1_10}
\end{eqnarray}
Imposing the following standard ansatz \cite{GK98}:
\begin{eqnarray}
\tilde{{\bf u}}_s(k,\eta)
=(-i\gamma^{\mu}\partial_{\mu}
-m_{\rm eff}a) \tilde{\psi}_k(t) W_{\pm}(k)\,,
\end{eqnarray}
where $W_{\pm}(k)$ are the eigenvectors of the helicity operator,
which satisfy the relation
$\gamma^0 W_{\pm}(k)=1$ and ${\bf k} \cdot \sum
W_{\pm}(k)= \pm 1$,
we obtain the mode equation for the $\tilde{\psi}_k$:
\begin{eqnarray}
\label{fermionequ}
\left[ \frac{\rd^2}{\rd x^2}+\kappa^2+f^2
-i \frac{\rd f}{\rd x} \right]\tilde{\psi}_k=0\,,
\end{eqnarray}
where
\begin{equation}
f \equiv \frac{m_{\rm eff}a}{\sqrt{\lambda}\phi(0)},~~~
\kappa^2 \equiv \frac{k^2}{\lambda \phi^2(0)},~~~
x \equiv \sqrt{\lambda}\phi(0) \eta\,.
\end{equation}
Here $\phi(0)$ is the initial value of inflaton
at the onset of preheating.
Equation (\ref{fermionequ}) bears a striking resemblance
to the Klein-Gordon equation except for the appearance of
the complex term $i \rd f/\rd x$
in the effective mass. This is to be expected since the Klein-Gordon
equation expresses relativistic energy momentum conservation which
must also apply to fermions. The complex term appears as the enforcer
of the Pauli exclusion principle.

Equation (\ref{fermionequ}) has a WKB-form solution given in terms of creation
and annihilation operators by
\begin{eqnarray}
\tilde{\psi}_k &=& \alpha_k N_+
\exp \left(-i \int_0^t \Omega_k {\rm d}x \right)  \nonumber \\
& &+\beta_k N_-  \exp \left(+i \int_0^t \Omega_k {\rm d}x
\right)\,,
\end{eqnarray}
where $\Omega_k^2 \equiv \kappa^2+f^2$ and
$N_{\pm} \equiv 1/\sqrt{2\Omega_k(\Omega_k \pm f)}$.
The comoving number density of produced fermions is
given in terms of the  Bogoliubov coefficients
$\beta_k$ by \cite{GK98,Tsuji00}
\begin{eqnarray}
\label{number}
n_k \equiv |\beta_k|^2
=\frac12 -\frac{\kappa^2}{\Omega_k}
{\rm Im} \left(\tilde{\psi}_k \frac{d \tilde{\psi}_k^*}{dx} \right)
-\frac{f}{2\Omega_k}\,,
\end{eqnarray}
where $^*$ denotes complex
conjugation and $\mbox{Im}$ the imaginary part of the expression.
The initial conditions are chosen to be $\alpha_k(0)=1$, $\beta_k(0)=0$,
which corresponds to $n_k(0)=0$. The Bogoliubov coefficients
satisfy the relation $|\alpha_k(t)|^2 + |\beta_k(t)|^2 = 1$, which means
that the exclusion principle restricts the number density of fermions
to below unity, $n_k(t) \le 1$.
It is interesting to consider the limits of $\kappa \rightarrow 0$ in
Eq.~(\ref{number}). In this case $n_k \rightarrow 0$
irrespective of $q$, reinforcing our earlier discussion that the
fermion has no homogeneous component.

Fermions are non-adiabatically created when the effective masses
of fermions change rapidly\footnote{We caution the reader to
distinguish between the use of ``non-adiabatic'' here (where we use it
in the sense of  ``adiabatic invariants'') and its use in the discussion
of metric perturbations where it is used in conjunction with
entropy/isocurvature perturbations. The two uses are different.}.
This takes place around $m_{{\rm eff}}=0$ \cite{Giudice99},
corresponding to the inflaton value
\begin{eqnarray}
\label{phic}
\phi_{c}=-m_{\psi}/h\,.
\end{eqnarray}
When the condition, $m_\psi>|h\phi|$, is satisfied at the beginning of
reheating, the inflaton does not pass through the resonance point
(\ref{phic}). Therefore we require the condition, $m_{\psi}<|h\phi|$,
to lead to parametric excitation of fermions.
Since inflation ends around $\phi=0.3m_{\rm pl}$, it is possible
to generate heavy fermions whose masses are of order
$10^{17}$-$10^{18}$\,GeV \cite{Giudice99,PS00}.

When we consider supersymmetric theories such as supergravity,
gravitino production can provide us a useful tool to constrain
particle physics models in early universe.
In a perturbative theory of reheating the thermal production of
gravitinos places a constraint $T_{{\rm rh}} \lesssim 10^9$ GeV,
on the reheating temperature \cite{KM95,moroi95}.
On the other hand non-thermal production of gravitinos
during preheating has been extensively studied by many authors
\cite{MM99,Giudice99v2,Kallosh99,Kallosh00,Nilles01,Nilles01v2,GKM03}.
In particular it was found in Ref.~\cite{Nilles01} that gravitino
creation is suppressed relative to the superpartner of
the inflaton (inflatino) for a model of two scalar fields
including a supersymmetry breaking field.

A similar conclusion has been reached in Ref.~\cite{GKM03}
for a more realistic supergravity inflation model.
While these results show that gravitino over-production
can be avoided during preheating, further studies of how the mixing
occurs between fermionic fields
for the full Lagrangian derived from supergravity is required for a
complete understanding of the problem.

\subsection{Instant preheating}

Non-perturbative parametric or stochastic resonance is not the only
way that a changing effective mass can lead to interesting effects.
Consider a scalar field $\chi$ with a bare mass, $m_{\chi}$
coupled to the inflaton through a term $(1/2)g\phi^2\chi^2$.
The $\chi$ field has an effective
mass squared of: $m_{\chi}^2 + g \phi^2$.
If $g > 0$, the effective mass is always larger
than the bare mass. However, if $g < 0$, then
the effective mass vanishes when $\phi = \pm m_{\chi}/\sqrt{-g}$.
As a result the inflaton is kinematically allowed to decay there and the
corresponding $R_a$ diverges, see Eq.~(\ref{massiver}).

This insight was used to argue that in so-called distributed-mass
models where there are a large number of decay states with a spectrum
of masses (as occurs in string theory due to the exponential density
of states), the slowly rolling inflaton will successively make each
of the states massless and hence will lead to successive bursts of particle
production which may be strong enough to sustain warm inflation
\cite{warm}.

Coupling these insights to those of parametric resonance leads to
an interesting phenomenology \cite{instant}. For a coupling
$(1/2)g^2 \phi^2 \chi^2$ of some scalar $\sigma$
and for large $q \gg 1$ we have
shown that particle production only occurs in small bursts near $\phi
= 0$. Now imagine the $\chi$ field is also coupled to a fermion
field through an interaction $h\chi \overline{\psi}\psi$. Since
these are single body decays, one may use Eq.~(\ref{singlebody})
with $\phi$ replaced by $\chi$. We see that
\begin{eqnarray}
\Gamma_{\chi \rightarrow \overline{\psi}\psi} \simeq
\frac{h^2g|\phi|}{8\pi}\,.
\end{eqnarray}
Hence there is
massive resonant production of $\chi$ particles when $\phi \simeq 0$
(during which time $\Gamma_{\chi \rightarrow \overline{\psi}\psi}
\sim 0$) followed by $\phi$ oscillating to its maximum at which point
the $\chi$ bosons have swelled to maximum effective mass and
are most
likely to decay. In the first couple oscillations
$|\phi| \sim 0.1 m_{{\rm pl}}$ and hence the $\chi$ bosons are
kinematically allowed to decay to fermion pairs of mass up to
$\sim g|\phi| \simeq g m_{{\rm pl}}/10
\sim 10^{18}\,{\rm GeV}$ if we allow $g \sim 1$.

Production of particles near the Planck mass is difficult to
achieve even for $q \gg 1$ in standard parametric resonance but it is a
characteristic feature of instant reheating where large amounts
of energy are transferred into massive fermions within a couple of
oscillations. We note that instant preheating scenario may be applied
to a quintessential inflation in which the potential does not have a
minimum, see e.g., \cite{SS04}.

\subsection{Backreaction and rescattering}

So far we have only considered the production of secondary fields
($\chi$, $\psi$) through parametric resonance. Usually in cosmology
the perturbations depend on the background dynamics but not vice
versa. However, the rapid draining of energy due to the rampant
particle production soon affects the dynamics of the inflaton itself.
How can this be modelled? We can treat this first by considering the
Hartree, or mean-field, approximation \cite{KT2,KLS97}.
In this approximation all
effects of the amplification of the $\chi$ field (we consider only
the scalar case here) are mediated through the variance of $\chi$,
$\langle \chi^2 \rangle$,\footnote{One way to realise that this is a
significant simplification is to note that the variance is a single
real number which replaces the operator product $\chi^2$.} and the
homogeneous part of the inflaton now obeys the equation:
\beq
\label{Har}
\ddot{\phi} + 3 H \dot{\phi} + V_{\phi} +
g^2 \langle \chi^2 \rangle
\phi = 0\,,
\eeq
where the variance is defined to be:
\beq
\langle \chi^2 \rangle = \frac{1}{2\pi^2}
\int \rd k\,k^2 |\chi_k|^2\,.
\label{var}
\eeq
Crudely speaking the variance is correlated with the energy in the
$\chi$-fluctuations.

Let us consider the quadratic inflaton potential given by
Eq.~(\ref{massive}).
Initially the variance term is vanishingly small, but it grows
rapidly according to $\propto e^{2\mu m_{\phi} t}$ (where $\mu$ is
some suitable average Floquet index) during preheating and therefore
increases the effective mass of the inflaton. We can understand the
effect of this increase qualitatively through analogy with a simple
harmonic oscillator. Firstly the frequency of $\phi$ oscillations
increases and secondly, the approximate conservation of energy means
the amplitude of $\phi$ oscillations decreases roughly as $\bar{\phi}
\sim m_{\phi, {\rm eff}}^{-1}$.
This in turn rapidly decreases the resonance parameter $q \propto
\bar{\phi}^2/m_{\phi, {\rm eff}}^2$, which acts to shut off
the resonance, stopping the production of $\chi$ particles.

To estimate the maximum variance that can be achieved one can simple
equate the two mass terms in the equation of motion for the
condensate, $\phi$. These two terms are $m_{\phi}^2$
and $g^2 \langle \chi^2 \rangle$ and hence we generally
expect $\langle \chi^2
\rangle_{{\rm max}} \sim m_{\phi}^2/g^2$.
It is clear we are dealing with a
non-perturbative process since the coupling $g$ appears in the
denominator. When the backreaction of the $\chi$ fluctuations are as
large as the inflaton bare mass it is difficult for the resonance to
continue much further for the reasons discussed above. The time at
which this occurs can be estimated by writing $\langle \chi^2 \rangle
\propto e^{2\mu m_{\phi} t}$, hence
\beq
t_{{\rm end}} \sim \frac{1}{\mu m_{\phi}}
\ln \left(\frac{m_{\phi}}{g}\right)\,.
\label{end}
\eeq
Again notice the non-perturbative nature of this expression and the
logarithmic dependence on the couplings. This comes from the
exponential growth of fluctuations which means that the end of the
resonance is rather robust in these theories.  Nevertheless the
Hartree approximation certainly does not give a complete description
of preheating because it  neglects the fluctuations of the inflaton.

To go beyond the mean-field approximation let us examine the
equations of motion in real space. The equations of motion involve
products $\phi^2({\bf x},t) \chi({\bf x},t)$ and $\phi({\bf x},t)
\chi^2({\bf x},t)$. We transform them into Fourier space using the
convolution theorem which states that the Fourier transform (denoted
FT) of a product is the convolution (denoted $*$) of the individual
Fourier transforms, i.e.,
\beq
FT(f\times g) = FT(f) * FT(g)\,,
\label{conv}
\eeq
where the convolution in three dimensions is defined as
\beq
(f * g)(\bk) = \int d^3k' f(\bk) g(\bk - \bk')\,.
\eeq
We can recover the mean-field/Hartree equation (\ref{Har})
by assuming $\phi = \phi(t)$ only, with no spatial dependence.
In Fourier space
this corresponds to a delta-function at $k=0$ which collapses the
convolution immediately.

In general the inflation has quantum fluctuations (that give
rise to an adiabatic spectrum of perturbations) so $\phi = \phi_0(t)
+ \delta \phi({\bf x},t)$. In transforming to Fourier space the
convolutions do not collapse and we are left with complicated
integro-differential equations. For the effective potential
given by $V_{{\rm eff}} =
(1/2)m_{\phi}^2\phi^2 +(1/2)g^2 \phi^2 \chi^2$
the equations of motion in Fourier space for the Fourier modes of the
fluctuations $\delta\phi$ are:
\bea
\hspace*{-3.0em}& &
\ddot{\delta\phi}_k + 3H\dot{\delta\phi_k}
+ \left(\frac{k^2}{a^2}
+ m_{\phi}^2\right)\delta\phi_k  \nonumber\\
\hspace*{-3.0em}& &=
-\frac{g^2\phi_0(t)}{(2\pi )^3}\int \rd^3k' \chi_{\bk-\bk'}
\chi_{\bk'} \nonumber \\
\hspace*{-3.0em}&&
-\frac{g^2}{(2\pi )^3}\int \rd^3 k'  \rd^3 k''
\delta\phi_{\bk-\bk'+\bk''} \chi_{\bk'} \chi_{\bk''}\,,
\label{rescat_phi}
\eea
and those for $\chi_k$:
\bea
\hspace*{-3.0em} & &
\ddot{\chi_k} + 3H\dot{\chi_k}
+ \left[\frac{k^2}{a^2}
+ g^2\phi_0^2(t) \right] \chi_k \nonumber\\
\hspace*{-3.0em}&&=
-\frac{g^2\phi_0(t)}{(2\pi )^3}
\int \rd^3k' \chi_{\bk-\bk'}
\delta \phi_{\bk'} \nonumber \\
\hspace*{-3.0em}&& -\frac{g^2}{(2\pi )^3}
\int \rd^3 k'  \rd^3 k''
\chi_{\bk-\bk'+\bk''} \delta\phi_{\bk'} \delta\phi_{\bk''}\,.
\label{rescat_chi}
\eea

\begin{figure}
\includegraphics[scale=0.43]{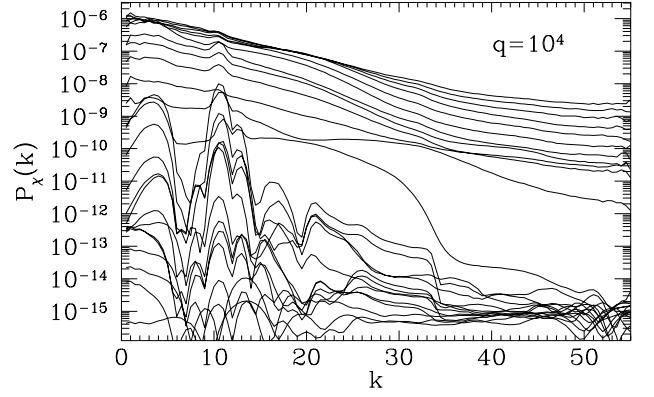}
\caption{
Lattice simulations for the temporal evolution of the spectrum of
the $\chi$ fluctuation  during preheating for the model
$V(\phi, \chi)=(1/2)m_\phi^2\phi^2+(1/2)g^2\phi^2\chi^2$ with
an initial resonance parameter $q=10^4$.
At early times resonance bands are visible but with the
subsequent rescattering the modes in between the resonance bands
are filled in and the spectrum tends towards
a featureless spectrum.
{}From Ref.~\cite{KT3}. }
\label{spec_evolve}
\end{figure}

The Hartree approximation corresponds to neglecting the scattering
between different Fourier modes.
Under this approximation only the remaining contribution
on the r.h.s. of Eq.~(\ref{rescat_phi})
is the second term with $\bk'=\bk''$, which gives rise
to the $g^2 \langle \chi^2 \rangle \delta \phi_k$ term.
Similarly we obtain the $g^2 \langle \delta \phi^2 \rangle \chi_k$
term from the r.h.s. of Eq.~(\ref{rescat_chi}).
Hence the perturbed field equations under the Hartree approximation are
\bea
\hspace*{-3.0em}
& &\ddot{\delta\phi}_k + 3H\dot{\delta\phi_k}
+ \left(\frac{k^2}{a^2}
+ m_{\phi}^2+g^2 \langle \chi^2 \rangle
\right)\delta\phi_k  =0\,, \\
\hspace*{-3.0em}
& & \ddot{\chi_k} + 3H\dot{\chi_k}
+ \left[\frac{k^2}{a^2}
+ g^2(\phi_0^2(t)+\langle \delta \phi^2 \rangle)
\right]\chi_k=0\,.
\eea
Since the scattering of different momentum modes especially
becomes important at the nonlinear stage of preheating,
the use of the mean-field approximation obviously shows
a limitation to estimate the final variance correctly.
The transition from linear to nonlinear stages of preheating can
be clearly seen in Fig.~\ref{spec_evolve} that shows
the evolution of the power spectrum
${\cal P}_\chi(k)=\langle |\chi_{\bk}|^2 \rangle/V$
($V$ is a normalisation spatial volume).

%%%%%%%%%%
%%%%%%%%%%

%%%%%%%%%%
\begin{figure}
\includegraphics[scale=0.43]{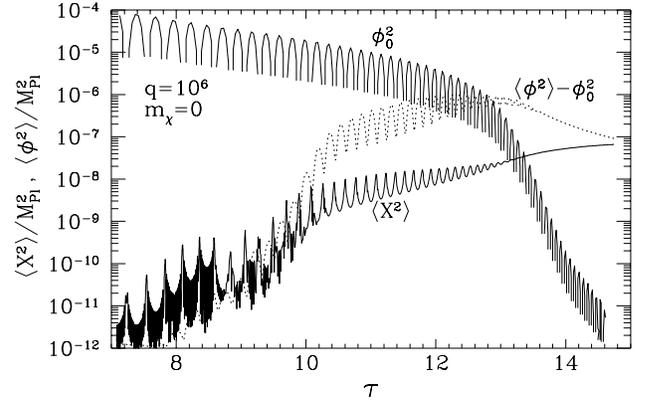}
\caption{
Lattice simulations for the evolution of the field variances
in the $\phi$ and $\chi$ fields
during preheating for the model
$V(\phi, \chi)=(1/2)m_\phi^2\phi^2+(1/2)g^2\phi^2\chi^2$ with
an initial resonance parameter $q=10^6$.
Here $\tau$ is a conformal time and
$\langle X^2 \rangle$ corresponds to $\langle \chi^2 \rangle$ in our notation.
Note that the $\chi$ fluctuation begins to grow
initially through parametric resonance but this is followed by the growth of the $\phi$
fluctuations through the nonlinear process of rescattering which is significantly
more rapid (with roughly double the Floquet index).
The backreaction shuts off the resonance in $\chi$ field earlier than
it does in the $\phi$ fluctuations which also dominate the final
variances showing how full lattice simulations are crucial to
a full understanding of the problem.
{}From Ref.~\cite{Kh97}.
}
\label{rescatt}
\end{figure}
%%%%%%%%%%

Equations (\ref{rescat_phi}) and (\ref{rescat_chi}) also
explain an observation of the early lattice
simulations that after the initial resonance in the $\chi$ field,
there is suddenly very rapid amplification of the fluctuations in the
inflaton. This can be seen by looking at the first term on the r.h.s. of
Eq.~(\ref{rescat_phi}). This is a term independent of $\delta\phi_k$
which grows as $\exp(2\mu m_{\phi} t)$ since each factor of $\chi$ is
growing exponentially with Floquet index $\mu$. Hence this provides a
rapidly growing source term for $\delta\phi_k$ fluctuations.
The mode-mode coupling between different momentum modes
is dubbed {\it rescattering} in Ref.~\cite{KT1,KLS97}.

The general solution to such an inhomogeneous equation is the
solution to the homogeneous part (in this case just simple
oscillations) plus a temporal integral over the source term
multiplied by the appropriate Green's function, in this case $\sin
(\sqrt{k^2 + m_{\phi}^2}(t-t'))$ \cite{KLS97}. Rather robustly
therefore one predicts $\delta\phi_k \propto e^{2\mu m_{\phi} t}$ and
hence that $\delta\phi_k$ fluctuations will grow basically with twice
the Floquet index of the $\chi$ fluctuations,
as is visible in Fig.~\ref{rescatt}.

If the homogeneous part of the field $\chi$ is non-vanishing,
the first terms on the r.h.s. of  Eqs.~(\ref{rescat_phi})
and (\ref{rescat_chi}) give rise to mixing terms:
$g^2\phi_0(t)\chi_0(t) \delta \chi_k$ and
$g^2\phi_0(t)\chi_0(t) \delta \phi_k$, by setting
$\bk=\bk'$.
This leads to an additional instability associated with
chaos other than parametric resonance \cite{PS02}.
For the quadratic inflaton potential this effect is vanishingly small
since the quasi-homogeneous field $\chi$ is strongly suppressed
during inflation. However the signature of chaos can be seen
for the model $V(\phi, \chi)=(1/4)\lambda \phi^4+
(1/2)g^2\phi^2\chi^2$ when the coupling $g^2/\lambda$
is not much larger than unity \cite{PS02,YT04}.

The importance of backreaction and rescattering during preheating
has been explored
in other ways, including the interesting idea that the large variances
may effectively restore broken symmetries \cite{KLS96}.
If the bare mass squared is negative, large variances can make
the effective mass positive, leading to the possibility of
restoring GUT symmetries and dangerous
topological defects when the GUT symmetries are re-broken
once preheating ends \cite{KK97,TKKL98}.

\subsection{Thermalisation}

One of the problems with preheating is that it is an extremely
non-thermal process, as we have discussed. The longest wavelength modes are amplified
preferentially and, in the periodic case, in resonance bands. This
means that the approach to equilibrium is non-perturbative and the
simple estimates for the time it takes to equilibrate and the reheat
temperature we made in Eq.~(\ref{trh}) can be wrong.

\begin{figure}
\includegraphics[scale=0.48]{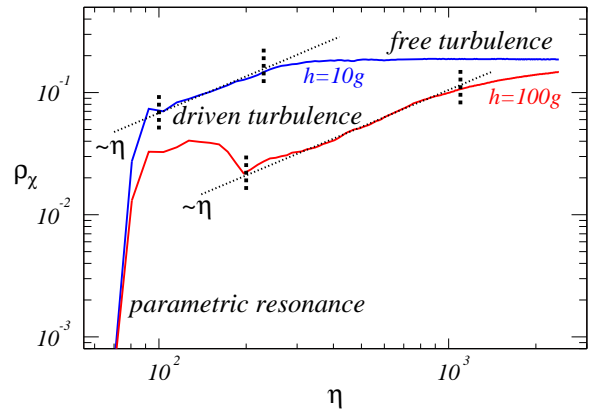}
\caption{
Different phases of evolution of the energy density in
the $\chi$ field with conformal time $\eta$
from the initial parametric resonance through driven turbulence
and free turbulence (characterised by the end
of growth of $\rho_{\chi}$). The model is given
 by an effective potential
$V(\phi, \chi)=\lambda [(1/4)\phi^4+(1/2)g\phi^2\chi^2
+(1/4)h\chi^4]$.
{}From Ref.~\cite{MT04}.
}
\label{rho_evolution}
\end{figure}

Classical numerical simulations show that after the initial resonance
and rescattering phases the system is followed by driven and free
turbulent regimes \cite{MT04} which makes it difficult to estimate
the reheat temperature (see Fig.~\ref{rho_evolution}).
Recent progress using kinetic theory shows
that the evolution of occupation numbers is self-similar.
These methods allow one to estimate the time-scale
for thermalisation, $\tau^{th}$, as \cite{MT04}:
\beq
\tau^{th} \sim \left(\frac{k_f}{k_i}\right)^{1/p}\,.
\eeq
Here $k_{f}$ is the momentum where occupation numbers
drop to of order unity $n_{k_f} \simeq 1$ and $k_i$ is the
initial momentum where energy was injected into the system, corresponding
to the initial parametric resonance
which generically is less than $k_i \sim m_{\phi}$.
The index $p$  determines the rapidity with which the
distribution function moves over momentum space with a numerical value of
$p \sim 1/5$ found from both scaling arguments and numerical simulations \cite{MT04}.
The reheat temperature can then be estimated using conservation of
energy - a long thermalisation timescale implies
a lower reheat temperature and vice versa.

%%%%%%%%%%
%%%%%%%%%%

% v2 DW: my interpretation of Anupam's argument
Recent work \cite{Allahverdi} has suggested that if
number-violating processes are suppressed relative to
number-conserving interactions after preheating then the universe
could enter a ``quasi-thermal" phase which relaxes to a kinetic
equilibrium for some period before number-violating interactions
establish full chemical equilibrium. This could occur if the gauge
bosons mediating the number-violating process acquire a large
mass, suppressing these interactions and further reducing the
temperature when full thermal equilibrium is reached.

\subsection{Interesting applications of preheating}

The nature of preheating lends itself to a number of novel effects which we discuss briefly.

\subsubsection{Non-thermal symmetry restoration and phase transitions}

It was pointed out by Kofman {\em et al.}\cite{KLS96} that the growth
of fluctuations can lead to non-equilibrium
restoration of broken symmetries. Although the notion of an effective potential
is not well-defined far from equilibrium the basic ideas are most easily understood using the concept.
Consider a typical broken symmetry potential
\beq
V(\phi,\chi) = \frac{\lambda}{4}(\phi^2 - v^2)^2 + \frac{g^2}{2}\phi^2\chi^2\,.
\eeq
If $\langle \chi \rangle = 0$ then the only minima correspond to $\chi=0; \phi = \pm v$ and the curvature
of the potential at $\phi = \chi = 0$ is: $V_{\phi\phi} = -\lambda v^2 < 0$.
However, if there is parametric resonance then the effective mass
of $\phi$ at the point $\phi=0$ is:
\beq
m_{\phi,{\rm eff}}^2 = -\lambda v^2 +
3\lambda \langle \delta\phi^2 \rangle + g^2 \langle \chi^2 \rangle\,.
\eeq
As a result, there is the exciting possibility that the growth of fluctuations can make
$m_{\phi,{\rm eff}}^2 > 0$ with the associated possibility of restoring GUT
symmetries. When backreaction ends the resonant
growth and the subsequent expansion of the universe causes the variances to decay
and $m_{\phi,{\rm eff}}^2 \rightarrow -\lambda v^2 < 0$,
breaking the symmetry again.

The danger of this symmetry breaking is the possibility of producing topological
defects that inflation was designed to take care of!
The non-equilibrium nature of the symmetry restoration makes it
difficult to know when defects will be produced and to estimate their density \cite{KK97,TKKL98,RC90}.
In general defect densities during non-equilibrium phase transitions are determined
by the correlation length of the field at the moment when the relaxation timescale of the inflaton
is equal to the time left before the phase transition \cite{zurek1,zurek2},
rather than the Hubble scale as in the usual Kibble mechanism.
Preheating provides a wonderful laboratory for studying non-equilibrium
phase transitions with potentially important implications for inflation model building.

\subsubsection{Amplification of vector fields}

So far we have focused on the amplification of scalar and fermion fields.
It is also possible to resonantly amplify vector fields.
One of the most interesting cases is the amplification of a $U(1)$ gauge field
like electromagnetism with vector potential $A_{\mu}$.
The minimal approach is to couple $A_{\mu}$ to a complex scalar field
$\sigma$ via the covariant derivative: $D_{\mu} \equiv \nabla_{\mu} + ie \sigma A_{\mu}$,
where $e$ is the usual charge. The kinetic term $D_{\mu} \sigma (D^{\mu} \sigma)^*$
leads to an effective mass for the ``photon'' of
\beq
m_A^2 = 2e^2|\sigma|^2\,,
\eeq
which breaks the $U(1)$ symmetry when $\sigma$ condenses.
The Fourier modes, $A_k$, of the spatial part of the vector potential
then obey \cite{Bassett01b,FG01,RC90}
\beq
A_k'' + (k^2 + 2e^2 a^2 |\sigma|^2)A_k
= -\sigma_c a A_k'\,,
\label{maga}
\eeq
where $'$ again denotes derivative with respect to conformal time and
$\sigma_c$ is the electrical conductivity of the medium.
Initially during preheating $\sigma_c \simeq 0$ and choosing a quartic
potential for $\sigma$ implies that this equation becomes formally identical
to Eq.~(\ref{Lame}) and there is exponential growth of $A_k$ fluctuations
within the appropriate resonance bands.

Applying this elegant mechanism to the generation of the observed large scale magnetic fields of order $10^{-6} {\rm G}$ is complicated by several factors - (1) the $U(1)$ symmetry does not exist in this simple form above the electroweak symmetry energy but is unified with the weak force. Hence one needs to study the full theory. (2) The growth of conductivity that must generically occur during preheating provides a strong damping term to Eq.~(\ref{maga}) which means that any predictions are very model dependent and difficult to make. (3) Resonance can only amplify an existing seed field. (4) The coupling to the $\sigma$ field can make the spectrum of $A_k$ very red if $e$ is to large.
Nevertheless, preheating remains a very promising era for generating large-scale magnetic fields, see e.g., \cite{mag1,mag2}.

%*%

%%%%%%%%%%%%%%%%%%%%%%%%%%%%%%%%%%%
\section{The evolution of metric perturbations during reheating}\label{metricpre}
%%%%%%%%%%%%%%%%%%%%%%%%%%%%%%%%%%%

Until now we have neglected metric perturbations in the dynamics of
all fields just as we initially neglected backreaction and
rescattering. The neglect of metric perturbations is technically
incorrect (it violates the Einstein field equations) but is sometimes
a good approximation, sometimes not, in the sense that in some cases
their inclusion can cause fundamental changes to the dynamics of the
fields. A simple example is provided by the ekpyrotic universe.
If we neglect metric perturbations, the perturbation in $\varphi$
corresponding to the separation of two branes exhibits a nearly
scale-invariant spectrum \cite{Khoury01,Kallosh01}.
The inclusion of metric perturbations leads to a blue-tilted spectrum
given by Eq.~(\ref{speekp})
\cite{Wands98,Lyth02ekp1,Lyth02ekp2,Bran01,Hwang02eky,Tsujiekp,Allen04}.

Another example is provided by a single inflaton,
non-minimally coupled to gravity with an interaction
$V_{\rm int}(\phi) = (1/2) \xi R \phi^2$ with vanishing bare mass
of $\phi$. The effective mass of the inflaton is then
$m_{{\rm eff}}^2 = \xi R$.
Hence with the appropriate sign of $\xi$ the effective mass
squared is negative, the field is tachyonic and we should expect
exponential, runaway growth of the fluctuations for all modes
satisfying $k^2/a^2 < |\xi| R$, i.e., the very long-wavelength modes.
In the absence of metric perturbations this is indeed what happens.
The field fluctuations exhibit exponential increase for $k \sim 0$.
Note that beyond a critical wavenumber $k_{{\rm crit}} \sim
\sqrt{|\xi| R a^2}$ there is no negative instability since the effective
mass becomes positive due to the momentum of the modes.

This picture changes completely if we consistently include metric
perturbations however \cite{BT}. The equations of motion for
$\delta\phi_k$ now are coupled to the linear metric perturbations
(e.g. $\Phi$ and $\Psi$ in the longitudinal gauge). Since the field
is non-minimally coupled the anisotropic stress does not vanish and
$\Phi \neq \Psi$.

However, despite the apparent complexity of the resulting equations
they are actually made integrable by the addition of metric
perturbations and one can show analytically that the long-wavelength
solution ($k \to 0$) is \cite{BT}
\bea
\label{delphiana}
\delta \phi=-\frac{\dot{\phi}_{0}(t)}{aF}
\left(c_1-2c_2 \int aF \rd t \right)\,,
\eea
where $\phi_0(t)$ is the homogeneous part of the inflaton,
$F \equiv 1-8\pi \xi \phi_0^2(t)/m_{\rm pl}^2$ and
$c_1$ and $c_2$ are integration constants.
Equation (\ref{delphiana}) shows that the long-wavelength modes
{\em do not grow at all}, in complete
contrast to the naive estimate without including metric perturbations.
This is not too surprising since there is no relative entropy
perturbation for a single field and the intrinsic entropy
perturbation is proportional to $k^2 \Psi_k$
so is negligible on large scales.
As a result ${\cal R}$ or $\zeta$ is conserved in the
$k \to 0$ limit.

\begin{figure}
\includegraphics[scale=0.55]{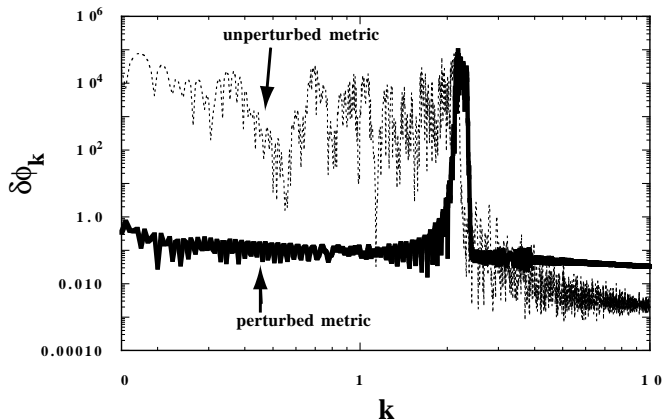}
\caption{The spectrum of field fluctuations for
$\xi=-100$ for both the perturbed and unperturbed metrics
in the case of the quartic inflaton potential
$V(\phi)=(1/4)\lambda \phi^4$ with a non-minimal
coupling $(1/2)\xi R \phi^2$ \cite{BT}.
When metric perturbations are included, the
super-Hubble resonance disappears and is replaced
by a single, sub-Hubble band.
}
\label{spectrum}
\end{figure}

The contrast between the results in the two
cases is illustrated in Fig.~\ref{spectrum} which shows
the spectrum of $\delta\phi_k$ at the end of reheating
for $\xi = -100$ both with and without the inclusion of
metric perturbations. The two spectra differ by five
orders of magnitude at $k \sim 0$.

%%%%%%%%%%
%%%%%%%%%%

This non-minimally coupled inflaton model
clearly illustrates that there are cases
where neglecting metric perturbations gives a wrong
picture of preheating. Our aim here is to discuss when such cases can
be expected and what the implications of including metric
perturbations is in general, both on large scales and small scales
(relevant to black hole formation).

A fundamental question related to preheating is whether it can affect
the evolution of super-Hubble metric perturbations. This is crucial
since inflationary models are tested against the CMB and hence if
reheating affects these predictions it will make model verification
and falsification significantly more complex\footnote{We note that
non-Gaussianities generated in preheating may also be important
to distinguish between different inflationary models.
See Refs.~\cite{Enqvist05,Enqvist05JCAP} for details.}.

%-v2 I added the footnote. by S.T.

\subsection{The criterion for the growth of metric perturbations}

After a significant amount of work
\cite{KH96,NT96,TN98,PE98,Bassett99p1,Bassett99p2,Bassett00p1,Bassett00p2,Finelli99,Finelli00,PE98,EP99},
it has become clear that in order for preheating to affect super-Hubble metric
perturbations there are certain criteria which need to be satisfied
\cite{Tsuji02}.
The most important of these is that there must be an
entropy/isocurvature perturbation mode which is not suppressed on
very large scales \cite{JS00,Ivanov00,Liddle99}
(i.e., that has a power spectrum that it not too blue).
When the effective mass of entropy field perturbation
$\delta s$ is light relative to the Hubble rate
$H$ during inflation, i.e.,
\beq
\mu_s^2 \equiv V_{ss}+3\dot{\theta}^2~\lsim~H^2\,,
\label{sup}
\eeq
$\delta s$ is {\it not} suppressed on
super-Hubble scales during inflation
[See Eq.~(\ref{eq:entropyeom})].
Then during preheating if
$\delta s$ is resonantly amplified due to a time-dependent
effective mass, this can lead to the growth of ${\cal R}$ on large
scales by Eq.~(\ref{dotR}), thereby altering the power spectrum generated during
inflation. In contrast, if the entropy perturbation is
heavy during inflation, ($\mu_{s}^2 \gg H^2 $) then $|\delta s|
\sim a^{-3/2}$ and the growth during preheating means that the
change of ${\cal R}$ is negligible before backreaction ends the
resonance.

\subsubsection{Quadratic potential}

Let us first consider the simple case corresponding to two fields
with effective potential:
\beq
\label{quad}
V(\phi,\chi) = \frac12 m_{\phi}^2\phi^2 +
\frac12 g^2 \phi^2 \chi^2\,.
\eeq
Since $|\tan \theta|=|\dot{\chi}/\dot{\phi}| \ll 1$ during inflation,
one can estimate the effective mass (\ref{sup}), as
\beq
\mu_s^2 \simeq V_{\chi \chi}=g^2\phi^2\,,
\eeq
where we used Eq.~(\ref{Vdd}).
In order for preheating to occur, we require a large resonance
parameter, $q_{i}=g^2\phi_{i}^2/(4m_{\phi}^2) \gg 1$,
at the beginning of reheating,
which translates into the condition $g \gg 10^{-5}$
(where we used $\phi_i=0.2m_{\rm pl}$
and $m_\phi=10^{-6}m_{\rm pl}$).
By using the slow-roll approximation,
$H^2 \sim m_{\phi}^2 \phi^2/m_{\rm pl}^2$, one can show that
the effective mass $\mu_{s}$ is much larger than $H$ during inflation
when preheating occurs:
\beq
\frac{\mu_s^2}{H^2} \sim \left( g\frac{m_{\rm pl}}
{m_{\phi}} \right)^2 \gg 1\,.
\eeq
Hence the long-wavelength modes of $\delta s$ are
exponentially suppressed during inflation and the entropy perturbation has a
highly blue tilted spectrum. To see this note that for
$k \sim 0$ $\delta s \simeq a^{-3/2}$ while for the perturbations
deep inside the Hubble radius ($k \gg aH$),
the modes evolve as in the vacuum state in Minkowski spacetime.
Hence the long-wavelengths suffer
suppression while the short wavelengths do not, leading to a blue
spectrum. Even if $\delta s$ is amplified by a factor $10^5$-$10^6$
during preheating, the suppression of the entropy perturbation in the
preceding  inflationary stage is too strong to give rise to
the variation of curvature perturbations on super-Hubble scales {\em by the time} backreaction
ends the resonance due to the growth of sub-Hubble scale fluctuations.
As a result the existence of the preheating stage does not affect the
CMB power spectrum for the model given by Eq.~(\ref{quad}).

\subsubsection{Quartic potential}

One elegant case in which the entropy field perturbation is not
necessarily suppressed is the conformal model with potential
\beq
\label{quar}
V(\phi,\chi) = \frac14 \lambda \phi^4 +
\frac12 g^2 \phi^2 \chi^2\,.
\eeq
In the linear regime of preheating where $|\chi| \ll |\phi|$
is satisfied, one has $|\theta| \ll 1$ and
$V_{ss} \simeq V_{\chi \chi}=g^2\phi^2$
in Eq.~(\ref{eq:entropyeom}).
Then Eq.~(\ref{eq:entropyeom}) approximately reads
\beq
\label{quarapp}
\frac{\rd^2}{\rd x^2} \delta \tilde{s}+
\left[\kappa^2+\frac{g^2}{\lambda}{\rm cn}^2
\left( x; \frac{1}{\sqrt{2}}\right) \right]
\delta \tilde{s}=0\,,
\eeq
where $\delta \tilde{s} \equiv a \delta s$,
$x$ and $\kappa^2$ are defined in the same way
as in Eq.~(\ref{Lame}).
Here we used the solution (\ref{cn}) and also
neglected the terms which include $\dot{\theta}$.
We find that Eq.~(\ref{quar}) is the same equation as
the one that the perturbation $\delta \chi$ obeys.
Hence we can use the stability-instability chart of Lam\'e
equation for $\delta \tilde{s}$ at the linear regime of preheating.
When the field $\chi$ grows comparable to $\phi$,
the field trajectory becomes curved ($\dot{\theta} \ne 0$)
with $\theta$ of order 1. Then the approximate equation
(\ref{quarapp}) can no longer be used at this stage.

As we already explained in the previous section,
the long wavelength modes of the perturbation $\delta s$
are amplified for the parameter range
$n(2n-1)<g^2/\lambda<n(2n+1)$ with integer $n$.
While $\delta s$ is exponentially suppressed during inflation
for $g^2/\lambda \gg 1$, this suppression does not occur
for $g^2/\lambda \sim 1$ because of the light
effective mass \cite{Bassett00p1,Finelli00}.

Since the super-Hubble modes of $\delta s$ grow exponentially
during preheating, we can expect that large-scale curvature
perturbations may be enhanced for the parameter range around
$g^2/\lambda \sim 2$.

\begin{figure}
\includegraphics[scale=0.55]{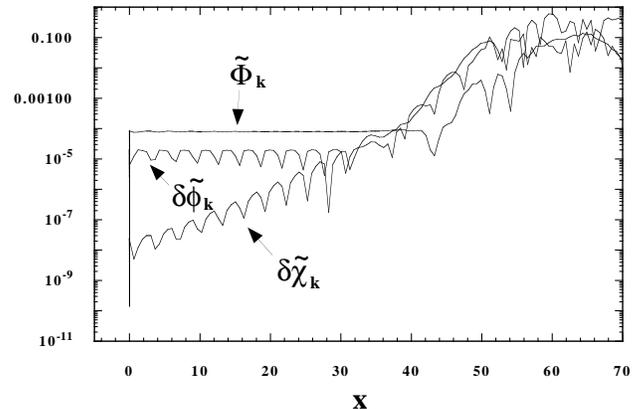}
\caption{Evolution of the gravitational potential
$\tilde{\Phi}_k \equiv k^{3/2} \Phi_k$ together with
the field perturbations $\delta\tilde{\chi}_k
\equiv k^{3/2}{\chi}_k/m_{\rm pl}$, and $\delta\tilde{\phi}_k
\equiv k^{3/2}{\phi}_k/m_{\rm pl}$  on cosmological scales
during inflation and preheating for $g^2/\lambda=2$.
With the use of the Hartree approximation we find that
super-Hubble metric perturbations are amplified
during preheating. }
\label{mpreheating}
\end{figure}

Numerical simulations based upon the Hartree approximation
shows that this actually happens for $g^2/\lambda \sim 2$
around the end of preheating once the entropy field fluctuation
is sufficiently amplified \cite{Zibin00,Tsuji00}\footnote{Although see also \cite{TB03}.},
see Fig.~\ref{mpreheating}.
However large-scale curvature perturbations do not exhibit
parametric amplification for $g^2/\lambda \gtrsim 8$ due to the suppression
during inflation within the
mean-field approximation \cite{Zibin00}.
If we take into account a negative non-minimal coupling $(1/2)\xi R
\chi^2$, curvature perturbations can grow
even for $g^2/\lambda \sim 50$ \cite{Tsuji02}.

Is the amplification of large-scale metric perturbations
consistent with causality?
The resolution of this is that there is no
transfer of energy over super-Hubble scales.
Instead, the {\em pre-existing} entropy perturbation
(which for $g^2/\lambda < {\cal O}(1)$ is light) is
amplified by the resonance. The power of isocurvature perturbations to
alter super-Hubble adiabatic perturbations is well known
\cite{Bassett99p2}.
The large-scale entropy perturbations do drive variation in ${\cal R}$
and there is no violation of causality.

%%%%%%%%%%
%%%%%%%%%%

When the entropy perturbations are sufficiently enhanced, one can
expect that the correlation between adiabatic and isocurvature
should be very strong. This should alter the shape of the CMB power
spectrum as we already have seen in Sec.~\ref{correlation}
(see Fig.~\ref{cospectra}).
It provides a proof that reheating can affect the
predictions of the model.

It is certainly of interest to see whether or not
the parameter region $g^2/\lambda \sim 2$
is compatible with the CMB constraints.
This requires a full numerical study of non-linear perturbation
dynamics which includes the decay of  scalar fields.
Another interesting model which leads to the enhancement of
curvature perturbations is the hybrid (double) inflation model
with a tachyonic instability \cite{Tsuji02}.
However almost all other two-field models
have cosmological predictions that are not violently affected by the
details of preheating.

\subsection{Production of particles and magnetic fields through
metric perturbations}

The growth of metric perturbation during preheating means that on
certain scales the assumption of FRW metric is bad.
Including the metric perturbations one now has
a background with no symmetries and
non-vanishing Weyl tensor, which as a result is not conformally flat.
All the standard results about conformally invariant fields no longer hold.

A simple example is provided by massless fermion fields. In a
flat FRW background there is no production due to the expansion of the
universe since the equation of motion can be brought into Minkowski
form by suitable rescaling. However, in the presence of metric
perturbations we can treat fermions in an external field given by the
metric perturbations \cite{CV90,CV92,Frieman89},
and hence there will be particle production because of this external field.
In the massless case, a direct computation \cite{BT10} shows that in
the massless limit the number of particles produced is:
\beq
N_T = \frac{1}{160\pi a^3}\int
\frac{{\rm d}^4p}{(2\pi)^4}
\theta(p^0)\theta(p^2) |
\tilde{C}^{abcd}(p)|^2\,,
\label{nconf}
\eeq
where $p$ denotes four-momentum, $\theta$ is the Heaviside step
function and $\tilde{C}^{abcd}$ is the Fourier transform
of the Weyl tensor:
\beq
\tilde{C}^{abcd}(p) \equiv \int
{\rm d}^4 x e^{ip_{\mu}x^{\mu}} C^{abcd}\,,
\eeq
which is the trace-free part of the full Riemann tensor. In
conformally flat spacetimes $C^{abcd} = 0$, but when we include
metric perturbations the Weyl tensor does not vanish
but depends linearly on the metric perturbations \cite{BT10}. As a
general rule, the number of particle produced as a result of the
breaking of conformal flatness is therefore small, being quadratic in
the metric perturbations. However, in preheating, with the
enhancement of metric fluctuations in certain wavelengths
the production can be significant. In Ref.~\cite{BT10} it was shown
that this metric perturbation-driven production of fermions
dominates the homogeneous production of fermions
in the standard chaotic inflation model with the quartic potential for
fermion masses below $10^5\,{\rm GeV}$ \cite{BT10}.

A similar discussion can be made for Maxwell fields whose equation of
motion is conformally invariant. The observed large-scale,
large-amplitude magnetic fields are therefore somewhat of a mystery
since they are not amplified by the expansion of the universe.
Allowing for the perturbative breaking of conformal flatness of FRW
by metric perturbations offers a generic and viable mechanism for the
generation of magnetic fields during
inflation \cite{Bassett00p2,Maroto01,Bassett01b,MMNR05}.

\subsection{Primordial black hole formation during preheating}

Although we have shown that typically (at least in two-field
models) large wavelength fluctuations
do not cause a large change in
${\cal R}$ or $\zeta$ except for certain cases, and hence
do not usually affect the
predictions of inflation for the CMB, preheating can lead to the
growth of metric fluctuations on smaller scales, around the Hubble
scale, $k=aH$ which is many orders of magnitude smaller than
CMB-relevant modes at preheating. As a result one might expect that
preheating may lead to copious over-production of primordial black
holes (PBH) \cite{GM00,Bassett01a,FK01,FK02}.

The PBH density can be constrained in a number of ways. When they
evaporate via Hawking radiation they release entropy and high-energy
products that, depending on when they evaporate, can destroy the
predictions of nucleosynthesis or predict $\gamma$-ray flux in excess
of that observed today. PBH can also simply overclose the universe
since their energy density scales as $a^{-3}$ compared with the
radiation scaling law of $a^{-4}$.

However, there are several robust reasons to believe that resonant
preheating does not
significantly alter the abundance of PBH's. First, the peak in the
power spectrum of density perturbations (the relevant quantity for
PBH production), is always at sub-Hubble wavelengths since the
maximum momentum amplified is
\beq
\frac{k_{{\rm max}}}{H} \sim \frac{m_{\rm pl}}
{\phi}q^{1/4} \gg 1\,,
\eeq
where the final inequality follows from requiring strong preheating
$q \gg 1$ and noting that $\phi < m_{\rm pl}$ at the start of preheating
in all known inflationary models. Three dimensional lattice
simulations \cite{PBHkyoto} using a modified version of
LATTICEEASY \cite{latticeeasy} show
that the peak value of the density
perturbation satisfies $\delta_H < 1$, and occurs at scales
significantly smaller than the Hubble scale which is relevant to PBH
production. Since the resulting density power spectrum is typically
$\propto k^3$ on these scales \cite{PBHkyoto} the value of
$\delta_H \sim 10^{-4}$ is typical at the Hubble scale,
implying no excess PBH production.

In the case of tachyonic preheating, long-wavelength modes
are amplified during the tachyonic (spinodal) phase.
In these cases the peak of the spectrum can be at scales
close to the Hubble scale and approach the threshold for
over-production of PBH \cite{PBHkyoto2}. Nevertheless, all
simulations to date have been limited by numerical resolution (in
three dimensions) and have not consistently included metric
perturbations or have not covered all the relevant length scales of
the problem.

%%%%%%%%%%%%%%%%%%%%%%%%%%%%%%%%%%
\section{Curvaton}
\label{curvaton}
%%%%%%%%%%%%%%%%%%%%%%%%%%%%%%%%%%

The original inflation models assumed that the field that drives
inflation is also the field responsible for the origin of structure
in our Universe. This seems an economical approach, but recently
several authors have begun to reconsider whether this is necessarily
so. Might it be possible that perturbations in some field other than
the inflaton could be responsible for the primordial density
perturbation? If so, we need to interpret observational constraints
upon the dynamics of inflation quite differently.

Consider a weakly-coupled, massive scalar field, $\chi$, that decays
some time after inflation has ended. There are many such scalar
degrees of freedom in supersymmetric theories and if they are too
weakly coupled, and their lifetime is too long, this may lead to the
moduli or Polonyi problem. Assuming the field is displaced from the
minimum of its effective potential at the end of inflation, the
field evolves little until the Hubble rate drops below its effective
mass. Then it oscillates, with a time-averaged equation of state for
a pressureless fluid, $P_\chi=0$, (or, equivalently, a collection of
non-relativistic particles). It would eventually come to dominate
the energy density of the Universe, so to avoid disrupting the
successful ``hot big bang'' model of the early universe and, in
particular, to preserve the successful radiation-dominated model of
primordial nucleosynthesis, we require that such fields decay into
radiation before $t\sim 1$~second. For a weakly-coupled field that
decays with only gravitational strength, $\Gamma\sim
m_\chi^3/m_{\rm pl}^2$, this requires $m_\chi>100$~TeV.
Indeed such
late-decaying scalar fields may be no bad thing. Late-entropy
production reduces the minimum duration of inflation required to
produce the total entropy of our observed universe and dilutes other
dangerous relics such as gravitinos, primordial black holes or
monopoles.

But there is a further important feature of late-decaying scalar
fields that has only recently received serious consideration. If the
field is inhomogeneous then it could lead to an inhomogeneous
radiation density after it decays
\cite{Mollerach90,LindeMukhanov96}. This is the basis of the
curvaton scenario \cite{Enqvist02,Lyth02,Moroi01}.

If the curvaton field is light ($m<H$) during inflation then
small-scale quantum fluctuations will lead to a spectrum of
large-scale perturbations, whose initial amplitude at Hubble-exit is
given by Eq.~(\ref{H2pi}). When the Hubble rate drops and the field
begins oscillating after inflation, this leads to a primordial
density perturbation in the $\chi$-field:
\begin{equation}
 \label{defzetachi}
\zeta_\chi = -\psi + \frac{\delta\rho_\chi}{3\rho_\chi} \,,
\end{equation}
where $\rho_\chi=m_\chi^2\chi^2/2$. $\zeta_\chi$ remains constant
for the oscillating curvaton field on large scales, so long as we
can neglect its energy-transfer, i.e., before it decays. Using
Eq.~(\ref{H2pi}) for the field fluctuations at Hubble-exit and
neglecting any non-linear evolution of the $\chi$-field after
inflation (consistent with our assumption that the field is weakly
coupled), we have
\begin{equation}
\label{Pzetachi} {\cal P}_{\zeta_\chi} \simeq \left(
\frac{H}{6\pi\chi} \right)^2_{k=aH} \,.
\end{equation}
The total density perturbation (\ref{zetasum}), considering
radiation, $\gamma$, and the curvaton, $\chi$, is given by
\begin{equation}
\zeta = \frac{4\rho_\gamma\zeta_\gamma+3\rho_\chi\zeta_\chi}
{4\rho_\gamma+3\rho_\chi}\,.
\end{equation}
Thus if the radiation generated by the decay of the inflaton at the
end of inflation is unperturbed (${\cal P}_{\zeta_\gamma}^{1/2}\ll
10^{-5}$) the total curvature perturbation grows as the density of
the $\chi$-field grows relative to the radiation:
$\zeta\sim\Omega_\chi\zeta_\chi$.

Ultimately the $\chi$-field must decay (when $H\sim\Gamma$) and
transfer its energy density and, crucially, its perturbation to the
radiation and/or other matter fields. In the simplest case that the
non-relativistic $\chi$-field decays directly to radiation a full
analysis \cite{Malik02,Gupta} of the coupled evolution equation
gives the primordial radiation perturbation (after the decay)
\begin{equation}
 \label{zetagamma}
\zeta_\gamma = r(p) \zeta_\chi \,,
\end{equation}
where $p\equiv [\Omega_\chi/(\Gamma/H)^{1/2}]_{\rm initial}$ is a
dimensionless parameter which determines the maximum value of
$\Omega_\chi$ before it decays, and empirically we find \cite{Gupta}
\begin{equation}
 \label{defrp}
r(p) \simeq 1- \left( 1+ \frac{0.924}{1.24}p \right)^{-1.24}  \,.
\end{equation}
For $p\gg1$ the $\chi$-field dominates the total energy density
before it decays and $r\sim1$, while for $p\ll1$ we have
$r\sim0.924p\ll1$.

Finally combining Eqs.~(\ref{Pzetachi}) and (\ref{zetagamma}) we
have
\begin{equation}
 \label{Pzetagamma}
{\cal P}_{\zeta_\gamma} \simeq r^{\,2}(p) \left( \frac{H}{6\pi\chi}
 \right)^2_{k=aH} \,.
\end{equation}
In contrast to the inflaton scenario the final density perturbation
in the curvaton scenario is a very much dependent upon the physics
after the field perturbation was generated during inflation. For
instance, if the curvaton lifetime is too short then it will decay
before it can significantly perturb the total energy density and
${\cal P}_{\zeta_\gamma}^{1/2}\ll 10^{-5}$.
The observational constraint on
the amplitude of the primordial perturbations gives a single
constraint upon both the initial fluctuations during inflation and
the post-inflationary decay time. This is in contrast to the
inflaton scenario where the primordial perturbation gives a direct
window onto the dynamics of inflation, independently of the physics
at lower energies. In the curvaton scenario there is the possibility
of connecting the generation of primordial perturbations to other
aspects of cosmological physics. For instance, it may be possible to
identify the curvaton with fields whose late-decay is responsible
for the origin of the baryon asymmetry in the universe, in
particular with sneutrino models of leptogenesis (in which an
initial lepton asymmetry is converted into a baryon asymmetry at the
electroweak transition) \cite{sneutrino}.

The curvaton scenario has re-invigorated attempts to embed models of
inflation in the very early universe within minimal supersymmetric
models of particle physics constrained by experiment
\cite{Enqvist03cur,mssm,EKM03,Postma02,Postma03,Hamaguchi03,Dimo03,McDonald03}.
It may be possible that
the inflaton field driving inflation can be completely decoupled
from visible matter if the dominant radiation in the universe today
comes from the curvaton decay rather than reheating at the end of
inflation. Indeed the universe need not be radiation-dominated at
all until the curvaton decays if instead the inflaton fast-rolls at
the end of inflation.

The curvaton offers a new range of theoretical possibilities, but
ultimately we will require observational and/or experimental
predictions to decide whether the curvaton or inflaton generated the
primordial perturbation.

% added 20/4/05

\subsection{Non-Gaussianity}

The best way to distinguish between different scenarios for the
origin of structure could be the statistical properties of the
primordial density perturbation.
All the inflationary models we have discussed start with small-scale
vacuum fluctuations of an effectively free scalar field, described
by a Gaussian random field, with vanishing three-point function.
Deviations from Gaussianity in the curvaton scenario can be
parameterised by a dimensionless parameter $f_{nl}$ \cite{Komatsu03}
defined in Eq.~(\ref{deffnl1}).
In terms of the initial curvaton density perturbation
(\ref{defzetachi}) on spatially-flat hypersurfaces we have, from
Eq.~(\ref{zetagamma}),
\begin{equation}
\zeta_\gamma = \frac{r}{3} \left( \frac{\delta\rho_\chi}{\rho_\chi}
 \right)_{\psi=0} \,.
\end{equation}
When the curvaton field begins oscillating about a quadratic minimum
of its potential we have $\rho_\chi=m_\chi^2\chi^2/2$, and thus, in
terms of the Newtonian potential on large scales in the
matter-dominated era, $\Phi=-3\zeta/5$, this gives
\begin{equation}
\Phi = - \frac{r}{5} \left( \frac{2\chi\delta\chi +
    \delta\chi^2}{\chi^2} \right)
\,.
\end{equation}
Identifying $\Phi_{\rm Gauss}=-(2r/5)\delta\chi/\chi$ and
substituting into (\ref{deffnl1}) we obtain \cite{Lyth03}
\begin{equation}
 \label{nongau}
f_{nl} = - \frac{5}{4r} \,.
\end{equation}
In other words, the smaller the fraction of the radiation density
due to curvaton decay, the larger the non-Gaussianity of the
primordial density perturbation.
Eq.~(\ref{nongau}) corrects a sign error in the expression
for $f_{NL}$ given in Ref.~\cite{Lyth03}.
We note that $f_{nl}$ is subject to a modification
when gravitational second-order corrections are taken into
account, but it reproduces Eq.~(\ref{nongau}) in the
limit $r \ll 1$ \cite{Bartolo04cur}.

Current bounds from the WMAP satellite require $-58<f_{nl}<134$ at
the 95\% confidence limit \cite{Komatsu03}, and hence require
$r>0.021$ but future experiments such as Planck could detect
$f_{nl}$ as small as around $5$.

By contrast in the inflaton scenario the inflaton field fluctuations
at horizon-crossing determine the large-scale curvature perturbation
$\zeta$ which will remain constant on super-Hubble scales. One can
estimate the amplitude of the three-point correlation function by
noting that the local amplitude of fluctuations will depend on local
variations in the Hubble rate. This gives a robust estimate of the
primordial non-Gaussianity in the inflaton scenario
\cite{Maldacena02}
\begin{equation}
f_{nl} \sim \frac{n_s-1}{4} \,,
\end{equation}
where $n_s-1$ is the scale-dependence of the primordial power
spectrum. Note that this estimate relies on the adiabaticity of the
perturbations in the inflaton scenario which ensures that there
exists a non-linearly conserved density perturbation on large
scales, from Hubble-exit during inflation until last-scattering of
the CMB photons. Any detection of primordial non-Gaussianity
$f_{nl}>1$ would therefore rule out this inflaton scenario.

\subsection{Residual isocurvature perturbations}

In multi-field scenarios such as the curvaton scenario the initial
perturbation is supposed to be a non-adiabatic perturbation and
hence can in principle leave behind a residual non-adiabatic
component.  In the curvaton scenario, perturbations in just one
field, the curvaton, would be responsible for both the total
primordial density perturbation and any isocurvature mode and hence
there is the clear prediction that the two should be completely
correlated, corresponding to $\cos\Delta=\pm1$ in
Eq.~(\ref{defDelta}), or $A_r/A_s\to0$ in Eq.~(\ref{defArAs}) and
$n_2=n_3=n_c$.

Using $\zeta_i$ for different matter components it is easy to see
how the curvaton could leave residual isocurvature perturbations
after the curvaton decays. If any fluid has decoupled before the
curvaton contributes significantly to the total energy density that
fluid remains unperturbed with $\zeta_i\simeq0$, whereas after the
curvaton decays the photons perturbation is given by
(\ref{Pzetagamma}). Thus a residual isocurvature perturbation
(\ref{defSi}) is left
\begin{equation}
S_i = -3 \zeta_\gamma \,,
\end{equation}
which remains constant for decoupled perfect fluids on large scales.

The observational bound on isocurvature matter perturbations
completely correlated with the photon perturbation, is
\cite{Gordon02}
\begin{equation}
-0.53 <\frac{S_B+(\rho_{\rm c}/\rho_B)S_{\rm c}} {\zeta_\gamma} <
0.43\,.
\end{equation}
In particular if the baryon asymmetry is generated while the total
density perturbation is still negligible then the residual baryon
isocurvature perturbation, $S_B=-3\zeta_\gamma$ would be much larger
than the observational bound and such models are thus ruled out. The
observational bound on CDM isocurvature perturbations are stronger
by the factor $\rho_{\rm c}/\rho_B$ although CDM is usually assumed
to decouple relatively late.

An interesting amplitude of residual isocurvature perturbations
might be realised if the decay of the curvaton itself is the
non-equilibrium event that generates the baryon asymmetry. In this
case the net baryon number density directly inherits the
perturbation $\zeta_B=\zeta_\chi$ while the photon perturbation
$\zeta_\gamma\leq\zeta_\chi$ may be diluted by pre-existing
radiation and is given by Eq.~(\ref{zetagamma}). Note that so long
as the net baryon number is locally conserved it defines a conserved
perturbation on large scales, even though it may still be
interacting with other fluids and fields. Hence the primordial
baryon isocurvature perturbation (\ref{defSi}) in this case is given
by
\begin{equation}
S_B = 3 (1-r) \zeta_\chi = \frac{3(1-r)}{r} \zeta_\gamma \,.
\end{equation}

But there is no lower bound on the predicted amplitude of residual
non-adiabatic modes and the non-detection of primordial isocurvature
density perturbations cannot be used to rule out all alternative
scenarios. For instance if after the curvaton decays at sufficiently
high temperature and all the particles produced relax to an thermal
equilibrium abundance, characterised by a common temperature and
vanishing chemical potential then no residual isocurvature
perturbations survive. In this case there is a unique attractor
trajectory in phase-space and only adiabatic perturbations (along
this trajectory) survive on large scales.

%%%%%%%%%%%%%%%%%%%%%%%%%%%%%%%%%%%%%%%%%%%%%%%%%%%%%%%%%%%%%%%%%%%%%%
\section{Modulated reheating}
%%%%%%%%%%%%%%%%%%%%%%%%%%%%%%%%%%%%%%%%%%%%%%%%%%%%%%%%%%%%%%%%%%%%%%

The curvaton scenario is one possible way in which a light scalar
field during inflation can influence the primordial density
perturbation on large scales after inflation. An alternative
possibility \cite{Dvali03_1,Kofman03} is that the primordial
perturbation could be generated by the spatial variation of the
inflaton decay rate, $\Gamma$, at the end of inflation.

The decay rate of the inflaton $\phi$ in the ``old'' reheating
scenario is given by $\Gamma \sim \lambda^2 m$, where $m$ is the
inflaton mass and $\lambda$ a dimensionless coupling of the inflaton
to other light fields. When $\Gamma$ is less than the Hubble rate at
the end of inflation, the reheating temperature $T_{{\rm rh}}$
is estimated as
\begin{equation}
T_{{\rm rh}} \sim \sqrt{\Gamma m_{\rm pl}} \sim
\lambda \sqrt{mm_{\rm pl}}\,.
\end{equation}
If the local coupling strength $\lambda$ is dependent on the local
value of another scalar field, $\chi$, this can give rise to
fluctuations in the reheating temperature:
\begin{equation}
\frac{\delta T_{{\rm rh}}}
{T_{{\rm rh}}} \sim \frac{\delta \Gamma}{\Gamma} \sim
\frac{\delta \lambda}{\lambda}\,.
\end{equation}
Thus density perturbations after inflation are sourced by local
fluctuations in $\chi$.
If $\chi$ is light during inflation then it can acquire an almost
scale-invariant spectrum of perturbations at Hubble exit, which are
then imprinted on the radiation field during reheating.
It is natural to consider perturbed couplings, since the coupling
``constants'' of the low-energy effective action in string theory
are generally functions of the vacuum expectation values of light
moduli fields \cite{Kofman03}. In what follows we refer to this
scenario as ``modulated reheating''.

Let us estimate the curvature perturbation in the radiation fluid
generated after inflation in the modulated reheating scenario. In
doing so we recall that the curvature perturbation $\zeta$, defined
in Eq.~(\ref{defzeta}), can be interpreted as the dimensionless
density perturbation on spatially flat hypersurfaces, which are
separated by a uniform expansion. In a region with a larger local
decay rate, $\Gamma+\delta\Gamma$, the local energy density on
spatially flat hypersurfaces differs with respect to the average by
an amount $\delta\rho=H\rho\delta t$ due to the earlier change from
matter to radiation equation of state. Since the average decay time
is given by $t=\Gamma^{-1}$, the perturbation of the local decay
rate corresponds to the perturbed decay time $t+\delta t$ with
$\delta t=-\delta \Gamma/\Gamma^2$. Then by using
Eq.~(\ref{eq:defzetai}), the curvature perturbation in radiation
fluid after the decay of inflaton is found to be
\begin{equation}
    \label{zetaap}
\zeta_\gamma = \frac14 \left( \frac{\delta\rho}
{\rho}\right)_{\psi=0}
 = - \frac14 \frac{H\delta\Gamma}{\Gamma^2}
 = - \frac16 \frac{\delta\Gamma}{\Gamma}\,,
\end{equation}
where in the last equality we used $t=(2/3)H^{-1}=\Gamma^{-1}$ at
the decay time.

The validity of this estimate, which assumes a sudden decay of the
inflaton, can be investigated numerically using the coupled
evolution equations for perturbations with interacting
fluids~\cite{Malik04}. Here we treat the inflaton is treated as a
pressureless fluid decaying to a radiation fluid with a perturbed
decay rate~\cite{Mazumdar03_2,Mata03,Ver03}.
The background energy
density of the inflaton field $\rho_\phi$ and radiation
$\rho_\gamma$ satisfy
\begin{eqnarray}
\label{drhop}
\dot\rho_{\phi}
&=&-3H\rho_{\phi} +Q_{\phi}\,, \\
\label{drhogam}
\dot\rho_{\gamma} &=&
-4H\rho_{\gamma}+Q_{\gamma}\,.
\end{eqnarray}
Here the energy transfer from inflaton to radiation is characterized
by $Q_{\phi}=-\Gamma \rho_{\phi}$ and $Q_\gamma=\Gamma \rho_{\phi}$.
Introducing dimensionless quantities
\begin{eqnarray}
\Omega_{\phi} \equiv \frac{\rho_{\phi}}
{\rho_{\phi}+\rho_{\gamma}}\,,~~ \Omega_{\gamma} \equiv
\frac{\rho_{\gamma}} {\rho_{\phi}+\rho_{\gamma}}\,,~~ g \equiv
\frac{\Gamma}{H}\,,
\end{eqnarray}
the background equations (\ref{drhop}) and (\ref{drhogam}) may be
rewritten as
\begin{eqnarray}
\label{backOme1} & & \frac{\rd \Omega_{\phi}}{\rd N}=
\Omega_{\phi}(1-\Omega_{\phi}-g)\,, \\
\label{backOme2} & & \frac{\rd g}{\rd N}=
\frac{g}{2}(4-\Omega_{\phi}) +\frac{g}{\Gamma}\frac{\rd \Gamma}{\rd
N}\,,
\end{eqnarray}
with $N$ is the number of e-foldings and
$\Omega_{\phi}+\Omega_{\gamma}=1$.
We assume that the scalar field whose local expectation value leads
to the spatial variation of the decay rate $\Gamma$, makes a
negligible contribution to the energy density.

The perturbed energy transfer is given by $\delta Q_{\phi}=-\Gamma
\delta \rho_{\phi}- \delta \Gamma \rho_{\phi}$ and $\delta
Q_{\gamma}=\Gamma \delta \rho_{\phi}+ \delta \Gamma \rho_{\phi}$.
The perturbation equations on wavelengths larger than the Hubble
radius are \cite{Malik02,Mata03}
\begin{eqnarray}
& &\dot{\delta \rho}_i+3H(\delta \rho_i+\delta P_i)
-3(\rho_i+P_i) \dot{\psi} \nonumber \\
& &=Q_iA+\delta Q_i\,,
\end{eqnarray}
where $i=\phi, \gamma$ for inflaton and radiation.

The gauge-invariant curvature perturbation for each component
($i=\phi, \gamma$) are defined by Eq.~(\ref{eq:defzetai}). Then the
total curvature perturbation, $\zeta$, is given by
\begin{eqnarray}
\label{zeta} \zeta &=& \frac{\dot{\rho}_{\phi}}{\dot{\rho}}
\zeta_{\phi}+\frac{\dot{\rho}_{\gamma}}
{\dot{\rho}}\zeta_{\gamma}\,, \\
&=& f\zeta_\phi+(1-f)\zeta_{\gamma}\,,
\end{eqnarray}
where
\begin{eqnarray}
 f \equiv
\frac{(3+g)\Omega_{\phi}}{(4-\Omega_\phi)}\,.
\end{eqnarray}
Note that this corrects Eq.~(2.38) of Ref.~\cite{Mata03}.

We define a relative entropy perturbation (\ref{defSi}) as
\begin{eqnarray}
{\cal S}_{\phi \gamma} \equiv 3(\zeta_{\phi}-\zeta_{\gamma})\,,
\end{eqnarray}
which leads to a non-adiabatic pressure perturbation
(\ref{defPnad}). Then we find that the evolution equations for
$\zeta_{\phi}$ and $\zeta_{\gamma}$ are \cite{Mata03}
\begin{eqnarray}
\label{dotzetacurv} \dot\zeta_\phi &=& -\frac{\Gamma}{6}
\frac{\rho_\phi}{\rho} \frac{\dot{\rho}_{\gamma}}{\dot{\rho}_{\phi}}
{\cal S}_{\phi\gamma}+
H\frac{\rho_\phi}{\dot\rho_{\phi}}\delta\Gamma_\phi^{\rm GI} \,, \\
\label{dotzetarad} \dot\zeta_\gamma &=& \frac{\Gamma}{3}
\frac{\dot{\rho}_{\phi}}{\dot{\rho}_{\gamma}}
\left(1-\frac{\rho_\phi}{2\rho}\right) {\cal S}_{\phi\gamma} -
H\frac{\rho_\phi}{\dot\rho_{\gamma}} \delta\Gamma_\gamma^{\rm GI}
\,,
\end{eqnarray}
where the gauge-invariant perturbation $\delta\Gamma_i^{\rm GI}$ is
defined by
\begin{eqnarray}
\delta\Gamma_i^{\rm GI}=\delta\Gamma-\dot\Gamma
\frac{\delta\rho_i}{\dot\rho_i}\,.
\end{eqnarray}
This describes a non-adiabatic energy transfer \cite{Malik04} which
can source the entropy perturbation $\S_{\phi\gamma}$. We include
any time variation of the background decay rate $\dot{\Gamma}$ in
order to construct the gauge-invariant energy transfer.

Eqs.~(\ref{dotzetacurv}) and (\ref{dotzetarad}) show that the
presence of the entropy perturbation ${\cal S}_{\phi \gamma}$ and
the gauge-invariant perturbation $\delta\Gamma_i^{\rm GI}$ leads to
the variation of $\zeta_{\phi}$ and $\zeta_{\gamma}$. In what
follows we shall consider a situation in which the time variation of
$\Gamma$ is neglected. Then by using Eq.~(\ref{zeta}) together with
Eqs.~(\ref{dotzetacurv}) and  (\ref{dotzetarad}), we find
\cite{Mazumdar03_2}
\begin{eqnarray}
\label{dotzeta} \frac{\rd \zeta}{\rd N} &=&
-\frac{(3+f)\Omega_{\phi}}{4-\Omega_{\phi}}
(\zeta-\zeta_{\phi})\,, \\
\label{dotphi} \frac{\rd \zeta_{\phi}}{\rd N} &=&
\frac{(4-\Omega_{\phi})f}{2(3+f)}(\zeta-\zeta_{\phi}) -\frac{f}{3+f}
\frac{\delta \Gamma}{\Gamma}\,.
\end{eqnarray}

\begin{figure}
\includegraphics[scale=0.45]{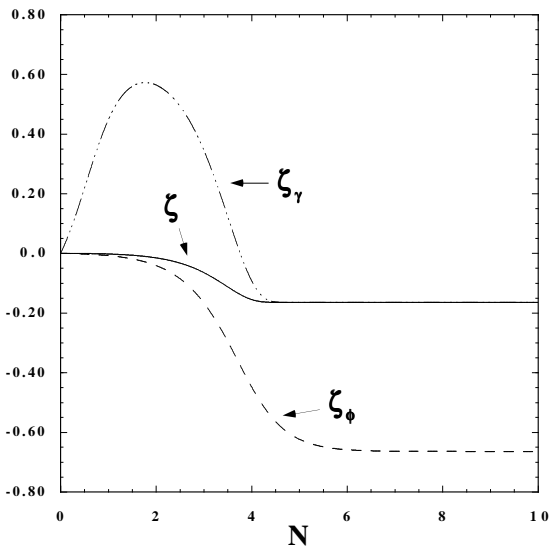}
\caption{ The evolution of $\zeta$, $\zeta_{\phi}$ and
$\zeta_{\gamma}$ (normalized by $\delta \Gamma/\Gamma$)
in the modulated reheating scenario
with initial conditions $\Omega_{\phi}(0)=0.99$,
$\Omega_{\gamma}(0)=0.01$, $g=0.01$,
$\zeta(0)=0.0$ and $\zeta_{\phi}(0)=0.0$. }
 \label{zetamodu}
\end{figure}

We numerically solve the perturbation equations (\ref{dotzeta}) and
(\ref{dotphi}) together with the background equations
(\ref{backOme1}) and (\ref{backOme2}), see Fig.~\ref{zetamodu}.
Under the initial conditions $\zeta(0)=0.0$, $\zeta_{\phi}(0)=0.0$,
$\Omega_{\phi}(0)=0.99$ and $g(0)=0.01$, we find that both $\zeta$
and $\zeta_{\gamma}$ approach the analytic value given by
Eq.~(\ref{zetaap}) at late times. As long as $|\zeta_{\phi}(0)|
\lesssim 10^{-3}\delta \Gamma/\Gamma$, the evolution of $\zeta$ is
similar to the one shown in Fig.~\ref{zetamodu}. When
$|\zeta_{\phi}(0)| \gtrsim 10^{-3}\delta \Gamma/\Gamma$, the final
value of $\zeta$ exhibits some deviation from the analytic value
(\ref{zetaap}) \cite{Mazumdar03_2}. For example one has
$\zeta=-0.118\delta \Gamma/\Gamma$ for $\zeta_{\phi}(0)=0.05\delta
\Gamma/\Gamma$.

Thus the final curvature perturbation is generated by the perturbed
coupling in addition to any initial inflaton fluctuation. In
Ref.~\cite{Tsuji03per} the spectra of primordial perturbations were
evaluated numerically for the system in which perturbations of both
the inflaton and the decay rate coexist. It was shown that even in
low energy-scale inflation a nearly scale-invariant spectra, with an
overall amplitude set by observations, can be obtained through the
conversion of $\chi$ fluctuations into adiabatic density
perturbations.
A model for fluctuating inflaton coupling was proposed in
Ref.~\cite{Mazumda03_1} using (s)neutrinos as a source
for adiabatic perturbations.
In Ref.~\cite{Ackerman04} the generation of
density perturbations in preheating was studied for the model
in which the coupling $g$ between inflaton and decay products
is perturbed by another scalar field.
In Ref.~\cite{BKU} modulated fluctuations  from hybrid inflation
was studied in the case where $\lambda$ and $g$ in
Eq.~(\ref{hybridpo})
depend upon a light scalar field.
It was shown in Ref.~\cite{BGS05} that if heavy particles are
in thermal equilibrium until they become relativistic,
perturbations in the annihilation cross section of this particle
receive additional sources of fluctuations.

Finally we note that an alternative mechanism combining features of
both modulated reheating and the curvaton scenario could arise if
after inflation and homogeneous reheating the universe becomes
dominated by an oscillating (curvaton-type) field. Primordial
perturbations are produced if the decay-rate and/or mass of this
curvaton-type field varies \cite{Dvali03_2}, even if the curvaton
itself has no perturbations, due to spatial variations in the VEV of
another light field, $\phi$. Vernizzi \cite{Ver03} studied the
coupled perturbation equations in this three fluid case finding the
resulting primordial perturbation is given by
\begin{equation}
\zeta = r(p) \left[ - \frac43 \frac{\delta M}{M} - \frac16
\frac{\delta\Gamma}{\Gamma} \right]  \,,
\end{equation}
where $r(p)$ for the curvaton-type field is given in
Eq.~(\ref{defrp}).

The non-Gaussianity of primordial perturbations in the modulated
reheating scenarios was studied in Refs.~\cite{Ver03,Za03}.
Unlike the curvaton scenario there is an additional parameter
$\alpha_\Gamma$ in the model-dependent transfer coefficient
\begin{equation}
\zeta = r \alpha_\Gamma \zeta_\chi \,.
\end{equation}
A simple estimation of the parameter $f_{nl}$ defined in
Eq.~(\ref{deffnl2}) gives \cite{Ver03}
\begin{eqnarray}
f_{nl}=-\frac{5}{6r \alpha_{\Gamma}} \, .
\end{eqnarray}
The curvaton scenario corresponds to $\alpha_{\Gamma}=-2/3$, see
Eq.~(\ref{nongau}). When the field $\phi$ completely dominates the
universe before it decays, one has $r=1$ and $\alpha_{\Gamma}=1/6$
in the modulated reheating scenario, thus giving $f_{nl}=-5$.
Hence this scenario has a possibility to confront with the future
observations of Planck satellite whose sensitivity should reach to
the level $|f_{nl}| \sim 5$.

%%%%%%%%%%%%%%%%%%%%
%%%%%%%%%%%%%%%%%%%

%%%%%%%%%%%%%%%%%%%%%%%%%%%%%%%%%%%%%%%%%%%%%%%%%%%%%%%%%%%%%%%%%%%%%
\section{Conclusions}
%%%%%%%%%%%%%%%%%%%%%%%%%%%%%%%%%%%%%%%%%%%%%%%%%%%%%%%%%%%%%%%%%%%%

The building of realistic inflationary models will inevitably imply a
phase in which multiple fields are studied. Whether we are close to
building ``realistic" models of inflation is not clear. At present
single field models are still the best fit to the data precisely because
of their simplicity and lack of free parameters.

As with the search for dark energy dynamics, one of the major
challenges for observational cosmology is the hunt for a signature of
dynamics. At present all observations are consistent with an exactly
scale-invariant Harrison-Zel'dovich primordial spectrum and a universe
which today is dominated by a pure cosmological constant. We await
conclusive evidence that the Universe is, or ever was dominated by one
or more light scalar fields. If such evidence arrives, it will require
model-building in ernest.

In science there is a natural resonance between data and theory
% v2* typo - in
which drives the complexity of models to roughly match the amount
and quality of data available to test them. Until now there has
been little motivation - beyond exploring the range of what is
possible - for considering models of inflation with many fields
which are light compared with the Hubble constant. Entering the
era of data-driven cosmology this situation is changing rapidly
and although single field models are still arguably the best fits
to current CMB and large scale structure data, their dominance
must be challenged by making accurate and detailed predictions for
models with more than one light field.

Beyond this practical motivation to study multi-field models there is a
deep conceptual reason: in single field models the cosmological
predictions of inflation are robust and depend very little on the
dynamics of reheating. When there are more than one light field this
is no longer the case and the full details of reheating and
post-inflation dynamics must be considered when comparing
the model to the data.
On the one hand this opens up the exciting possibility that
we may be able to probe details of particle physics beyond the standard
model
with cosmology while on the other hand it introduces much larger
parameter spaces and removes some of the elegant
model-independence of single-field inflation.

What experimental progress can we expect in the next decade or
two? The Planck satellite should fix
% v2* typo
$\Omega_m h^2$
to high precision which, combined with polarisation
measurements will help significantly in constraining multi-field
inflationary models. Upcoming large galaxy surveys such as those
with WFMOS/KAOS, LSST, DES and PANSTARRS will help pin down the
matter power spectrum with exquisite accuracy, culminating in the
Square Kilometre Array which will find redshifts of around one
billion galaxies, giving the ultimate measurement of the power
spectrum at $z < 1$.

By that time we should have an excellent understanding of galaxy and matter
biasing and will be able to study the time-evolution of the power spectrum.
Leveraging the CMB and the matter power spectrum (together with
traditional distance measurements via type Ia supernovae and baryon acoustic
oscillations)
should allow us to distinguish between the effects of dark energy
(late time evolution) and the effects associated with multi-field
inflation. Hopefully we will find that the cosmos exhibits some of the
interesting possibilities offered by the dynamics of multiple field
inflation.

%%%%%%%%%%%%%%%%%%%%%%%%%%%%%%%%%
\begin{acknowledgments}
  We are grateful to Laura Allen, Luca Amendola, Nicola Bartolo,
  Mariam Bouhmadi Lopez, Robert Brandenberger, Helen A. Bridgman,
  Gianluca Calcagni, Edmund Copeland, Fabio Finelli,
  Juan Garcia-Bellido, Mohammad R.~Garousi,
  Chris Gordon, Burin Gumjudpai, Sujata Gupta,
  Imogen Heard, Yoshida Jin, David I. Kaiser, Sugumi Kanno,
  Kazuya Koyama, Hideaki Kudoh, David Langlois, Sam Leach,
  Andrew Liddle, Andrei Linde, David Lyth,
  Roy Maartens, Kei-ichi Maeda, Karim Malik, Sabino Matarrese,
  Shuntaro Mizuno, David Parkinson, Marco Peloso,
  Yun-Song Piao, Giuseppe Pollifrone, Toni Riotto,
  M.~Sami, Misao Sasaki, Parampreet Singh,  Jiro Soda,
  Lorenzo Sorbo, Alexei Starobinsky, Teruaki Suyama,
  Fabrizio Tamburini, Takahiro Tanaka, Takashi Torii,
  Carlo Ungarelli, Fermin Viniegra, Hiroki Yajima, Jun'ichi Yokoyama
  and Xin-min Zhang for fruitful collaborations about inflation and
  cosmological perturbations over recent years.
We also thank Anupam Mazumdar, Edmund Copeland, Eiichiro Komatsu,
David Lyth and T.\,Padmanabhan for useful correspondence and
for insightful comments on the draft.
We are grateful to Sergei Khlebnikov and Igor Tkachev
for permission to include figures from their papers.
The work of B.\,B. and D.\,W. was supported in part by PPARC grant
PPA/G/S/2000/00115.
S.\,T. was supported by JSPS (Grant No.\,30318802).
\end{acknowledgments}
%%%%%%%%%%%%%%%%%%%%%%%%%%%%%%%%%


\begin{thebibliography}{9999}

\bibitem{abbott82}
L.~F.~Abbott, E.~Fahri and M.~Wise
Phys.\ Lett.\ B {\bf 117}, 29 (1982).

\bibitem{AS03}
L.~R.~Abramo and L.~J.~Sodre,
%``Can the Local Supercluster explain
%de low CMB multipoles?,''
arXiv:astro-ph/0312124.

\bibitem{Ackerman04}
L.~Ackerman, C.~W.~Bauer, M.~L.~Graesser and M.~B.~Wise,
%``Light scalars and the generation of density perturbations
%during preheating or inflaton decay,''
Phys.\ Lett.\ B {\bf 611}, 53 (2005).

\bibitem{Acqu03}
V.~Acquaviva, N.~Bartolo, S.~Matarrese and A.~Riotto,
%``Second-order cosmological perturbations from inflation,''
Nucl.\ Phys.\ B {\bf 667}, 119 (2003).

\bibitem{ARS97}
J.~A.~Adams, G.~G.~Ross and S.~Sarkar,
%``Multiple inflation,''
Nucl.\ Phys.\ B {\bf 503}, 405 (1997).

\bibitem{Akaike74}
H.~Akaike, IEEE Trans. Auto. Control, {\bf 19}, 716 (1974).

\bibitem{Albre82}
A.~Albrecht and P.~Steinhardt,
Phys.~Rev.~Lett. {\bf 48}, 1220 (1982).

\bibitem{Allahverdi}
R.~Allahverdi and A.~Mazumdar,
%``Quasi-thermal universe: From cosmology to colliders,''
arXiv:hep-ph/0505050.

\bibitem{Allen04}
L.~E.~Allen and D.~Wands,
%``Cosmological perturbations through a simple bounce,''
Phys.\ Rev.\ D {\bf 70}, 063515 (2004).

\bibitem{Amendola02}
L.~Amendola, C.~Gordon, D.~Wands and M.~Sasaki,
Phys.\ Rev.\ Lett.\  {\bf 88},
211302 (2002).

\bibitem{Arkani03}
N.~Arkani-Hamed, P.~Creminelli, S.~Mukohyama and
M.~Zaldarriaga,
%``Ghost inflation,''
JCAP {\bf 0404}, 001 (2004).

\bibitem{Picon99}
C.~Armendariz-Picon, T.~Damour and V.~Mukhanov,
%``k-inflation,''
Phys.\ Lett.\ B {\bf 458}, 209 (1999).

\bibitem{Baacke97}
J.~Baacke, K.~Heitmann and C.~Patzold,
%``Renormalization of nonequilibrium dynamics
%in FRW cosmology,''
Phys.\ Rev.\ D {\bf 56}, 6556 (1997).

\bibitem{Baacke98}
J.~Baacke, K.~Heitmann and C.~Patzold,
%``Nonequilibrium dynamics of fermions in a spatially
%homogeneous background field,''
Phys.\ Rev.\ D {\bf 58}, 125013 (1998).

\bibitem{Bardeen80}
J.~M.~Bardeen,
%``Gauge Invariant Cosmological Perturbations,''
Phys.\ Rev.\ D {\bf 22}, 1882 (1980).

\bibitem{Bardeen88}
J.~M.~Bardeen,
{\it Lectures given at 2nd Guo Shou-jing Summer School on Particle
Physics and Cosmology, Nanjing, China, Jul 1988.}

\bibitem{Bardeen83}
J.~M.~Bardeen, P.~J.~Steinhardt and M.~S.~Turner,
Phys.\ Rev.\ D {\bf 28}, 679 (1983).

\bibitem{Barger03}
V.~Barger, H.~S.~Lee, and D.~Marfatia,
%``WMAP and inflation,''
Phys.\ Lett.\ B{\bf 565}, 33 (2003).

\bibitem{Bartolo01p1}
N.~Bartolo, S.~Matarrese and A.~Riotto,
%``Oscillations during inflation and
%cosmological density perturbations,''
Phys.\ Rev.\ D {\bf 64}, 083514 (2001).

\bibitem{Bartolo01p2}
N.~Bartolo, S.~Matarrese and A.~Riotto,
%``Adiabatic and isocurvature perturbations
%from inflation: Power spectra  and
%consistency relations,''
Phys.\ Rev.\ D {\bf 64}, 123504 (2001).

\bibitem{Bartolo02}
N.~Bartolo, S.~Matarrese and A.~Riotto,
%``Non-Gaussianity from inflation,''
Phys.\ Rev.\ D {\bf 65}, 103505 (2002).

\bibitem{Bartolo04cur}
N.~Bartolo, S.~Matarrese and A.~Riotto,
%``On nonGaussianity in the curvaton scenario,''
Phys.\ Rev.\ D {\bf 69}, 043503 (2004).

\bibitem{BKMR04}
N.~Bartolo, E.~Komatsu, S.~Matarrese and A.~Riotto,
%``Non-Gaussianity from inflation:
% Theory and observations,''
Phys.\ Rept.\  {\bf 402}, 103 (2004).

\bibitem{BL97}
B.~A.~Bassett and S.~Liberati,
%``Geometric reheating after inflation,''
Phys.\ Rev.\ D {\bf 58}, 021302 (1998)
[Erratum-ibid.\ D {\bf 60}, 049902 (1999)].

\bibitem{BB98}
B.~A.~Bassett,
Phys.\ Rev.\ D {\bf 58}, 021303 (1998).

\bibitem{BT98}
B.~A.~Bassett and F.~Tamburini,
%``Inflationary reheating in grand
%unified theories,''
Phys.\ Rev.\ Lett.\  {\bf 81}, 2630 (1998).

\bibitem{Bassett99p1}
B.~A.~Bassett, D.~I.~Kaiser and R.~Maartens,
%``General relativistic preheating after inflation,''
Phys.\ Lett.\ B {\bf 455}, 84 (1999).

\bibitem{Bassett99p2}
B.~A.~Bassett, F.~Tamburini, D.~I.~Kaiser and R.~Maartens,
%``Metric preheating and limitations of
%linearized gravity. II,''
Nucl.\ Phys.\ B {\bf 561}, 188 (1999).

\bibitem{Bassett00p1}
B.~A.~Bassett and F.~Viniegra,
%``Massless metric preheating,''
Phys.\ Rev.\ D {\bf 62}, 043507 (2000).

\bibitem{Bassett00p2}
B.~A.~Bassett, C.~Gordon, R.~Maartens and D.~I.~Kaiser,
%``Restoring the sting to metric preheating,''
Phys.\ Rev.\ D {\bf 61}, 061302 (2000).

\bibitem{Bassett01a}
B.~A.~Bassett and S.~Tsujikawa,
%``Inflationary preheating and primordial black holes,''
Phys.\ Rev.\ D {\bf 63}, 123503 (2001).

\bibitem{Bassett01b}
B.~A.~Bassett, G.~Pollifrone, S.~Tsujikawa and F.~Viniegra,
%``Preheating as cosmic magnetic dynamo,''
Phys.\ Rev.\ D {\bf 63}, 103515 (2001).

\bibitem{BT10}
B.~A.~Bassett, M.~Peloso, L.~Sorbo and S.~Tsujikawa,
%``Fermion production from preheating-amplified
%metric perturbations,''
Nucl.\ Phys.\ B {\bf 622}, 393 (2002).

\bibitem{BFM03}
M.~Bastero-Gil, K.~Freese and L.~Mersini-Houghton,
Phys.\ Rev.\ D {\bf 68}, 123514 (2003).

\bibitem{BG05}
T.~J.~Battefeld and G.~Geshnizjani,
%``Perturbations in a regular bouncing universe,''
arXiv:hep-th/0503160.

% v2 - added citation - response to Bauer

\bibitem{BGS05}
C.~W.~Bauer, M.~L.~Graesser and M.~P.~Salem,
%``Fluctuating annihilation cross sections
%and the generation of density perturbations,''
Phys.\ Rev.\ D {\bf 72}, 023512 (2005).

\bibitem{Beltran04}
M.~Beltran, J.~Garcia-Bellido, J.~Lesgourgues and A.~Riazuelo,
%``Bounds on CDM and neutrino isocurvature
%perturbations from CMB and LSS data,''
Phys.\ Rev.\ D {\bf 70}, 103530 (2004).

\bibitem{Beltran05}
M.~Beltran, J.~Garcia-Bellido, J.~Lesgourgues,
A.~R.~Liddle and A.~Slosar,
%``Bayesian model selection and isocurvature perturbations,''
Phys.\ Rev.\ D {\bf 71}, 063532 (2005).

\bibitem{BB89}
M.~C.~Bento and O.~Bertolami,
%``String Generated Gravity Models
%With Cubic Curvature Terms,''
Phys.\ Lett.\ B {\bf 228}, 348 (1989).

\bibitem{warm}
A.~Berera,
%``Interpolating the stage of exponential expansion
%in the early universe:
%A possible alternative with no reheating,''
Phys.\ Rev.\ D {\bf 55}, 3346 (1997).

\bibitem{warm2}
A.~Berera and T.~W.~Kephart,
%``Ubiquitous inflaton in string-inspired models,''
Phys.\ Rev.\ Lett.\  {\bf 83}, 1084 (1999).

\bibitem{Bernar02}
F.~Bernardeau and J.~P.~Uzan,
%``Non-Gaussianity in multi-field inflation,''
Phys.\ Rev.\ D {\bf 66}, 103506 (2002).

\bibitem{BKU}
F.~Bernardeau, L.~Kofman and J.~P.~Uzan,
%``Modulated fluctuations from hybrid inflation,''
Phys.\ Rev.\ D {\bf 70}, 083004 (2004).

\bibitem{BDL}
P.~Binetruy, C.~Deffayet and D.~Langlois,
%``Non-conventional cosmology from a brane-universe,''
Nucl.\ Phys.\ B {\bf 565}, 269 (2000).

\bibitem{Blanco04}
J.~J.~Blanco-Pillado {\it et al.},
%``Racetrack inflation,''
JHEP {\bf 0411}, 063 (2004).

\bibitem{Boy96}
D.~Boyanovsky, H.~J.~de Vega, R.~Holman and J.~F.~J.~Salgado,
%``Analytic and numerical study of
%preheating dynamics,''
Phys.\ Rev.\ D {\bf 54}, 7570 (1996).

\bibitem{Boy97}
D.~Boyanovsky, D.~Cormier, H.~J.~de Vega,
R.~Holman, A.~Singh and M.~Srednicki,
%``Scalar field dynamics in Friedman
%Robertson Walker spacetimes,''
Phys.\ Rev.\ D {\bf 56}, 1939 (1997).

\bibitem{Bozza02}
V.~Bozza, M.~Gasperini, M.~Giovannini and G.~Veneziano,
%``Constraints on pre-big bang parameter
%space from CMBR anisotropies,''
Phys.\ Rev.\ D {\bf 67}, 063514 (2003).

\bibitem{Bozza05}
V.~Bozza and G.~Veneziano,
%``Regular two-component bouncing cosmologies and perturbations therein,''
arXiv:gr-qc/0506040.

\bibitem{Robert85}
R.~H.~Brandenberger,
%``Quantum Field Theory Methods And
% Inflationary Universe Models,''
Rev.\ Mod.\ Phys.\  {\bf 57}, 1 (1985).

\bibitem{Bran01}
R.~Brandenberger and F.~Finelli,
%``On the spectrum of fluctuations in an effective
%field theory of the ekpyrotic universe,''
JHEP {\bf 0111}, 056 (2001).

\bibitem{BVD}
P.~Brax, C.~van de Bruck and A.~C.~Davis,
%``Brane world cosmology,''
Rept.\ Prog.\ Phys.\  {\bf 67}, 2183 (2004).

\bibitem{BMW}
H.~A.~Bridgman, K.~A.~Malik and D.~Wands,
%``Cosmological perturbations in the bulk and on the brane,''
Phys.\ Rev.\ D {\bf 65}, 043502 (2002).

\bibitem{Bru94}
R.~Brustein, M.~Gasperini, M.~Giovannini,
V.~F.~Mukhanov and G.~Veneziano,
%``Metric perturbations in dilaton driven inflation,''
Phys.\ Rev.\ D {\bf 51}, 6744 (1995).

\bibitem{Bru97}
R.~Brustein and R.~Madden,
%``A model of graceful exit in string cosmology,''
Phys.\ Rev.\ D {\bf 57}, 712 (1998).

\bibitem{Bucher99}
M.~Bucher, K.~Moodley and N.~Turok,
%``The general primordial cosmic perturbation,''
Phys.\ Rev.\ D {\bf 62}, 083508 (2000).

\bibitem{Bunn96}
E.~F.~Bunn, A.~R.~Liddle and M.~J.~White,
%``Four-year COBE normalization of inflationary cosmologies,''
Phys.\ Rev.\ D {\bf 54}, 5917 (1996).

\bibitem{Bur04}
C.~P.~Burgess, J.~M.~Cline, H.~Stoica and F.~Quevedo,
%``Inflation in realistic D-brane models,''
JHEP {\bf 0409}, 033 (2004).

\bibitem{Cal03}
G.~Calcagni,
%``Consistency equations in Randall-Sundrum cosmology:
%A test for braneworld inflation,''
JCAP {\bf 0311}, 009 (2003).

\bibitem{Cal03v2}
G.~Calcagni,
%``Degeneracy of consistency equations
% in Randall-Sundrum inflation,''
JCAP {\bf 0406}, 002 (2004).

\bibitem{Calshinji}
G.~Calcagni and S.~Tsujikawa,
%``Observational constraints on patch inflation
% in noncommutative spacetime,''
Phys.\ Rev.\ D {\bf 70}, 103514 (2004).

\bibitem{CTS05}
G.~Calcagni, S.~Tsujikawa and M.~Sami,
%``Dark energy and cosmological solutions
%in second-order string gravity,''
arXiv:hep-th/0505193.

\bibitem{mag2}
E.~A.~Calzetta and A.~Kandus,
%``Self consistent estimates of magnetic
%  fields from reheating,''
Phys.\ Rev.\ D {\bf 65}, 063004 (2002).

\bibitem{CV92}
A.~Campos and E.~Verdaguer,
Phys. Rev. D {\bf 45}, 4428 (1992).

\bibitem{Cartier99}
C.~Cartier, E.~J.~Copeland and R.~Madden,
%``The graceful exit in string cosmology,''
JHEP {\bf 0001}, 035 (2000).

\bibitem{Cartier01}
C.~Cartier, J.~c.~Hwang and E.~J.~Copeland,
%``Evolution of cosmological perturbations
%in non-singular string cosmologies,''
Phys.\ Rev.\ D {\bf 64}, 103504 (2001).

\bibitem{Cartier03}
C.~Cartier, R.~Durrer and E.~J.~Copeland,
%``Cosmological perturbations and the transition
%from contraction to expansion,''
Phys.\ Rev.\ D {\bf 67}, 103517 (2003).

\bibitem{Cartier04}
C.~Cartier,
%``Scalar perturbations in an alpha'-regularised
%cosmological bounce,''
arXiv:hep-th/0401036.

\bibitem{CV90}
J.~Cespedes and E.~Verdaguer,
Phys. Rev. D {\bf 41}, 1022 (1990).

\bibitem{CNM}
  T.~Charters, A.~Nunes and J.~P.~Mimoso,
  %``Phase dynamics and particle production in preheating,''
  Phys.\ Rev.\ D {\bf 71}, 083515 (2005)
  [arXiv:hep-ph/0502053].
  %%CITATION = HEP-PH 0502053;%%

\bibitem{Chiba97}
T.~Chiba, N.~Sugiyama and J.~Yokoyama,
%``Imprints of the metrically coupled
%dilaton on density perturbations  in
%inflationary cosmology,''
Nucl.\ Phys.\ B {\bf 530}, 304 (1998).

\bibitem{mcmc}
N.~Christensen and R.~Meyer,
arXiv:astro-ph/0006401.

\bibitem{mcmc2}
N.~Christensen, R.~Meyer, L.~Knox and B.~Luey,
Class.\ Quant.\ Grav.\  {\bf 18}, 2677 (2001).

\bibitem{Contaldi03}
C.~R.~Contaldi, M.~Peloso, L.~Kofman and A.~Linde,
%``Suppressing the lower Multipoles in the CMB Anisotropies,''
JCAP {\bf 0307}, 002 (2003).

\bibitem{CKLLreview}
E.~J.~Copeland, E.~W.~Kolb, A.~R.~Liddle and J.~E.~Lidsey,
%``Reconstructing the inflation potential,
%in principle and in practice,''
Phys.\ Rev.\ D {\bf 48}, 2529 (1993).

\bibitem{Cope94}
E.~J.~Copeland, A.~R.~Liddle, D.~H.~Lyth, E.~D.~Stewart
and D.~Wands,
%``False vacuum inflation with Einstein gravity,''
Phys.\ Rev.\ D {\bf 49}, 6410 (1994).

\bibitem{Copeland77}
E.~J.~Copeland, R.~Easther and D.~Wands,
%``Vacuum fluctuations in axion-dilaton cosmologies,''
Phys.\ Rev.\ D {\bf 56}, 874 (1997).

\bibitem{Cope01}
E.~J.~Copeland, A.~R.~Liddle and J.~E.~Lidsey,
%``Steep inflation: Ending braneworld inflation
%by gravitational particle production,''
Phys.\ Rev.\ D {\bf 64}, 023509 (2001).

\bibitem{Cormier99}
D.~Cormier and R.~Holman,
%``Spinodal inflation,''
Phys.\ Rev.\ D {\bf 60}, 041301 (1999).

\bibitem{Cormier00}
D.~Cormier and R.~Holman,
%``Spinodal decomposition and inflation:
%Dynamics and metric  perturbations,''
Phys.\ Rev.\ D {\bf 62}, 023520 (2000).

\bibitem{CNZ05}
P.~Creminelli, A.~Nicolis and M.~Zaldarriaga,
%``Perturbations in bouncing cosmologies:
%Dynamical attractor vs scale invariance,''
Phys.\ Rev.\ D {\bf 71}, 063505 (2005).

\bibitem{Crotty03}
P.~Crotty, J.~Garcia-Bellido, J.~Lesgourgues and A.~Riazuelo,
%``Bounds on isocurvature perturbations
% from CMB and LSS data,''
Phys.\ Rev.\ Lett.\  {\bf 91}, 171301 (2003).

\bibitem{Der95}
N.~Deruelle and V.~F.~Mukhanov,
%``On matching conditions for cosmological perturbations,''
Phys.\ Rev.\ D {\bf 52}, 5549 (1995).

\bibitem{DiMarco03}
F.~Di Marco, F.~Finelli and R.~Brandenberger,
%``Adiabatic and Isocurvature Perturbations for
%Multifield Generalized Einstein Models,''
Phys.\ Rev.\ D {\bf 67}, 063512 (2003).

\bibitem{mag1}
K.~Dimopoulos, T.~Prokopec, O.~Tornkvist and A.~C.~Davis,
%``Natural magnetogenesis from inflation,''
Phys.\ Rev.\ D {\bf 65}, 063505 (2002).

\bibitem{Dimo03}
K.~Dimopoulos, D.~H.~Lyth, A.~Notari and A.~Riotto,
%``The curvaton as a Pseudo-Nambu-Goldstone boson,''
JHEP {\bf 0307}, 053 (2003).

\bibitem{DL82}
A.~D.~Dolgov and A.~D.~Linde,
Phys.\ Lett.\ B {\bf 116}, 329 (1982).

\bibitem{Dolgov82}
A.~D.~Dolgov and D.~P.~Kirilova,
Sov.~J.~Nucl.~Phys., {\bf 51}, 172 (1990).

\bibitem{Dolgov02}
A.~D.~Dolgov, S.~H.~Hansen, S.~Pastor,
S.~T.~Petcov, G.~G.~Raffelt and D.~V.~Semikoz,
%``Cosmological bounds on neutrino degeneracy
%improved by flavor oscillations,''
Nucl.\ Phys.\ B {\bf 632}, 363 (2002).

\bibitem{cmbeasy}
M.~Doran,
%``CMBEASY
arXiv:astro-ph/0302138.

\bibitem{Doran}
M.~Doran, C.~M.~Muller, G.~Schafer and C.~Wetterich,
%``Gauge-invariant initial conditions and early time
%perturbations in quintessence universes,''
Phys.\ Rev.\ D {\bf 68}, 063505 (2003).

\bibitem{cmbeasy2}
M.~Doran and C.~M.~Mueller,
JCAP {\bf 0409}, 003 (2004).

\bibitem{DLMS04}
J.~F.~Dufaux, J.~E.~Lidsey, R.~Maartens and M.~Sami,
%``Cosmological perturbations from brane
%inflation with a Gauss-Bonnet term,''
Phys.\ Rev.\ D {\bf 70}, 083525 (2004).

\bibitem{Durrer02}
R.~Durrer and F.~Vernizzi,
%``Adiabatic perturbations in pre big bang models:
%Matching conditions and scale invariance,''
Phys.\ Rev.\ D {\bf 66}, 083503 (2002).

\bibitem{Dvali99}
G.~R.~Dvali and S.~H.~H.~Tye,
%``Brane inflation,''
Phys.\ Lett.\ B {\bf 450}, 72 (1999).

\bibitem{Dvali03_1}
G.~Dvali, A.~Gruzinov and M.~Zaldarriaga,
%``A new mechanism for generating
%density perturbations from inflation,''
Phys.\ Rev.\ D {\bf 69}, 023505 (2004).

\bibitem{Dvali03_2}
G.~Dvali, A.~Gruzinov and M.~Zaldarriaga,
%``Cosmological perturbations from
%inhomogeneous reheating, freezeout, and
%mass domination,''
Phys.\ Rev.\ D {\bf 69}, 083505 (2004).

\bibitem{EP99}
R.~Easther and M.~Parry,
%``Gravity, parametric resonance and chaotic inflation,''
Phys.\ Rev.\ D {\bf 62}, 103503 (2000).

\bibitem{Efs03}
G.~Efstathiou,
%``The Statistical Significance of
%the Low CMB Multipoles,''
Mon.\ Not.\ Roy.\ Astron.\ Soc.\  {\bf 346}, L26 (2003).

\bibitem{ellis84}
J.~R.~Ellis, J.~S.~Hagelin, D.~V.~Nanopoulos,
K~A.~Olive and M.~Srednicki,
Nucl.~Phys.~B{\bf 238} 453 (1984).

\bibitem{ellis99}
J.~R.~Ellis, N.~Kaloper, K.~A.~Olive and J.~Yokoyama,
%``Topological R**4 inflation,''
Phys.\ Rev.\ D {\bf 59}, 103503 (1999).

\bibitem{Enqvist02}
K.~Enqvist and M.~S.~Sloth,
%``Adiabatic CMB perturbations in pre big
%bang string cosmology,''
Nucl.\ Phys.\ B {\bf 626}, 395 (2002).

\bibitem{mssm}
K.~Enqvist and A.~Mazumdar,
%``Cosmological consequences of MSSM flat directions,''
Phys.\ Rept.\  {\bf 380}, 99 (2003).

\bibitem{EKM03}
K.~Enqvist, S.~Kasuya and A.~Mazumdar,
%``Adiabatic density perturbations and matter
% generation from the MSSM,''
Phys.\ Rev.\ Lett.\  {\bf 90}, 091302 (2003).

\bibitem{Enqvist03cur}
K.~Enqvist, A.~Jokinen, S.~Kasuya and A.~Mazumdar,
%``MSSM flat direction as a curvaton,''
Phys.\ Rev.\ D {\bf 68}, 103507 (2003).

% v2 - added citation - response to Anupam

\bibitem{Enqvist05}
K.~Enqvist, A.~Jokinen, A.~Mazumdar,
T.~Multamaki and A.~Vaihkonen,
%``Non-Gaussianity from Preheating,''
Phys.\ Rev.\ Lett.\  {\bf 94}, 161301 (2005).

\bibitem{Enqvist05JCAP}
K.~Enqvist, A.~Jokinen, A.~Mazumdar,
T.~Multamaki and A.~Vaihkonen,
%``Non-gaussianity from instant and tachyonic preheating,''
JCAP {\bf 0503}, 010 (2005).

% v2 - added citation - response to Anupam

\bibitem{FT02}
M.~Fairbairn and M.~H.~G.~Tytgat,
%``Inflation from a tachyon fluid?,''
Phys.\ Lett.\ B {\bf 546}, 1 (2002).

\bibitem{Fe02}
A.~Feinstein,
%``Power-law inflation from the rolling tachyon,''
Phys.\ Rev.\ D {\bf 66}, 063511 (2002).

\bibitem{instant}
G.~N.~Felder, L.~Kofman and A.~D.~Linde,
%``Instant preheating,''
Phys.\ Rev.\ D {\bf 59}, 123523 (1999).

\bibitem{Felder01p1}
G.~N.~Felder, J.~Garcia-Bellido, P.~B.~Greene,
L.~Kofman, A.~D.~Linde and I.~Tkachev,
%``Dynamics of symmetry breaking and tachyonic preheating,''
Phys.\ Rev.\ Lett.\  {\bf 87}, 011601 (2001).

\bibitem{latticeeasy}
G.~N.~Felder and I.~Tkachev,
%``LATTICEEASY: A program for lattice simulations of
%scalar fields in an expanding universe,''
arXiv:hep-ph/0011159.

\bibitem{Felder01p2}
G.~N.~Felder, L.~Kofman and A.~D.~Linde,
%``Tachyonic instability and dynamics
%of spontaneous symmetry breaking,''
Phys.\ Rev.\ D {\bf 64}, 123517 (2001).

\bibitem{Feng03}
B.~Feng and X.~Zhang,
%``Double inflation and the low CMB quadrupole,''
Phys.\ Lett.\ B {\bf 570}, 145 (2003).

\bibitem{FRV04}
F.~Ferrer, S.~Rasanen and J.~Valiviita,
%``Correlated isocurvature perturbations from
%mixed inflaton-curvaton decay,''
JCAP {\bf 0410}, 010 (2004).

\bibitem{Finelli99}
F.~Finelli and R.~H.~Brandenberger,
%``Parametric amplification of gravitational
%fluctuations during  reheating,''
Phys.\ Rev.\ Lett.\  {\bf 82}, 1362 (1999).

\bibitem{Finelli00}
F.~Finelli and R.~H.~Brandenberger,
%``Parametric amplification of metric fluctuations
%during reheating in two field models,''
Phys.\ Rev.\ D {\bf 62}, 083502 (2000).

\bibitem{FG01}
F.~Finelli and A.~Gruppuso,
%``Resonant amplification of gauge
% fields in expanding universe,''
Phys.\ Lett.\ B {\bf 502}, 216 (2001).

\bibitem{FK01}
F.~Finelli and S.~Khlebnikov,
%``Large metric perturbations from rescattering,''
Phys.\ Lett.\ B {\bf 504}, 309 (2001).

\bibitem{FK02}
F.~Finelli and S.~Khlebnikov,
%``Metric perturbations at reheating:
%The use of spherical symmetry,''
Phys.\ Rev.\ D {\bf 65}, 043505 (2002).

\bibitem{Finelli02a}
F.~Finelli and R.~Brandenberger,
%``On the generation of a scale-invariant spectrum of adiabatic
%fluctuations in cosmological models with a contracting phase,''
Phys.\ Rev.\ D {\bf 65}, 103522 (2002).

\bibitem{Foffa99}
S.~Foffa, M.~Maggiore and R.~Sturani,
%``Loop corrections and graceful exit in string cosmology,''
Nucl.\ Phys.\ B {\bf 552}, 395 (1999).

\bibitem{Freese90}
K.~Freese, J.~A.~Frieman and A.~V.~Olinto,
%``Natural Inflation With Pseudo - Nambu-Goldstone Bosons,''
Phys.\ Rev.\ Lett.\  {\bf 65}, 3233 (1990).

\bibitem{chain}
K.~Freese and D.~Spolyar,
  %``Chain inflation: 'Bubble bubble toil and trouble',''
  arXiv:hep-ph/0412145.

\bibitem{Frieman89}
J.~A.~Frieman, Phys.\,Rev.\,D {\bf 39}, 389 (1989).

\bibitem{F96}
H.~Fujisaki, K.~Kumekawa, M.~Yamaguchi and  M.~Yoshimura,
%``Particle production and dissipative cosmic field,''
Phys.\ Rev.\ D {\bf 53}, 6805 (1996).

\bibitem{Garcia95}
J.~Garcia-Bellido and D.~Wands,
%``Constraints from inflation on scalar - tensor gravity theories,''
Phys.\ Rev.\ D {\bf 52}, 6739 (1995).

\bibitem{Garcia96}
J.~Garcia-Bellido and D.~Wands,
%``Metric perturbations in two-field inflation,''
Phys.\ Rev.\ D {\bf 53}, 5437 (1996).

\bibitem{Garcia96hybrid}
J.~Garcia-Bellido, A.~D.~Linde and D.~Wands,
%``Density perturbations and black hole formation
%in hybrid inflation,''
Phys.\ Rev.\ D {\bf 54}, 6040 (1996).

\bibitem{Garousi04}
M.~R.~Garousi, M.~Sami and S.~Tsujikawa,
Phys.\ Rev.\ D {\bf 70}, 043536 (2004).

\bibitem{Gas93}
M.~Gasperini and G.~Veneziano,
%``Pre - big bang in string cosmology,''
Astropart.\ Phys.\  {\bf 1} (1993) 317.

\bibitem{Gas96}
M.~Gasperini, M.~Maggiore and G.~Veneziano,
%``Towards a non-singular pre-big bang cosmology,''
Nucl.\ Phys.\ B {\bf 494}, 315 (1997).

\bibitem{Gasperini03}
M.~Gasperini, M.~Giovannini and G.~Veneziano,
%``Perturbations in a non-singular bouncing universe,''
Phys.\ Lett.\ B {\bf 569}, 113 (2003).

\bibitem{Gas02}
M.~Gasperini and G.~Veneziano,
%``The pre-big bang scenario in string cosmology,''
Phys.\ Rept.\  {\bf 373}, 1 (2003).

\bibitem{Gasperini04}
M.~Gasperini, M.~Giovannini and G.~Veneziano,
%``Cosmological perturbations across a curvature bounce,''
Nucl.\ Phys.\ B {\bf 694}, 206 (2004).

\bibitem{Gio05}
M.~Giovannini,
%``Theoretical tools for the physics
%of CMB anisotropies,''
Int.\ J.\ Mod.\ Phys.\ D {\bf 14}, 363 (2005).

\bibitem{Giudice99}
G.~F.~Giudice, M.~Peloso, A.~Riotto and I.~Tkachev,
%``Production of massive fermions at preheating and leptogenesis,''
JHEP {\bf 9908}, 014 (1999).

\bibitem{Giudice99v2}
G.~F.~Giudice, A.~Riotto and I.~Tkachev,
%``Thermal and non-thermal production
%of gravitinos in the early universe,''
JHEP {\bf 9911}, 036 (1999).

\bibitem{Giudice00}
G.~F.~Giudice, E.~W.~Kolb and A.~Riotto,
%``Largest temperature of the radiation era
%and its cosmological implications,''
Phys.\ Rev.\ D {\bf 64}, 023508 (2001).

\bibitem{Gordon00}
C.~Gordon, D.~Wands, B.~A.~Bassett and R.~Maartens,
%``Adiabatic and entropy perturbations from inflation,''
Phys.\ Rev.\ D {\bf 63}, 023506 (2001).

\bibitem{Gordon02}
C.~Gordon and A.~Lewis,
%``Observational constraints on the curvaton model of inflation,''
Phys.\ Rev.\ D {\bf 67}, 123513 (2003).

\bibitem{GordonHu}
C.~Gordon and W.~Hu,
%``A Low CMB Quadrupole from Dark Energy
%Isocurvature Perturbations,''
Phys.\ Rev.\ D {\bf 70}, 083003 (2004).

\bibitem{stocres}
I.~Goychuk and P.~Hanggi, Phys. Rev. Lett. {\bf 91}, 070601 (2003).

\bibitem{GM00}
A.~M.~Green and K.~A.~Malik,
%``Primordial black hole production due to preheating,''
Phys.\ Rev.\ D {\bf 64}, 021301 (2001).

\bibitem{negci}
B.~R.~Greene, T.~Prokopec and T.~G.~Roos,
% Inflaton Decay and Heavy Particle Production
%with Negative Coupling
Phys.\ Rev.\ D {\bf 56}, 6484 (1997).

\bibitem{GKLS97}
P.~B.~Greene, L.~Kofman, A.~D.~Linde and A.~A.~Starobinsky,
%``Structure of resonance in preheating after inflation,''
Phys.\ Rev.\ D {\bf 56}, 6175 (1997).

\bibitem{GK98}
P.~B.~Greene and L.~Kofman,
%``Preheating of fermions,''
Phys.\ Lett.\ B {\bf 448}, 6 (1999).

\bibitem{GKM03}
P.~B.~Greene, K.~Kadota and H.~Murayama,
%``Supergravity inflation free from harmful relics,''
Phys.\ Rev.\ D {\bf 68}, 043502 (2003).

\bibitem{Groot01}
S.~Groot Nibbelink and B.~J.~W.~van Tent,
%``Scalar perturbations during multiple
% field slow-roll inflation,''
Class.\ Quant.\ Grav.\  {\bf 19}, 613 (2002).

\bibitem{Gupta}
S.~Gupta, K.~A.~Malik and D.~Wands,
%``Curvature and isocurvature perturbations
% in a three-fluid model of  curvaton decay,''
Phys.\ Rev.\ D {\bf 69}, 063513 (2004).

\bibitem{Guth81}
A.~H.~Guth,
%``The Inflationary Universe: A Possible Solution
%To The Horizon And Flatness Problems,''
Phys.\ Rev.\ D {\bf 23}, 347 (1981).

\bibitem{Guth82}
A.~H.~Guth and S.~Y.~Pi,
%``Fluctuations In The New Inflationary Universe,''
Phys.\ Rev.\ Lett.\  {\bf 49} (1982) 1110.

\bibitem{sneutrino}
K.~Hamaguchi, H.~Murayama and T.~Yanagida,
%``Leptogenesis from sneutrino-dominated early universe,''
Phys.\ Rev.\ D {\bf 65}, 043512 (2002).

\bibitem{Hamaguchi03}
K.~Hamaguchi, M.~Kawasaki, T.~Moroi and F.~Takahashi,
%``Curvatons in supersymmetric models,''
Phys.\ Rev.\ D {\bf 69}, 063504 (2004).

\bibitem{HY05}
T.~Hattori and K.~Yamamoto,
%``Non-Gaussianity in Multi-field Stochastic
%Inflation with the Scaling Approximation,''
arXiv:astro-ph/0506373.

\bibitem{Hawking82}
S.~W.~Hawking,
Phys.\ Lett.\ B {\bf 115}, 295 (1982).

\bibitem{HHR01}
S.~W.~Hawking, T.~Hertog and H.~S.~Reall,
%``Trace anomaly driven inflation,''
Phys.\ Rev.\ D {\bf 63}, 083504 (2001).

\bibitem{HW02}
I.~P.~C.~Heard and D.~Wands,
%``Cosmology with positive and
%negative exponential potentials,''
Class.\ Quant.\ Grav.\  {\bf 19}, 5435 (2002).

\bibitem{Huey01}
G.~Huey and J.~E.~Lidsey,
%``Inflation, braneworlds and quintessence,''
Phys.\ Lett.\ B{\bf 514}, 217 (2001).

\bibitem{subir}
P.~Hunt and S.~Sarkar,
  %``Multiple inflation and the WMAP 'glitches',''
  Phys.\ Rev.\ D {\bf 70}, 103518 (2004)

\bibitem{Hwang96}
J.~c.~Hwang and H.~Noh,
Phys.\ Rev.\ D {\bf 54}, 1460 (1996).

\bibitem{Hwang00}
J.~c.~Hwang and H.~Noh,
%``Cosmological perturbations with multiple scalar fields,''
Phys.\ Lett.\ B {\bf 495}, 277 (2000).

\bibitem{Hwang02eky}
J.~c.~Hwang,
%``Cosmological structure problem in the ekpyrotic scenario,''
Phys.\ Rev.\ D {\bf 65}, 063514 (2002).

\bibitem{Hwang02eky2}
J.~Hwang and H.~Noh,
%``Identification of perturbation modes and
%controversies in ekpyrotic perturbations,''
Phys.\ Lett.\ B {\bf 545}, 207 (2002).

\bibitem{Ivanov00}
P.~Ivanov,
Phys.\ Rev.\ D {\bf 61}, 023505 (2000).

\bibitem{JS00}
K.~Jedamzik and G.~Sigl,
%``On metric preheating,''
Phys.\ Rev.\ D {\bf 61}, 023519 (2000).

\bibitem{YT04}
Y.~Jin and S.~Tsujikawa,
%``Chaotic dynamics in preheating after inflation,''
arXiv:hep-ph/0411164.

\bibitem{Kachru03}
S.~Kachru, R.~Kallosh, A.~Linde, J.~Maldacena,
L.~McAllister and S.~P.~Trivedi,
%``Towards inflation in string theory,''
JCAP {\bf 0310}, 013 (2003).

\bibitem{Kadota03}
K.~Kadota and E.~D.~Stewart,
%``Successful modular cosmology,''
JHEP {\bf 0307}, 013 (2003).

\bibitem{Kadota03v2}
K.~Kadota and E.~D.~Stewart,
%``Inflation on moduli space and
%cosmic perturbations,''
JHEP {\bf 0312}, 008 (2003).

\bibitem{Kaiser97}
D.~I.~Kaiser,
%``Preheating in an expanding universe:
%Analytic results for the massless case,''
Phys.\ Rev.\ D {\bf 56}, 706 (1997).

\bibitem{Kaiser98}
D.~I.~Kaiser,
%``Resonance structure for preheating with massless fields,''
Phys.\ Rev.\ D {\bf 57}, 702 (1998).

\bibitem{Kallosh99}
R.~Kallosh, L.~Kofman, A.~D.~Linde and A.~Van Proeyen,
%``Gravitino production after inflation,''
Phys.\ Rev.\ D {\bf 61}, 103503 (2000).

\bibitem{Kallosh01}
R.~Kallosh, L.~Kofman and A.~D.~Linde,
%``Pyrotechnic universe,''
Phys.\ Rev.\ D {\bf 64}, 123523 (2001).

\bibitem{Kallosh00}
R.~Kallosh, L.~Kofman, A.~D.~Linde and A.~Van Proeyen,
%``Superconformal symmetry, supergravity and cosmology,''
Class.\ Quant.\ Grav.\  {\bf 17}, 4269 (2000)
[Erratum-ibid.\  {\bf 21}, 5017 (2004)].

\bibitem{Kana00p1}
T.~Kanazawa, M.~Kawasaki, N.~Sugiyama and T.~Yanagida,
%``Double inflation in supergravity and
%the large scale structure,''
Phys.\ Rev.\ D {\bf 61}, 023517 (2000).

\bibitem{KSW05}
S.~Kanno, J.~Soda and D.~Wands,
%``Braneworld flux inflation,''
arXiv:hep-th/0506167.

\bibitem{KK97}
S.~Kasuya and M.~Kawasaki,
%``Can topological defects be formed during preheating?,''
Phys.\ Rev.\ D {\bf 56}, 7597 (1997).

\bibitem{Kazanas80}
D.~Kazanas,
%``Dynamics Of The Universe And
%Spontaneous Symmetry Breaking,''
Astrophys.\ J.\  {\bf 241} L59 (1980).

\bibitem{Kawai99}
S.~Kawai and J.~Soda,
%``Structure formation from non-singular kinetic inflation,''
arXiv:gr-qc/9906046.

\bibitem{KM95}
M.~Kawasaki and T.~Moroi,
%``Gravitino production in the inflationary
%universe and the effects on big
%bang nucleosynthesis,''
Prog.\ Theor.\ Phys.\  {\bf 93}, 879 (1995).

\bibitem{Kawasaki01}
M.~Kawasaki and F.~Takahashi,
%``Adiabatic and isocurvature fluctuations of Affleck-Dine
% field in D-term inflation model,''
Phys.\ Lett.\ B {\bf 516}, 388 (2001).

\bibitem{Kawasaki03}
M.~Kawasaki, M.~Yamaguchi and J.~Yokoyama,
%``Inflation with a running spectral index in supergravity,''
Phys.\ Rev.\ D {\bf 68}, 023508 (2003).

\bibitem{KT1}
S.~Y.~Khlebnikov and I.~I.~Tkachev,
%``Classical decay of inflaton,''
Phys.\ Rev.\ Lett.\  {\bf 77}, 219 (1996).

\bibitem{KT2}
S.~Y.~Khlebnikov and I.~I.~Tkachev,
%``The universe after inflation:
% The wide resonance case,''
Phys.\ Lett.\ B {\bf 390}, 80 (1997).

\bibitem{KT3}
S.~Y.~Khlebnikov and I.~I.~Tkachev,
%``Resonant decay of Bose condensates,''
Phys.\ Rev.\ Lett.\  {\bf 79}, 1607 (1997).

\bibitem{Kh97}
S.~Khlebnikov,
%``Non-linear dynamics of reheating,''
arXiv:hep-ph/9708313.

\bibitem{Khoury01}
J.~Khoury, B.~A.~Ovrut, P.~J.~Steinhardt and N.~Turok,
%``The ekpyrotic universe: Colliding branes
%and the origin of the hot big bang,''
Phys.\ Rev.\ D {\bf 64}, 123522 (2001).

\bibitem{Khoury02}
J.~Khoury, B.~A.~Ovrut, N.~Seiberg, P.~J.~Steinhardt and N.~Turok,
%``From big crunch to big bang,''
Phys.\ Rev.\ D {\bf 65}, 086007 (2002).

\bibitem{KOST02}
J.~Khoury, B.~A.~Ovrut, P.~J.~Steinhardt and N.~Turok,
%``Density perturbations in the ekpyrotic scenario,''
Phys.\ Rev.\ D {\bf 66}, 046005 (2002).

\bibitem{Kinney04}
W.~H.~Kinney, E.~W.~Kolb, A.~Melchiorri and A.~Riotto,
%``WMAPping inflationary physics,''
Phys.\ Rev.\ D {\bf 69}, 103516 (2004).

\bibitem{Kodama84}
H.~Kodama and M.~Sasaki,
%``Cosmological Perturbation Theory,''
Prog.\ Theor.\ Phys.\ Suppl.\  {\bf 78}, 1 (1984).

\bibitem{KH96}
H.~Kodama and T.~Hamazaki,
Prog.\ Theor.\ Phys.\  {\bf 96}, 949 (1996).

\bibitem{Kodama00}
H.~Kodama, A.~Ishibashi and O.~Seto,
%``Brane world cosmology: Gauge-invariant
%formalism for perturbation,''
Phys.\ Rev.\ D {\bf 62}, 064022 (2000).

\bibitem{Kofman87}
L.~A.~Kofman and A.~D.~Linde,
%``Generation Of Density Perturbations
%In The Inflationary Cosmology,''
Nucl.\ Phys.\ B {\bf 282}, 555 (1987).

\bibitem{Kofman88}
L.~A.~Kofman and D.~Y.~Pogosian,
%``Nonflat Perturbations In Inflationary Cosmology,''
Phys.\ Lett.\ B {\bf 214}, 508 (1988).

\bibitem{KLS94}
L.~Kofman, A.~D.~Linde and A.~A.~Starobinsky,
%``Reheating after inflation,''
Phys.\ Rev.\ Lett.\  {\bf 73}, 3195 (1994).

\bibitem{KLS96}
L.~Kofman, A.~D.~Linde and A.~A.~Starobinsky,
%``NTPT after inflation,''
Phys.\ Rev.\ Lett.\  {\bf 76},  1011 (1996).

\bibitem{KLS97}
L.~Kofman, A.~D.~Linde and A.~A.~Starobinsky,
%``Towards the theory of reheating after inflation,''
Phys.\ Rev.\ D {\bf 56}, 3258 (1997).

\bibitem{Kofman02}
L.~Kofman and A.~Linde,
%``Problems with tachyon inflation,''
JHEP {\bf 0207}, 004 (2002).

\bibitem{Kofman03}
L.~Kofman,
%``Probing string theory with modulated
% cosmological fluctuations,''
arXiv:astro-ph/0303614.

\bibitem{Kolbbook}
E.~W.~Kolb and Turner,
{\em The Early Universe},
Addison-Wesley (1990).

\bibitem{Kolb99}
E.~W.~Kolb, arXiv:hep-ph/9910311 (1999).

\bibitem{Kolb03}
E.~W.~Kolb, A.~Notari and A.~Riotto,
%``On the reheating stage after inflation,''
Phys.\ Rev.\ D {\bf 68}, 123505 (2003).

\bibitem{KF99}
E.~Komatsu and T.~Futamase,
Phys.\ Rev.\ D {\bf 59}, 064029 (1999).

\bibitem{Komatsu01}
E.~Komatsu and D.~N.~Spergel,
Phys.\ Rev.\ D {\bf 63}, 063002 (2001).

\bibitem{Komatsu03}
E.~Komatsu {\it et al.},
%``First Year Wilkinson Microwave Anisotropy
%Probe (WMAP) Observations:
%Tests of Gaussianity,''
Astrophys.\ J.\ Suppl.\  {\bf 148}, 119 (2003).

\bibitem{Koyama00}
K.~Koyama and J.~Soda,
%``Evolution of cosmological perturbations in the brane world,''
Phys.\ Rev.\ D {\bf 62}, 123502 (2000).

\bibitem{Koyama03}
K.~Koyama,
%``CMB anisotropies in brane worlds,''
Phys.\ Rev.\ Lett.\  {\bf 91}, 221301 (2003).

\bibitem{KMW05}
K.~Koyama, S.~Mizuno and D.~Wands,
%``Slow-roll corrections to inflaton
% fluctuations on a brane,''
arXiv:hep-th/0506102.

\bibitem{Kurki-Suonio04}
H.~Kurki-Suonio, V.~Muhonen and J.~Valiviita,
%``Correlated Primordial Perturbations
% in Light of CMB and LSS Data,''
Phys.\ Rev.\ D {\bf 71}, 063005 (2005).

\bibitem{Lahiri}
J.~Lahiri and G.~Bhattacharya,
%``Cosmological perturbations in multiple-field inflation,''
arXiv:astro-ph/0506334.

\bibitem{Langlois99}
D.~Langlois,
%``Correlated Adiabatic And Isocurvature
%Perturbations From Double Inflation,''
Phys.\ Rev.\ D {\bf 59}, 123512 (1999).

\bibitem{Langlois00}
D.~Langlois and A.~Riazuelo,
%``Correlated Mixtures Of Adiabatic And Isocurvature
%Cosmological Perturbations,''
Phys.\ Rev.\ D {\bf 62}, 043504 (2000).

\bibitem{LMW00}
D.~Langlois, R.~Maartens and D.~Wands,
%``Gravitational waves from inflation on the brane,''
Phys.\ Lett.\ B {\bf 489}, 259 (2000).

\bibitem{Leach01}
S.~M.~Leach, M.~Sasaki, D.~Wands and A.~R.~Liddle,
%``Enhancement of superhorizon scale
%inflationary curvature perturbations,''
Phys.\ Rev.\ D {\bf 64}, 023512 (2001).

\bibitem{Leach02}
S.~M.~Leach, A.~R.~Liddle, J.~Martin and D.~J.~Schwarz,
%``Cosmological parameter estimation and
%the inflationary cosmology,''
Phys.\ Rev.\ D {\bf 66}, 023515 (2002).

\bibitem{Leach03}
S.~M.~Leach and A.~R.~Liddle,
%``Constraining slow-roll inflation with WMAP and 2dF,''
Phys.\ Rev.\ D {\bf 68}, 123508 (2003).

\bibitem{julien}
J.~Lesgourgues,
%``Double D-term inflation,''
Phys.\ Lett.\ B {\bf 452}, 15 (1999).

\bibitem{antony00}
A.~Lewis, A.~Challinor, and A.~Lasenby,
%``Efficient Computation of CMB
%anisotropies in closed FRW models,''
Astrophys.\ J.\  {\bf 538}, 473 (2000).

\bibitem{antony02}
A.~Lewis and S.~Bridle,
%``Cosmological parameters from CMB and other data:
% a Monte-Carlo approach,''
Phys.\ Rev.\ D{\bf 66}, 103511 (2002).

\bibitem{Liddle93}
A.~R.~Liddle and D.~H.~Lyth,
%``The Cold dark matter density perturbation,''
Phys.\ Rept.\  {\bf 231}, 1 (1993).

\bibitem{Liddleetal94}
A.~R.~Liddle, P.~Parsons and J.~D.~Barrow,
%``Formalizing the slow roll approximation in inflation,''
Phys.\ Rev.\ D {\bf 50}, 7222 (1994).

\bibitem{LLbook}
A.~R.~Liddle and D.~H. ~Lyth, {\em Cosmological inflation and
large-scale structure}, Cambridge University Press (2000).

\bibitem{Liddle99}
A.~R.~Liddle, D.~H.~Lyth, K.~A.~Malik and D.~Wands,
%``Super-horizon perturbations and preheating,''
Phys.\ Rev.\ D {\bf 61}, 103509 (2000).

\bibitem{LiddleTaylor}
A.~R.~Liddle and A.~N.~Taylor,
%``Inflaton potential reconstruction in the braneworld scenario,''
Phys.\ Rev.\ D {\bf 65}, 041301 (2002).

\bibitem{Liddle03efold}
A.~R.~Liddle and S.~M.~Leach,
%``How long before the end of inflation
%were observable perturbations produced?,''
Phys.\ Rev.\ D {\bf 68}, 103503 (2003).

\bibitem{Liddle03}
A.~R.~Liddle and A.~J.~Smith,
%``Observational constraints on braneworld chaotic inflation,''
Phys.\ Rev.\ D {\bf 68}, 061301 (2003).

\bibitem{Liddle04}
A.~R.~Liddle,
%``How many cosmological parameters?,''
Mon.\ Not.\ Roy.\ Astron.\ Soc.\  {\bf 351}, L49 (2004).

\bibitem{LLKCreview}
J.~E.~Lidsey, A.~R.~Liddle, E.~W.~Kolb, E.~J.~Copeland,
%``Reconstructing the inflaton potential: An overview,''
Rev.\ Mod.\ Phys.\  {\bf 69}, 373 (1997).

\bibitem{LWCreview}
J.~E.~Lidsey, D.~Wands and E.~J.~Copeland,
%``Superstring cosmology,''
Phys.\ Rept.\  {\bf 337}, 343 (2000).

\bibitem{Lidsey03}
J.~E.~Lidsey and N.~J.~Nunes,
%``Inflation in Gauss-Bonnet brane cosmology,''
Phys.\ Rev.\ D {\bf 67}, 103510 (2003).

\bibitem{LMMR04}
M.~Liguori, S.~Matarrese, M.~Musso and A.~Riotto,
JCAP {\bf 0408}, 011 (2004).

\bibitem{Linde82}
A.~D.~Linde,
Phys.\ Lett.\ B {\bf 108}, 389 (1982).

\bibitem{Linde83}
A.~D.~Linde,
%``Chaotic Inflation,''
Phys.\ Lett.\ B {\bf 129} 177 (1983).

\bibitem{Linde85}
A.~D.~Linde,
%``Generation Of Isothermal Density Perturbations
% In The Inflationary Universe,''
Phys.\ Lett.\ B {\bf 158}, 375 (1985).

\bibitem{Lindebook}
A.~Linde, {\em Particle Physics and Inflationary Cosmology},
Harwood, Chur (1990)
[arXiv:hep-th/0503203].

\bibitem{Linde94}
A.~D.~Linde,
%``Hybrid inflation,''
Phys.\ Rev.\ D {\bf 49}, 748 (1994).

\bibitem{Linde97a}
A.~D.~Linde and V.~Mukhanov,
%``Nongaussian isocurvature perturbations from inflation,''
Phys.\ Rev.\ D {\bf 56}, 535 (1997).

\bibitem{Linde97b}
A.~D.~Linde and A.~Riotto,
%``Hybrid inflation in supergravity,''
Phys.\ Rev.\ D {\bf 56}, 1841 (1997).

\bibitem{LindeMukhanov96}
A.~D.~Linde and V.~Mukhanov,
%``Nongaussian isocurvature perturbations from inflation,''
Phys.\ Rev.\ D {\bf 56}, 535 (1997).

\bibitem{LM85}
F.~Lucchin and S.~Matarrese,
%``Power Law Inflation,''
Phys.\ Rev.\ D {\bf 32}, 1316 (1985).

\bibitem{Lukash80}
V.~N.~Lukash,
%``Production Of Phonons In An Isotropic Universe,''
Sov.\ Phys.\ JETP {\bf 52}, 807 (1980).

\bibitem{Lyth85}
D.~H.~Lyth,
%``Large Scale Energy Density Perturbations And Inflation,''
Phys.\ Rev.\ D {\bf 31}, 1792 (1985).

\bibitem{LRreview}
D.~H.~Lyth and A.~Riotto,
%``Particle physics models of inflation and the
%cosmological density perturbation,''
Phys.\ Rept.\  {\bf 314}, 1 (1999).

\bibitem{Lyth02ekp1}
D.~H.~Lyth,
%``The primordial curvature perturbation in the ekpyrotic universe,''
Phys.\ Lett.\ B {\bf 524}, 1 (2002).

\bibitem{Lyth02ekp2}
D.~H.~Lyth,
%``The failure of cosmological perturbation
%theory in the new ekpyrotic scenario,''
Phys.\ Lett.\ B {\bf 526}, 173 (2002).

\bibitem{Lyth02}
D.~H.~Lyth and D.~Wands,
%``Generating the curvature perturbation without an inflaton,''
Phys.\ Lett.\ B {\bf 524}, 5 (2002).

\bibitem{Lyth03}
D.~H.~Lyth, C.~Ungarelli and D.~Wands,
%``The primordial density perturbation in the curvaton scenario,''
Phys.\ Rev.\ D {\bf 67}, 023503 (2003).

\bibitem{Lyth03p1}
D.~H.~Lyth and D.~Wands,
%``Conserved cosmological perturbations,''
Phys.\ Rev.\ D {\bf 68}, 103515 (2003).

\bibitem{Lyth03p2}
D.~H.~Lyth and D.~Wands,
%``The CDM isocurvature perturbation
% in the curvaton scenario,''
Phys.\ Rev.\ D {\bf 68}, 103516 (2003).

\bibitem{Lyth05}
D.~H.~Lyth, K.~A.~Malik and M.~Sasaki,
%``A general proof of the conservation of
% the curvature perturbation,''
JCAP {\bf 0505}, 004 (2005).

\bibitem{Rodriguez05}
D.~H.~Lyth and Y.~Rodriguez,
%``The inflationary prediction for primordial non-gaussianity,''
arXiv:astro-ph/0504045.

\bibitem{Maa99}
R.~Maartens, D.~Wands, B.~A.~Bassett and I.~Heard,
%``Chaotic inflation on the brane,''
Phys.\ Rev.\ D {\bf 62}, 041301 (2000).

\bibitem{Maa03}
R.~Maartens,
%``Brane-world gravity,''
Living Rev.\ Rel.\  {\bf 7}, 1 (2004).

\bibitem{Mathieu}
N.~W.~Mac Lachlan, Theory and Applications
of Mathieu Functions (Dover, New York, 1961).

\bibitem{Maeda89}
K.~i.~Maeda,
%``Towards The Einstein-Hilbert Action Via
% Conformal Transformation,''
Phys.\ Rev.\ D {\bf 39}, 3159 (1989).

\bibitem{MO04}
K.~i.~Maeda and N.~Ohta,
%``Inflation from M-theory with fourth-order
% corrections and large extra dimensions,''
Phys.\ Lett.\ B {\bf 597}, 400 (2004).

\bibitem{Maldacena02}
J.~Maldacena,
%``Non-Gaussian features of primordial fluctuations
%in single field inflationary models,''
JHEP {\bf 0305}, 013 (2003).

\bibitem{Malik02}
K.~A.~Malik, D.~Wands and C.~Ungarelli,
%``Large-scale curvature and entropy perturbations
%for multiple fluids,''
Phys.\ Rev.\ D {\bf 67}, 063516 (2003).

\bibitem{Malik04}
K.~A.~Malik and D.~Wands,
%``Adiabatic and entropy perturbations
%with interacting fluids and fields,''
JCAP {\bf 0502}, 007 (2005).

\bibitem{Malquarti}
M.~Malquarti and A.~R.~Liddle,
%``Evolution of large-scale perturbations in quintessence models,''
Phys.\ Rev.\ D {\bf 66}, 123506 (2002).

\bibitem{MM99}
A.~L.~Maroto and A.~Mazumdar,
%``Production of spin 3/2 particles from vacuum fluctuations,''
Phys.\ Rev.\ Lett.\  {\bf 84}, 1655 (2000).

\bibitem{Maroto01}
A.~L.~Maroto,
%``Primordial magnetic fields from metric perturbations,''
Phys.\ Rev.\ D {\bf 64}, 083006 (2001).

\bibitem{Martin98}
J.~Martin and D.~J.~Schwarz,
%``The influence of cosmological transitions on
%the evolution of density perturbations,''
Phys.\ Rev.\ D {\bf 57}, 3302 (1998).

\bibitem{Mata03}
S.~Matarrese and A.~Riotto,
%``Large-scale curvature perturbations with
%spatial and time variations of  the inflaton decay rate,''
JCAP {\bf 0308}, 007 (2003).

\bibitem{MMNR05}
S.~Matarrese, S.~Mollerach, A.~Notari and A.~Riotto,
%``Large-scale magnetic fields from
%density perturbations,''
Phys.\ Rev.\ D {\bf 71}, 043502 (2005).

\bibitem{Mazumda03_1}
A.~Mazumdar,
%``A model for fluctuating inflaton coupling:
%(s)neutrino induced  adiabatic perturbations
%%and non-thermal leptogenesis,''
Phys.\ Rev.\ Lett.\  {\bf 92}, 241301 (2004).

\bibitem{Mazumdar03_2}
A.~Mazumdar and M.~Postma,
%``Evolution of primordial perturbations and
%a fluctuating decay rate
Phys.\ Lett.\ B {\bf 573}, 5 (2003)
[Erratum-ibid.\ B {\bf 585}, 295 (2004)].

\bibitem{McDonald03}
J.~McDonald,
%``Right-handed sneutrinos as curvatons,''
Phys.\ Rev.\ D {\bf 68}, 043505 (2003).

\bibitem{MT04}
R.~Micha and I.~I.~Tkachev,
%``Turbulent thermalization,''
Phys.\ Rev.\ D {\bf 70}, 043538 (2004).

\bibitem{Mollerach90}
S.~Mollerach,
%``Isocurvature Baryon Perturbations And Inflation,''
Phys.\ Rev.\ D {\bf 42}, 313 (1990).

\bibitem{Moodley04}
K.~Moodley, M.~Bucher, J.~Dunkley, P.~G.~Ferreira
and C.~Skordis,
%``Constraints on isocurvature models from
% the WMAP first-year data,''
Phys.\ Rev.\ D {\bf 70}, 103520 (2004).

\bibitem{moroi95}
T.~Moroi,
%``Effects of the gravitino on the inflationary universe,''
arXiv:hep-ph/9503210.

\bibitem{Moroi01}
T.~Moroi and T.~Takahashi,
%``Effects of cosmological moduli fields
%on cosmic microwave background,''
Phys.\ Lett.\ B {\bf 522}, 215 (2001)
[Erratum-ibid.\ B {\bf 539}, 303 (2002)].

\bibitem{MTde}
T.~Moroi and T.~Takahashi,
%``Correlated isocurvature fluctuation in quintessence
%and suppressed CMB anisotropies at low multipoles,''
Phys.\ Rev.\ Lett.\  {\bf 92}, 091301 (2004).

\bibitem{Mukhanov81}
V.~F.~Mukhanov and G.~V.~Chibisov,
%``Quantum Fluctuation And 'Nonsingular' Universe.''
JETP Lett.\  {\bf 33} (1981) 532
[Pisma Zh.\ Eksp.\ Teor.\ Fiz.\  {\bf 33} (1981) 549].

\bibitem{Mukhanov85}
V.~F.~Mukhanov,
%``Gravitational Instability Of The Universe
%Filled With A Scalar Field,''
JETP Lett.\  {\bf 41}, 493 (1985)
[Pisma Zh.\ Eksp.\ Teor.\ Fiz.\  {\bf 41}, 402 (1985)].

\bibitem{Mukhanov88}
V.~F.~Mukhanov,
%``Quantum Theory Of Gauge Invariant
% Cosmological Perturbations,''
Sov.\ Phys.\ JETP {\bf 67}, 1297 (1988)
[Zh.\ Eksp.\ Teor.\ Fiz.\  {\bf 94N7}, 1 (1988)].

\bibitem{Mukhanov90}
V.~F.~Mukhanov, H.~A.~Feldman and R.~H.~Brandenberger,
 %``Theory Of Cosmological Perturbations. Part 1.
 %Classical Perturbations. Part
%2. Quantum Theory Of Perturbations. Part 3. Extensions,''
Phys.\ Rept.\  {\bf 215}, 203 (1992).

\bibitem{Mukhanov97}
V.~F.~Mukhanov and P.~J.~Steinhardt,
%``Density perturbations in multifield inflationary models,''
Phys.\ Lett.\ B {\bf 422}, 52 (1998).

\bibitem{Mukoh00}
S.~Mukohyama,
%``Gauge-invariant gravitational perturbations
%of maximally symmetric spacetimes,''
Phys.\ Rev.\ D {\bf 62}, 084015 (2000).

\bibitem{NT96}
Y.~Nambu and A.~Taruya,
Prog.\ Theor.\ Phys.\  {\bf 97}, 83 (1997).

\bibitem{NP91}
J.~V.~Narlikar and T.~Padmanabhan,
%``Inflation For Astronomers,''
Ann.\ Rev.\ Astron.\ Astrophys.\
{\bf 29} (1991) 325.

\bibitem{Nilles01}
H.~P.~Nilles, M.~Peloso and L.~Sorbo,
%``Nonthermal production of gravitinos and inflatinos,''
Phys.\ Rev.\ Lett.\  {\bf 87}, 051302 (2001).

\bibitem{Nilles01v2}
H.~P.~Nilles, M.~Peloso and L.~Sorbo,
%``Coupled fields in external background with
%application to nonthermal production of gravitinos,''
JHEP {\bf 0104}, 004 (2001).

% v2 - added citation - response to Sergei

\bibitem{NOZ00}
S.~Nojiri, S.~D.~Odintsov and S.~Zerbini,
%``Quantum (in)stability of dilatonic AdS
%backgrounds and holographic
%renormalization group with gravity,''
Phys.\ Rev.\ D {\bf 62}, 064006 (2000).

\bibitem{NO00}
S.~Nojiri and S.~D.~Odintsov,
%``Brane world inflation induced by quantum effects,''
Phys.\ Lett.\ B {\bf 484}, 119 (2000).

\bibitem{NO03}
S.~Nojiri and S.~D.~Odintsov,
%``Modified gravity with negative and positive
%powers of the curvature:
%Unification of the inflation and
%of the cosmic acceleration,''
Phys.\ Rev.\ D {\bf 68}, 123512 (2003).

\bibitem{Notari02}
A.~Notari and A.~Riotto,
%``Isocurvature perturbations in the ekpyrotic universe,''
Nucl.\ Phys.\ B {\bf 644}, 371 (2002).

\bibitem{Paddy88}
T.~Padmanabhan,
%``Acceptable Density Perturbations From
% Inflation Due To Quantum Gravitational Damping,''
Phys.\ Rev.\ Lett.\  {\bf 60}, 2229 (1988).

\bibitem{Paddy02}
T.~Padmanabhan,
%``Accelerated expansion of the universe
%driven by tachyonic matter,''
Phys.\ Rev.\ D {\bf 66}, 021301 (2002).

\bibitem{PST05}
S.~Panda, M.~Sami and S.~Tsujikawa,
%``Inflation and dark energy arising
% from geometrical tachyons,''
arXiv:hep-th/0510112.

\bibitem{Parkinson04}
D.~Parkinson, S.~Tsujikawa, B.~A.~Bassett and L.~Amendola,
%``Testing for double inflation with WMAP,''
Phys.\ Rev.\ D {\bf 71}, 063524 (2005).

\bibitem{PE98}
M.~Parry and R.~Easther,
%``Preheating and the Einstein field equations,''
Phys.\ Rev.\ D {\bf 59}, 061301 (1999).

\bibitem{Peacock}
J.~A.~Peacock,
{\em Cosmological Physics}, Cambridge University Press
(1999).

\bibitem{PV99}
P.~J.~E.~Peebles and A.~Vilenkin,
%``Quintessential inflation,''
Phys.\ Rev.\ D {\bf 59}, 063505 (1999).

\bibitem{Peiris03}
H.~V.~Peiris {\it et al.},
%``First year Wilkinson Microwave Anisotropy
%Probe (WMAP) observations: Implications for inflation,''
Astrophys.\ J.\ Suppl.\  {\bf 148}, 213 (2003).

\bibitem{PS00}
M.~Peloso and L.~Sorbo,
%``Preheating of massive fermions after
%inflation: Analytical results,''
JHEP {\bf 0005}, 016 (2000).

\bibitem{2dF}
W.~J.~Percival {\it et al.},
%``The 2dF Galaxy Redshift Survey:
%The power spectrum and the matter content of the universe,''
Mon. Not. Roy. Astron. Soc. {\bf 327}, 1297 (2001).

\bibitem{Piao03}
Y.~S.~Piao, B.~Feng and X.~m.~Zhang,
%``Suppressing CMB quadrupole with a bounce
%from contracting phase to inflation,''
Phys.\ Rev.\ D {\bf 69}, 103520 (2004).

\bibitem{Piao04}
Y.~S.~Piao, S.~Tsujikawa and X.~m.~Zhang,
%``Inflation in string-inspired cosmology and
%suppression of CMB low multipoles,''
Class.\ Quant.\ Grav.\  {\bf 21}, 4455 (2004).

\bibitem{PS02}
D.~I.~Podolsky and A.~A.~Starobinsky,
%``Chaotic reheating,''
Grav.\ Cosmol.\ Suppl.\  {\bf 8N1}, 13 (2002).

\bibitem{PFKP05}
D.~I.~Podolsky, G.~N.~Felder, L.~Kofman and M.~Peloso,
%``Equation of state and beginning of
% thermalization after preheating,''
arXiv:hep-ph/0507096.

\bibitem{Polarski92}
D.~Polarski and A.~A.~Starobinsky,
%``Spectra of perturbations produced by
%double inflation with an intermediate
%matter dominated stage,''
Nucl.\ Phys.\ B {\bf 385}, 623 (1992).

\bibitem{Polarski94}
D.~Polarski and A.~A.~Starobinsky,
%``Isocurvature perturbations in multiple inflationary models,''
Phys.\ Rev.\ D {\bf 50}, 6123 (1994).

\bibitem{Postma02}
M.~Postma,
%``The curvaton scenario in supersymmetric theories,''
Phys.\ Rev.\ D {\bf 67}, 063518 (2003).

\bibitem{Postma03}
M.~Postma and A.~Mazumdar,
JCAP {\bf 0401}, 005 (2004).

\bibitem{PR97}
T.~Prokopec and T.~G.~Roos,
%``Lattice study of classical inflaton decay,''
Phys.\ Rev.\ D {\bf 55}, 3768 (1997).

\bibitem{Quevedo02}
F.~Quevedo,
%``Lectures on string / brane cosmology,''
Class.\ Quant.\ Grav.\  {\bf 19}, 5721 (2002).

\bibitem{RC90}
A.~Rajantie and E.~J.~Copeland,
%``Phase transitions from preheating in gauge theories,''
Phys.\ Rev.\ Lett.\  {\bf 85}, 916 (2000).

\bibitem{Ramirez03}
E.~Ramirez and A.~R.~Liddle,
%``Inflationary slow-roll formalism and
%perturbations in the  Randall-Sundrum
%type II braneworld,''
Phys.\ Rev.\ D {\bf 69}, 083522 (2004).

\bibitem{Randall96}
L.~Randall, M.~Soljacic and A.~H.~Guth,
%``Supernatural Inflation: Inflation from Supersymmetry
%with No (Very) Small Parameters,''
Nucl.\ Phys.\ B {\bf 472}, 377 (1996).

\bibitem{RSI}
L.~Randall and R.~Sundrum,
%``A large mass hierarchy from a small extra dimension,''
Phys.\ Rev.\ Lett.\  {\bf 83}, 3370 (1999).

\bibitem{RSII}
L.~Randall and R.~Sundrum,
%``An alternative to compactification,''
Phys.\ Rev.\ Lett.\  {\bf 83}, 4690 (1999).

\bibitem{Rey96}
S.~J.~Rey,
%``Back reaction and graceful exit in
% string inflationary cosmology,''
Phys.\ Rev.\ Lett.\  {\bf 77}, 1929 (1996).

\bibitem{Rho03}
C.~S.~Rhodes, C.~van de Bruck, P.~Brax and A.~C.~Davis,
%``CMB anisotropies in the presence of extra dimensions,''
Phys.\ Rev.\ D {\bf 68}, 083511 (2003).

\bibitem{Rigopoulos03}
G.~I.~Rigopoulos and E.~P.~S.~Shellard,
%``The separate universe approach and the evolution of
%nonlinear  superhorizon cosmological perturbations,''
Phys.\ Rev.\ D {\bf 68}, 123518 (2003).

\bibitem{Rigopoulos}
G.~Rigopoulos,
%``On second order gauge invariant
%perturbations in multi-field inflationary models,''
Class.\ Quant.\ Grav.\  {\bf 21}, 1737 (2004).

\bibitem{RS05}
G.~I.~Rigopoulos and E.~P.~S.~Shellard,
%``Stochastic fluctuations in multi-field inflation,''
J.\ Phys.\ Conf.\ Ser.\  {\bf 8} (2005) 145.

\bibitem{Riotto02}
A.~Riotto,
%``Inflation and the theory of cosmological perturbations,''
arXiv:hep-ph/0210162.

% v2 - added citation - response to Varun

\bibitem{Sahni90}
V.~Sahni,
%``The Energy Density Of Relic Gravity
%Waves From Inflation,''
Phys.\ Rev.\ D {\bf 42}, 453 (1990).

\bibitem{Salopek89}
D.~S.~Salopek, J.~R.~Bond and J.~M.~Bardeen,
%``Designing Density Fluctuation Spectra In Inflation,''
Phys.\ Rev.\ D {\bf 40}, 1753 (1989).

\bibitem{Sami02}
M.~Sami,
%``Implementing power law inflation with
%rolling tachyon on the brane,''
Mod.\ Phys.\ Lett.\ A {\bf 18}, 691 (2003).

\bibitem{SCQ02}
M.~Sami, P.~Chingangbam and T.~Qureshi,
%``Aspects of tachyonic inflation with exponential potential,''
Phys.\ Rev.\ D {\bf 66}, 043530 (2002).

\bibitem{SS04}
M.~Sami and V.~Sahni,
%``Quintessential inflation on the brane and
%the relic gravity wave background,''
Phys.\ Rev.\ D {\bf 70}, 083513 (2004).

\bibitem{Sasaki86}
M.~Sasaki,
%``Large Scale Quantum Fluctuations
% In The Inflationary Universe,''
Prog.\ Theor.\ Phys.\  {\bf 76}, 1036 (1986).

\bibitem{Sasaki95}
M.~Sasaki and E.~D.~Stewart,
%``A General analytic formula for the spectral index of the density
%perturbations produced during inflation,''
Prog.\ Theor.\ Phys.\  {\bf 95}, 71 (1996).

\bibitem{Sasaki98}
M.~Sasaki and T.~Tanaka,
%``Super-horizon scale dynamics of multi-scalar inflation,''
Prog.\ Theor.\ Phys.\  {\bf 99}, 763 (1998).

\bibitem{Sato81p1}
K.~Sato, Mon. Not. R. Astron. Soc. {\bf 195}, 467 (1981).

\bibitem{Sato81p2}
K.~Sato, Phys. Lett. {\bf 99B}, 66 (1981).

\bibitem{Schwarz78}
G.~Schwarz, Annals of Statistics, {\bf 5}, 461 (1978).

\bibitem{Seery05a}
D.~Seery and J.~E.~Lidsey,
%``Primordial non-gaussianities in single field inflation,''
arXiv:astro-ph/0503692.
%%CITATION = ASTRO-PH 0503692;%%

\bibitem{Seery05b}
D.~Seery and J.~E.~Lidsey,
%``Primordial non-gaussianities from multiple-field inflation,''
arXiv:astro-ph/0506056.
%%CITATION = ASTRO-PH 0506056;%%

\bibitem{SMS}
T.~Shiromizu, K.~i.~Maeda and M.~Sasaki,
%``The Einstein equations on the 3-brane world,''
Phys.\ Rev.\ D {\bf 62}, 024012 (2000).

\bibitem{STB94}
Y.~Shtanov, J.~H.~Traschen and R.~H.~Brandenberger,
%``Universe reheating after inflation,''
Phys.\ Rev.\ D {\bf 51}, 5438 (1995).

\bibitem{Son}
D.~T.~Son,
%``Reheating and thermalization in a simple scalar model,''
Phys.\ Rev.\ D {\bf 54}, 3745 (1996).

\bibitem{WMAP}
D.~N.~Spergel {\it et al.},
%``First Year Wilkinson Microwave Anisotropy Probe (WMAP)
%Observations:Determination of Cosmological Parameters,''
Astrophys.\ J.\ Suppl.\  {\bf 148}, 175 (2003).


\bibitem{zurek2}
G. J. Stephens, Luis M. A. Bettencourt, W. H. Zurek, Phys. Rev. Lett. {\bf 88}, 137004 (2002)

\bibitem{SP05}
L.~Sriramkumar and T.~Padmanabhan,
%``Initial state of matter fields and
%trans-Planckian physics: Can CMB
%observations disentangle the two?,''
Phys.\ Rev.\ D {\bf 71}, 103512 (2005).


\bibitem{Star79}
A.~A.~Starobinsky,
%``Spectrum Of Relict Gravitational Radiation And
%The Early State Of The Universe,''
JETP Lett.\  {\bf 30} (1979) 682
[Pisma Zh.\ Eksp.\ Teor.\ Fiz.\  {\bf 30} (1979) 719].

\bibitem{Sta80}
A.~A.~Starobinsky,
%``A New Type Of Isotropic Cosmological
%Models Without Singularity,''
Phys.\ Lett.\ B {\bf 91} (1980) 99.


\bibitem{Sta82}
A.~A.~Starobinsky,
%``Dynamics Of Phase Transition In The New Inflationary
%Universe Scenario And Generation Of Perturbations,''
Phys.\ Lett.\ B {\bf 117} (1982) 175.

\bibitem{Sta92}
A.~A.~Starobinsky,
%``Spectrum Of Adiabatic Perturbations In The Universe
%When There Are Singularities In The Inflation Potential,''
JETP Lett.\  {\bf 55} (1992) 489
[Pisma Zh.\ Eksp.\ Teor.\ Fiz.\  {\bf 55} (1992) 477].

\bibitem{Sta94}
A.~A.~Starobinsky and J.~Yokoyama,
%``Density fluctuations in Brans-Dicke inflation,''
arXiv:gr-qc/9502002.

\bibitem{Sta01}
A.~A.~Starobinsky, S.~Tsujikawa and J.~Yokoyama,
%``Cosmological perturbations from multi-field inflation
%in generalized Einstein theories,''
Nucl.\ Phys.\ B {\bf 610}, 383 (2001).

\bibitem{Stein02}
P.~J.~Steinhardt and N.~Turok,
%``Cosmic evolution in a cyclic universe,''
Phys.\ Rev.\ D {\bf 65}, 126003 (2002).

\bibitem{Stewart93}
E.~D.~Stewart and D.~H.~Lyth,
%``A More accurate analytic calculation
%of the spectrum of cosmological
%perturbations produced during inflation,''
Phys.\ Lett.\ B {\bf 302}, 171 (1993).

\bibitem{Stewart01}
E.~D.~Stewart,
%``The spectrum of density perturbations produced
%during inflation to  leading
%order in a general slow-roll approximation,''
Phys.\ Rev.\ D {\bf 65}, 103508 (2002).

\bibitem{PBHkyoto}
T.~Suyama, T.~Tanaka, B. A.~Bassett and H.~Kudoh,
%``Are black holes over-produced during preheating?,''
Phys.\ Rev.\ D {\bf 71}, 063507 (2005).

\bibitem{PBHkyoto2}
T.~Suyama, T.~Tanaka, B. A.~Bassett and H.~Kudoh,
{\em in preparation} (2005).

\bibitem{TB03}
T.~Tanaka and B.~A.~Bassett, Proc. 12th JGRG meeting,
arXiv:astro-ph/0302544.

\bibitem{TN98}
A.~Taruya and Y.~Nambu,
%``Cosmological perturbation with two scalar fields
%in reheating after inflation,''
Phys.\ Lett.\ B {\bf 428}, 37 (1998).

\bibitem{Tegmark03}
M.~Tegmark {\it et al.}  [SDSS Collaboration],
%``Cosmological parameters from SDSS and WMAP,''
Phys.\ Rev.\ D{\bf 69}, 103501 (2004).

\bibitem{Tegmarkdata}
M.~Tegmark {\it et al.}  [SDSS Collaboration],
%``The 3D power spectrum of galaxies from the SDSS,''
Astrophys.\ J.\  {\bf 606}, 702 (2004).

\bibitem{TW05}
S.~Thomas and J.~Ward,
%``Inflation from geometrical tachyons,''
Phys.\ Rev.\ D {\bf 72}, 083519 (2005).

\bibitem{TKKL98}
I.~Tkachev, S.~Khlebnikov, L.~Kofman and A.~D.~Linde,
%``Cosmic strings from preheating,''
Phys.\ Lett.\ B {\bf 440}, 262 (1998).

\bibitem{Tolley02}
A.~J.~Tolley and N.~Turok,
%``Quantum fields in a big crunch / big bang spacetime,''
Phys.\ Rev.\ D {\bf 66}, 106005 (2002).

\bibitem{Tolley03}
A.~J.~Tolley, N.~Turok and P.~J.~Steinhardt,
%``Cosmological perturbations in a big crunch / big bang space-time,''
Phys.\ Rev.\ D {\bf 69}, 106005 (2004).

\bibitem{TB90}
J.~H.~Traschen and R.~H.~Brandenberger,
Phys.\ Rev.\ D {\bf 42}, 2491 (1990).

\bibitem{Trotta01}
R.~Trotta, A.~Riazuelo and R.~Durrer,
%``Reproducing Cosmic Microwave Background
%anisotropies with mixed isocurvature perturbations,''
Phys.\ Rev.\ Lett.\  {\bf 87}, 231301 (2001).

\bibitem{TMT99}
S.~Tsujikawa, K.~i.~Maeda and T.~Torii,
Phys.\ Rev.\ D {\bf 60}, 063515 (1999).

\bibitem{Tsujispi}
S.~Tsujikawa and T.~Torii,
%``Spinodal effect in the natural inflation model,''
Phys.\ Rev.\ D {\bf 62}, 043505 (2000).

\bibitem{BT}
S.~Tsujikawa and B.~A.~Bassett,
%``A new twist to preheating,''
Phys.\ Rev.\ D {\bf 62}, 043510 (2000).

\bibitem{Tsuji00}
S.~Tsujikawa, B.~A.~Bassett and F.~Viniegra,
%``Multi-field fermionic preheating,''
JHEP {\bf 0008}, 019 (2000).

\bibitem{TMM01}
S.~Tsujikawa, K.~i.~Maeda and S.~Mizuno,
%``Brane preheating,''
Phys.\ Rev.\ D {\bf 63}, 123511 (2001).

\bibitem{Tsujiekp}
S.~Tsujikawa,
%``Density perturbations in the ekpyrotic universe
%and string-inspired generalizations,''
Phys.\ Lett.\ B {\bf 526}, 179 (2002).

\bibitem{Tsuji02}
S.~Tsujikawa and B.~A.~Bassett,
%``When can preheating affect the CMB?,''
Phys.\ Lett.\ B {\bf 536}, 9 (2002).

\bibitem{TBF02}
S.~Tsujikawa, R.~Brandenberger and F.~Finelli,
%``On the construction of nonsingular pre-big-bang
%and ekpyrotic cosmologies and the resulting
%density perturbations,''
Phys.\ Rev.\ D {\bf 66}, 083513 (2002).

\bibitem{Tsuji03}
S.~Tsujikawa, D.~Parkinson and B.~A.~Bassett,
%``Correlation-consistency cartography
%of the double inflation landscape,''
Phys.\ Rev.\ D {\bf 67}, 083516 (2003).

\bibitem{Tsuji03per}
S.~Tsujikawa,
%``Cosmological density perturbations from perturbed couplings,''
Phys.\ Rev.\ D {\bf 68}, 083510 (2003).

\bibitem{Tsuji03noncom}
S.~Tsujikawa, R.~Maartens and R.~Brandenberger,
%``Non-commutative inflation and the CMB,''
Phys.\ Lett.\ B {\bf 574}, 141 (2003).

\bibitem{Tsujiloop}
S.~Tsujikawa, P.~Singh and R.~Maartens,
%``Loop quantum gravity effects on inflation and the CMB,''
Class.\ Quant.\ Grav.\  {\bf 21}, 5767 (2004).

\bibitem{Tsuji04}
S.~Tsujikawa and A.~R.~Liddle,
%``Constraints on braneworld inflation from CMB anisotropies,''
JCAP {\bf 0403}, 001 (2004).

\bibitem{TsujiBu}
S.~Tsujikawa and B.~Gumjudpai,
%``Density perturbations in generalized Einstein scenarios
%and constraints on nonminimal couplings from the Cosmic
% Microwave Background,''
Phys.\ Rev.\ D {\bf 69}, 123523 (2004).

\bibitem{TSM04}
S.~Tsujikawa, M.~Sami and R.~Maartens,
%``Observational constraints on braneworld inflation:
%the effect of a Gauss-Bonnet term''
Phys.\ Rev.\ D {\bf 70}, 063525 (2004).

\bibitem{Valiviita03}
J.~Valiviita and V.~Muhonen,
%``Correlated adiabatic and isocurvature CMB fluctuations
%in the wake of WMAP,''
Phys.\ Rev.\ Lett.\  {\bf 91}, 131302 (2003).

\bibitem{vande00}
C.~van de Bruck, M.~Dorca, R.~H.~Brandenberger and A.~Lukas,
%``Cosmological perturbations in brane-world theories: Formalism,''
Phys.\ Rev.\ D {\bf 62}, 123515 (2000).

\bibitem{Vene91}
G.~Veneziano,
%``Scale factor duality for classical and quantum strings,''
Phys.\ Lett.\ B {\bf 265}, 287 (1991).

\bibitem{Ver03}
F.~Vernizzi,
%``Cosmological perturbations from varying masses and couplings,''
Phys.\ Rev.\ D {\bf 69}, 083526 (2004).

\bibitem{Wands98}
D.~Wands,
%``Duality invariance of cosmological
% perturbation spectra,''
Phys.\ Rev.\ D {\bf 60}, 023507 (1999).

\bibitem{Wands00}
D.~Wands, K.~A.~Malik, D.~H.~Lyth and A.~R.~Liddle,
%``A new approach to the evolution of cosmological
%perturbations on large scales,''
Phys.\ Rev.\ D {\bf 62}, 043527 (2000).

\bibitem{Wands02}
D.~Wands, N.~Bartolo, S.~Matarrese and A.~Riotto,
%``An observational test of two-field inflation,''
Phys.\ Rev.\ D {\bf 66}, 043520 (2002).

\bibitem{Weinberg04a}
S.~Weinberg,
%``Can non-adiabatic perturbations arise
% after single-field inflation?,''
Phys.\ Rev.\ D {\bf 70}, 043541 (2004).

\bibitem{Weinberg04b}
S.~Weinberg,
%``Must cosmological perturbations remain non-adiabatic
%after multi-field inflation?,''
Phys.\ Rev.\ D {\bf 70}, 083522 (2004).

\bibitem{Yokoyama88}
J.~Yokoyama and K.~i.~Maeda,
%``On The Dynamics Of The Power Law Inflation
%Due To An Exponential Potential,''
Phys.\ Lett.\ B {\bf 207}, 31 (1988).

\bibitem{Yokoyama99}
J.~Yokoyama,
%``Chaotic new inflation and primordial
%spectrum of adiabatic fluctuations,''
Phys.\ Rev.\ D {\bf 59}, 107303 (1999).

\bibitem{Yoshi}
M.~Yoshimura,
%``Catastrophic particle production
%under periodic perturbation,''
Prog.\ Theor.\ Phys.\  {\bf 94}, 873 (1995).

\bibitem{Za03}
M.~Zaldarriaga,
%``Non-Gaussianities in models with
%a varying inflaton decay rate,''
Phys.\ Rev.\ D {\bf 69}, 043508 (2004).

\bibitem{Zanchin98}
V.~Zanchin, A.~.~J.~Maia, W.~Craig and R.~H.~Brandenberger,
%``Reheating in the presence of noise,''
Phys.\ Rev.\ D {\bf 57}, 4651 (1998).

\bibitem{Zanchin99}
V.~Zanchin, A.~.~J.~Maia, W.~Craig and R.~H.~Brandenberger,
%``Reheating in the presence of inhomogeneous noise,''
Phys.\ Rev.\ D {\bf 60}, 023505 (1999).

\bibitem{Zibin00}
J.~P.~Zibin, R.~H.~Brandenberger and D.~Scott,
%``Backreaction and the parametric resonance
% of cosmological  fluctuations,''
Phys.\ Rev.\ D {\bf 63}, 043511 (2001).

\bibitem{zurek1}
W. H. Zurek, Phys. Rept. {\bf 276},  177 (1996).

\end{thebibliography}
\end{document}